\documentclass[%
 reprint,
 superscriptaddress,
 amsmath,amssymb,
 aps,
 prx
]{revtex4-2}

\usepackage{graphicx}% Include figure files
\usepackage{dcolumn}% Align table columns on decimal point
\usepackage{bm}% bold math
\usepackage[colorlinks=true, allcolors=blue]{hyperref}
\usepackage[capitalise]{cleveref}
\usepackage{braket}
\usepackage[table]{xcolor}
\usepackage{xfrac}
\usepackage{comment}
\usepackage[normalem]{ulem}

\newcommand{\dg}{^\dagger}
\newcommand{\gkp}{\text{gkp}}
\newcommand{\cat}{\text{cat}}

\newcommand{\tr}{\text{tr}}

\renewcommand{\Im}{\text{Im}\,}
\newcommand{\hc}{\text{H.c.}}
\newcommand{\diag}{\text{diag}}

\newcommand{\XL}{\bar X}
\newcommand{\YL}{\bar Y}
\newcommand{\ZL}{\bar Z}
\newcommand{\SX}{\hat S_X}

\newcommand{\SZ}{\hat S_Z}

\begin{document}

\title{Quantum Error Correction with the Gottesman-Kitaev-Preskill Code: A Perspective}
\author{Arne L. Grimsmo}
\affiliation{ARC Centre of Excellence for Engineered Quantum Systems, School of Physics, The University of Sydney, Sydney, NSW 2006, Australia.}

\author{Shruti Puri}
\affiliation{Department of Applied Physics, Yale University, New Haven, CT 06511, USA}

\date{\today}

\begin{abstract}
The Gottesman-Kitaev-Preskill (GKP) code was proposed in 2001 by Daniel Gottesman, Alexei Kitaev, and John Preskill as a way to encode a qubit in an oscillator. The GKP codewords are coherent superpositions of periodically displaced squeezed vacuum states. Because of the challenge of merely preparing the codewords, the GKP code was for a long time considered to be impractical. However, the remarkable developments in quantum hardware and control technology in the last two decades has made the GKP code a frontrunner in the race to build practical, fault-tolerant bosonic quantum technology. In this Perspective, we provide an overview of the GKP code with emphasis on its implementation in the circuit-QED architecture  and present our outlook on the challenges and opportunities for scaling it up for hardware-efficient, fault-tolerant quantum error correction.

\end{abstract}

\maketitle

%\tableofcontents

\section{\label{sec:intro}Introduction}

%\alg{These are just ideas, I like to get a rough intro down on paper to help me formulate the ideas, I'm perfectly happy to delete anything I have written here.}

In 2001, Gottesman, Kitaev, and Preskill published a proposal to encode discrete quantum information in a continuous variable quantum system, or in other words, ``a qubit in an oscillator''~\cite{Gottesman01}. The encoding is designed such that it is possible to correct small shifts in the position and momentum quadratures of the oscillator. This remarkable idea can safely be said to have been ahead of its time: It took almost twenty years before the first experimental realization of their proposal was made by the Home group using an oscillating trapped ion~\cite{Fluhmann:2019aa}. It was almost immediately followed by the experimental realization  in the Devoret group, using a microwave cavity and a circuit quantum electrodynamics (cQED) approach~\cite{Campagne2020}. More broadly, these two experiments are part of a flourishing effort to demonstrate robust encoding of quantum information using bosonic degrees of freedom, with the ultimate long term goal of building a fault-tolerant quantum computer~\cite{Ofek16,Hu:2019aa,heeres2017implementing,xu2020demonstration,Ma2020,gertler2021protecting,de2020error}.
%  It  took almost  twenty  years  before  the  first  experimental  realizations of their proposal, first by the Home group using an oscillating trapped ion [2], and not long after by the Devoret  group,  using  a  microwave  cavity  and  a  circuit quantum  electrodynamics  (cQED)  approach  [3].

The idea of encoding information in a continuous variable quantum system is in many ways very natural. Quantum harmonic oscillators abound in nature, and well-defined bosonic modes can be isolated from environmental noise in many quantum technology platforms.
Early proposals for bosonic error correcting codes were made already in the late 90s~\cite{ChuaLeunYama97,CochMilbMunr99}. The core idea behind these proposals is to encode $k$ logical qubits into $n$ bosonic modes, and attempt to exploit the large Hilbert space of each bosonic mode to achieve an efficient encoding with good error correcting properties for a small $n$. Successful bosonic codes are referred to as \emph{hardware efficient}. Remarkably, interesting bosonic codes, and the Gottesman-Kitaev-Preskill (GKP) code in particular, exist even for $k=n=1$. 

%The number of correctible errors in the single-mode GKP code depends on its size or average photon number. That is, it is possible to arbitrarily reduce the probability  of error in a GKP state encoded in a single bosonic mode by simply making a GKP state  with a larger average photon number. 
%The GKP encoding is not only hardware efficient, but the error syndromes are relatively simple to decode compared to qubit-based encodings. 

%Today, the GKP code is seen as one of the most promising bosonic encodings.
%In fact, as we will explain later, many of the other bosonic codes that have been introduced since can be seen as natural generalizations of the GKP code. The GKP code is therefore something of a ``canonical example'' of a bosonic code.
The primary requirement for implementing bosonic codes, and also why it took two decades of technological developments before the GKP states were realized, is control of a hiqh-quality harmonic oscillator mode with a sufficiently strong and high-quality ancillary non-linearity. This nonlinearity can be a discrete two-level system. In case of trapped-ions, a bosonic state is encoded in the harmonic motion of a single trapped ion by exploiting the strong coupling with the ancillary atomic pseudospin states. In case of cQED, the ancillary levels of a transmon have been used to realize a bosonic encoding in the microwave fields of a superconducting cavity or resonator. Harmonic modes with high quality factors, combined with easy access to strong non-linearities with minimal dissipation, lead to an unprecedented coherent control over the oscillator Hilbert space in these platforms~\cite{Ofek16,Hu:2019aa,heeres2017implementing,xu2020demonstration,Fluhmann:2018aa,Fluhmann:2019aa,Campagne2020,de2020error}.

Optical systems are also researched actively in the context of GKP codes~\cite{takeda2019toward,Walshe2020,bourassa2021blueprint,larsen2021fault}.
Some proposals are based on using optical nonlinearities or interaction between atoms and light to generate photonic GKP states~\cite{pirandola2004constructing,MoteBaraGilc17}. Several other proposals rely on photon number resolving detectors as the ancillary nonlinear resource~\cite{Eaton:2019aa,Tzitrin2020}.  However, because of photon loss, weak optical nonlinearities, and the need for complex multiplexing with high efficiency number resolving detectors, generation of the highly non-Gaussian GKP states have not yet been demonstrated in the optical domain.

Given that preparation of GKP encoded states has now been demonstrated in the lab---and it is not a big leap to imagine that gates between two encoded GKP qubits are right around the corner---it is natural to ask whether GKP encodings can become a competitive approach to large-scale, fault-tolerant quantum computing. While the GKP code can correct for small quadrature shifts in the oscillator, realistic noise in an experimental platform is more complex and can introduce uncorrectable errors. Therefore, in practice the suppression in the logical error rate with a  single-mode GKP code will be limited. A natural approach to ``scale up'' is to reduce errors as much as possible in the single-mode encoding and then concatenate a number of encoded GKP qubits to a second error correcting code, for example a surface code, for a total of one logical qubit across $n$ physical modes~\cite{Vuillot2018,Noh:2019aa,Terhal2020,noh2021low}. If this approach leads to a substantially lower logical error rate than using a comparable number of un-encoded physical qubits, one may achieve a better encoding with a similar hardware cost.

A first milestone towards this goal of resource-efficient fault-tolerance would be to demonstrate basic operations on the encoded GKP states, used to compose error-correction circuits, with fidelities that are comparable or better than the best physical qubits to date. These operations include state preparation, entangling gates between two GKP-encoded modes, and measurement. This is a challenging goal considering the high fidelity qubit operations in both trapped ions and superconducting qubits today.

%Once sufficiently high fidelity operations are achieved, further improvements in the performance of the error correcting code are possible, for example by using analog information~\cite{Fukui:2017aa} in or by tailoring the noise channel of the GKP encoding~\cite{hanggli2020enhanced}.

A fundamental obstacle to achieving high-fidelity operations on encoded GKP states is that practical constructions of the required interactions can ruin the protection offered by the bosonic encoding. For example, if a two-level system is used to control the oscillator mode, a single error on the two level system may propagate to a logical error on the mode~\cite{Fluhmann:2019aa,Campagne2020,de2020error}. This can prohibit the fidelity of encoded operations on the GKP states from being significantly better those of the unencoded two-level ancilla. How to best achieve fault-tolerance against such ancilla errors in a bosonic code architecture is an important open question, and we will touch on some of the possibilities that have been put forth towards this goal. %Note that, while a large majority of the work has focused on concatenating GKP code to well studied topological codes like the surface code, this is by no means the only path towards scalability. It is possible that a better scheme where $k$ logical qubits are encoded in $n$ physical modes more directly, i.e., without concatenation with a conventional binary code, could exist. This avenue is still largely unexplored.

In this Perspective, we will discuss the prospect of scalable, fault-tolerant quantum computing with GKP codes with special emphasis on its implementation in a cQED architecture. While there are several excellent review articles on GKP and other bosonic codes~\cite{cai2021bosonic,Terhal2020,ma2021quantum,joshi2021quantum}, here we provide an application-level perspective highlighting the outstanding practical challenges. We focus on cQED partly because the two authors are working in this field, but also because we believe the flexibility and scalability of superconducting circuits make this a particularly promising platform for the long term goal of constructing a large scale quantum computer based on bosonic encodings. 

With this in mind we begin with an overview of the GKP code in~\cref{sec:intro_gkp}, go on to discuss state preparation and error correction in~\cref{sec:stateprep}, and
address the question of fault-tolerant, scalable quantum computing with GKP codes in~\cref{sec:scaling}. Throughout this article, we emphasize not only the advances made towards GKP error correction but also the challenges that must be overcome to make practical fault-tolerance with GKP codes possible. These challenges and opportunities for future research are summarized in~\ref{sec:summary}.

\section{Introduction to Gottesman-Kitaev-Preskill codes \label{sec:intro_gkp}}

\subsection{\label{sec:gkpdefs}Basic definitions}
In general, GKP codes encode a $d$ dimensional logical subspace in $n$ bosonic modes~\cite{Gottesman01}. We will here focus exclusively on the simplest nontrivial case $d=2$ and $n=1$, i.e., a single logical qubit encoded in a single bosonic mode. To define a GKP code, it is first convenient to introduce the displacement operators $\hat D(\alpha) = e^{\alpha \hat a\dg - \alpha^* \hat a}$, where $[\hat a,\hat a\dg] = 1$ are the usual ladder operators of a harmonic oscillator and $\alpha$ is a complex number. The displacement operators satisfy the property
\begin{equation}\label{eq:dispcomm}
    \begin{aligned}
    \hat D(\beta) \hat D(\alpha)
    &= e^{(\beta \alpha^* - \beta^* \alpha )/2} \hat D(\alpha + \beta)\\
    &= e^{\beta \alpha^* - \beta^* \alpha} \hat D(\alpha) \hat D(\beta).
    \end{aligned}
\end{equation}
In other words, displacements commute ``up to a phase.'' In particular, if
\begin{equation}\label{eq:unitarea}
    \beta \alpha^* - \beta^* \alpha = i\pi,
\end{equation}
the two operators anti-commute, while if $\alpha\beta^* - \alpha \beta^* = 2i\pi$ they commute.

To define a GKP code, we first choose logical Pauli operators $\XL = \hat D(\alpha)$ and $\ZL = \hat D(\beta)$, where $\alpha$ and $\beta$ are any two complex numbers that satisfy~\cref{eq:unitarea}. This ensures that $\XL\ZL = - \ZL\XL$. To ensure that $\XL$, $\ZL$, and $\YL = i\XL\ZL = \hat D(\alpha + \beta)$ behave like the usual two-by-two Pauli matrices, they should also square to the identity on any state in the code subspace (codespace). We therefore \emph{define} the GKP logical codespace to be the simultaneous $+1$ eigenspace of the two operators
\begin{equation}\label{eq:gkp_stab_and_pauli}
    \SX = \XL^2 = \hat D(2\alpha), \quad \SZ = \ZL^2 = \hat D(2\beta).
\end{equation}
It follows from~\cref{eq:dispcomm} that these two operators commute with each other, and the logical Paulis. The set $\{\SX^k, \SZ^l\}$ for $k,l \in \mathbb{Z}$ form the stabilizer group of the GKP code.

We can write the GKP codewords explicitly in terms of sums of quadrature eigenstates. To this end, first define two generalized quadratures $\hat Q = i(\beta^* \hat a - \beta \hat a\dg)/\sqrt{\pi}$, $\hat P = -i(\alpha^* \hat a - \alpha \hat a\dg)/\sqrt{\pi}$, such that $[\hat Q, \hat P]=i$ and 
\begin{equation}\label{eq:paulis}
\XL = e^{-i\sqrt{\pi}\hat P},\, \ZL = e^{i\sqrt{\pi}\hat Q},\, \YL = e^{i\sqrt{\pi}(\hat Q - \hat P)}.
\end{equation}
%generates translations of the $\hat Q$ ($\hat P$) quadrature by $\sqrt\pi$.
It is straight forward to check that
\begin{subequations}\label{eq:gkpcodewords2}
\begin{align}
    \ket{0_L} &= \sum_{j=-\infty}^\infty \ket{2j \sqrt{\pi}}_{\hat Q},\\
    \ket{1_L} &= \sum_{j=-\infty}^\infty \ket{(2j+1) \sqrt{\pi}}_{\hat Q},
\end{align}
\end{subequations}
are $\pm 1$ eigenstates of $\ZL$, respectively, and $+1$ eigenstates of $\SX$ and $\SZ$. Here we use a notation where $\ket{x}_{\hat O}$ is an eigenstate of $\hat O$ with eigenvalue $x$.
We have analogous expressions in the dual basis: $\ket{+_L} = \sum_j \ket{2j\sqrt{\pi}}_{\hat P}$, $\ket{-_L} = \sum_j \ket{(2j+1)\sqrt{\pi}}_{\hat P}$.

An alternative expression for the codewords can be found by noting that the state
%\begin{equation}
$\ket{0_L} \propto \sum_{k,l=-\infty}^\infty \SX^k \ZL^{l} \ket 0$,
%\end{equation}
is a simultaneous $+1$ eigenstate of the two stabilizer generators $\SX, \SZ$ and logical $\ZL$ (in fact, the vacuum state $\ket 0$ can be replaced by an arbitrary state $\ket\psi$ with non-zero overlap with $\ket{0_L}$ in this expression). With the help of~\cref{eq:dispcomm}, it follows that the logical states can be written
\begin{subequations}\label{eq:gkpcodewords}
\begin{align}
    \ket{0_L} &\propto \sum_{k,l=-\infty}^\infty e^{-i\pi kl} \ket{2k\alpha + l \beta},\\
    \ket{1_L} &\propto \sum_{k,l=-\infty}^\infty e^{-i\pi (kl + l/2)} \ket{(2k+1)\alpha + l\beta},
\end{align}
\end{subequations}
where the kets on the right hand side are coherent states $\ket\zeta = \hat D(\zeta)\ket 0$. Analogous expressions can be found for the $\pm 1$ eigenstates of $\XL$ following the same approach.

\begin{figure}
    \centering
    \includegraphics{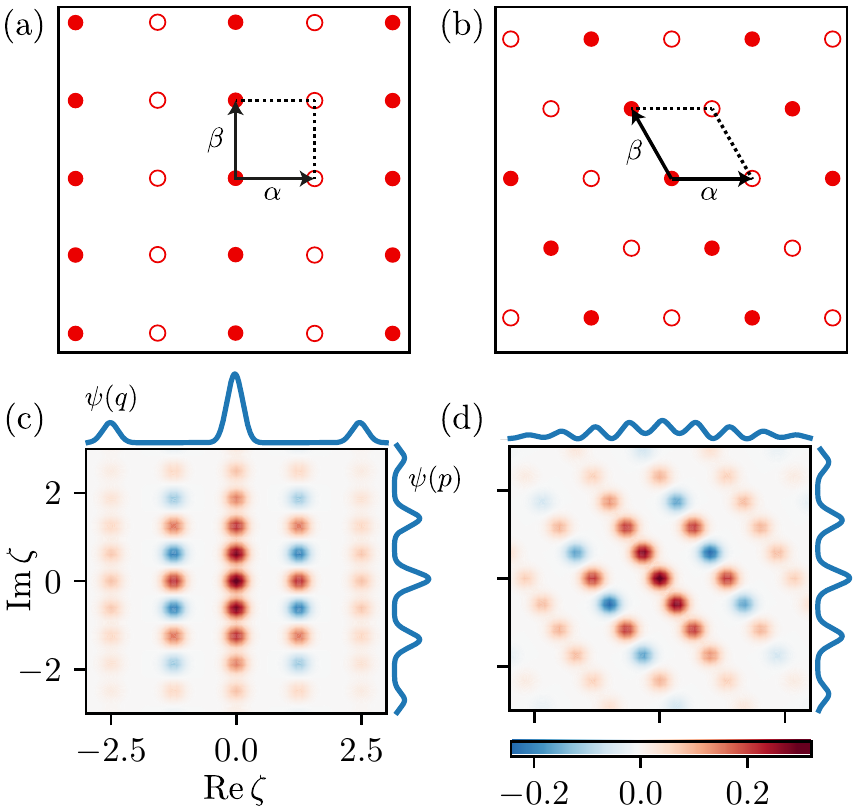}
    \caption{
    Square (a) and hexagonal (b) GKP lattices (adapted from Ref.~\cite{Grimsmo2020}). In~\cref{eq:gkpcodewords} coherent states are placed on the filled (empty) circles for the $\ket{0_L}$ ($\ket{1_L}$) state. The area of the lattice parallelogram is $\pi/2$, which ensures the correct anticommutation relation for the logical operators defined in~\cref{eq:gkp_stab_and_pauli}. Wigner functions $W(\zeta)$ for approximate $\ket{\tilde 0_L}$ states with $\Delta = 0.3$ are shown for square (c) and hexagonal (d) GKP codes. Position [$\hat q=(\hat a\dg + \hat a)/\sqrt 2$] and momentum [$\hat p = i(\hat a\dg - \hat a)/\sqrt 2$] wavefunctions are shown in blue.
    }
    \label{fig:gkplattices}
\end{figure}

Since any pair $\alpha,\beta$ that satisfy~\cref{eq:unitarea} is valid choice, there is an infinity of different GKP codes. The three most common choices are square, rectangular, and hexagonal codes, defined respectively by
\begin{subequations}
\begin{align}
\text{square: } &\alpha = \sqrt{\frac{\pi}{2}}, \quad \,\,\,\,\beta = i\sqrt{\frac{\pi}{2}},\\ %\sqrt{\frac{\pi}{2}},\\
\text{rect: } &\alpha = {\lambda}\sqrt{\frac{\pi}{2}}, \quad \,\,\beta = \frac{i}{{\lambda}}\sqrt{\frac{\pi}{2}}, \quad {\lambda} > 0,\label{eq:lattice_def}
\\
\text{hex: } &\alpha = \sqrt{\frac{\pi}{\sqrt 3}}, \quad \beta = e^{2i\pi/3}\sqrt{\frac{\pi}{\sqrt 3}}. %\sqrt{\frac{\pi}{\sqrt 3}}.
\end{align}
\end{subequations}
Note that for the square code, the generalized quadratures introduced above are just the usual position and momentum quadratures $\hat Q_\square = \hat q = (\hat a + \hat a\dg)/\sqrt 2$, $\hat P_\square = \hat p = -i(\hat a - \hat a\dg)/\sqrt 2$. The square and hexagonal GKP lattices are illustrated in~\cref{fig:gkplattices}.

\subsection{\label{sec:approxgkp}Approximate GKP codewords}

The GKP codespace is a rather abstract construction. The codewords in~\cref{eq:gkpcodewords2} are non-normalizable, and in general there is no physical process that can prepare a state lying entirely in the GKP codespace. In practice we have to make do with some type of approximation to~\cref{eq:gkpcodewords2,eq:gkpcodewords}.
Colloquially, we refer to any pair of normalized states $\ket{\tilde \mu_L}$, $\mu=0,1$ that satisfy $\hat S_P \ket{\tilde \mu_L} \to \ket{\tilde \mu_L}$ and $\ZL \ket{\tilde \mu_L} \to (-1)^\mu \ket{\tilde \mu_L}$ for $P=X,Z$, in some meaningful limit, as an approximate GKP code.

One natural way to define such an approximate code is~\cite{Menicucci14}
%to introduce a Gaussian envelope over the infinite sums in~\cref{eq:gkpcodewords}
%\begin{subequations}\label{eq:gkpcodewords_approx1}
%\begin{align}
%    \ket{\tilde 0_L} &\propto \sum_{k,l=-\infty}^\infty e^{-\Delta^2 |\zeta_{2k,l}|^2} e^{-i\pi kl} \ket{\zeta_{2k,l}},\\
%    \ket{\tilde 1_L} &\propto \sum_{k,l=-\infty}^\infty e^{-\Delta^2 |\zeta_{2k+1,l}|^2} e^{-i\pi (kl + l/2)} \ket{\zeta_{2k+1,l}},
%\end{align}
%\end{subequations}
\begin{equation}\label{eq:gkpcodewords_approx1}
    \ket{\tilde \mu_L} \propto e^{-\Delta^2 \hat a\dg \hat a} \ket{\mu_L},\\
    %\ket{\tilde 1_L} &\propto e^{-\Delta^2 \hat a\dg \hat a} \ket{1_L},
\end{equation}
%for $\mu=0,1$, and
where we ignore normalization constants, for simplicity. The ideal limit corresponds to $\Delta \to 0$.
\Cref{eq:gkpcodewords_approx1} introduces a Gaussian envelope over the infinite sums in~\cref{eq:gkpcodewords},
such that each coherent state $\ket\zeta$ is replaced by $e^{-\frac12\left(1-e^{-2\Delta^2}\right)|\zeta|^2}\ket{e^{-\Delta^2} \zeta} \simeq e^{-\Delta^2|\zeta|^2}\ket{e^{-\Delta^2} \zeta}$, which ensures the normalizability of the codewords. The $\ket{\tilde 0_L}$ codewords for square and hexagonal GKP codes with $\Delta=0.3$ are shown in~\cref{fig:gkplattices}. It is also possible to view the code defined in~\cref{eq:gkpcodewords_approx1} as resulting from a modificaiton of the stabilizers defining the codespace. More precisely, the codewords in~\cref{eq:gkpcodewords_approx1} are exact $+1$ eigenstates of the two commuting nonunitary operators
\begin{equation}
    S_{X,Z}^\Delta = e^{-\Delta^2 \hat a\dg \hat a} S_{X,Z} e^{\Delta^2 \hat a\dg \hat a},
    %\quad \SZ^\Delta = e^{-\Delta^2 \hat a\dg \hat a} \SZ e^{\Delta^2 \hat a\dg \hat a},
\end{equation}
with logical operators defined analogously, $\bar P^\Delta = e^{-\Delta^2 \hat a\dg \hat a} \bar P e^{\Delta^2 \hat a\dg \hat a}$ for $\bar P = \XL, \ZL$~\cite{Royer2020}.

Another common approximate form is found by applying weighted displacements to the ideal codewords
\begin{equation}\label{eq:gkpcodewords_approx2}
    \begin{aligned}
    \ket{\tilde \mu_L} 
    %&\propto \int d^2\zeta \, \eta(\sqrt{2}\Re\zeta,\sqrt{2}\Im\zeta) \hat D(\zeta) \ket{\mu_L},\\
    &= \int_{-\infty}^\infty du dv\, \eta_\Delta\left(u,v\right) e^{\frac{iuv}{2}} e^{-iu\hat P} e^{iv \hat Q} \ket{\mu_L},
    \end{aligned}
\end{equation}
where $\eta_\Delta(u,v)$ is concentrated around zero as $\Delta \to 0$.
\Cref{eq:gkpcodewords_approx1} is recovered with
$\eta_\Delta(x,y) = e^{-(x^2+y^2)/(4\tanh(\Delta^2/2)} / [\pi(1-e^{-\Delta^2})] \simeq e^{-(x^2+y^2)/2 \Delta^2} / \pi \Delta^2$~\cite{Royer2020}.
We can use~\cref{eq:gkpcodewords2} in~\cref{eq:gkpcodewords_approx2} and perform the integral over $v$ to find yet another approximation~\cite{Gottesman01}
\begin{equation}\label{eq:gkpcodewords_approx3}
    \begin{aligned}
    \ket{\tilde \mu_L}
    % &\simeq \sqrt{\frac{2}{\sqrt \pi}} \sum_{j=-\infty}^\infty e^{-\Delta^2 \pi (2j+\mu)^2 / 2} \\
    &\simeq \frac{1}{\sqrt{N_\mu}} \sum_{j=-\infty}^\infty e^{-\Delta^2 \pi (2j+\mu)^2 / 2} \\
    %&\propto \sum_{j=-\infty}^\infty e^{-\Delta^2 \pi (2j+\mu)^2 / 2} \\
    &{} \times \int_{-\infty}^\infty du e^{-\frac{u^2}{2\Delta^2}} \ket{(2j+\mu)\sqrt{\pi}+u}_{\hat Q},
    %&\simeq \frac{1}{\sqrt{N_\mu}}\int_{-\infty}^\infty du e^{-\Delta^2 u^2/2 + \cos(\sqrt\pi u)/\pi\Delta^2} \ket{(\mu\sqrt{\pi}+u}_{\hat Q}
    \end{aligned}
\end{equation}
where $N_\mu = \sqrt{\pi}/2 + \mathcal O\left(e^{-\pi/\Delta^2}\right)$ as $\Delta \to 0$.
This form has the physical interpretation of a comb of squeezed states with an overall Gaussian envelope.

It is convenient to introduce a metric to quantify how close an arbitrary state $\hat\rho$ is to an ideal GKP state. To this end, we introduce a modular ``squeezing'' parameter for each of the two stabilizers $\hat S_{X,Z}$~\cite{Duivenvoorden2017}
\begin{subequations}
\begin{align}
    \Delta_{X} ={}& \frac{1}{2|\alpha|}\sqrt{-\log(|\tr [\hat S_{X} \hat \rho]|^2)},\\
    \Delta_{Z} ={}& \frac{1}{{2|\beta|}}\sqrt{-\log(|\tr [\hat S_{Z} \hat \rho]|^2)}.
\end{align}
\end{subequations}
The squeezing parameters satisfy $\Delta_{X,Z} \ge 0$, and are zero if and only if the state is an eigenstate of the corresponding stabilizer. For the approximate GKP codewords introduced above, we have $\Delta_{X,Z} = \Delta$ as $\Delta \to 0$.
It is also conventional to measure the modular squeezing in dB
\begin{equation}\label{eq:squeezingdB}
    \mathcal S_{X,Z} = - 10  \log_{10}(\Delta_{X,Z}^2).
\end{equation}
For example, the square (hexagonal) approximate codeword shown in~\cref{fig:gkplattices} has $\mathcal S_X = \mathcal S_Z = 10.1$ dB ($9.48$ dB), corresponding to an average photon number of approximately $\braket{\hat n} = 4.6$. We will from now on drop the subscript $X,Z$ and simply write $\Delta$ and $\mathcal S$ when the two quadratures are approximately equally squeezed and the distinction is unimportant.

\subsection{\label{sec:qec_criteria}Error correcting properties}

The purpose of encoding a logical qubit in a GKP code is that it provides protection from noise through error correction. To see this,
consider first an error model consisting of small displacements applied to the oscillator. For an arbitrary displacement $\hat D(\zeta)$ we can write $\zeta = (u \alpha + v\beta)/\sqrt \pi$, where $u,v$ are real, and consequently
%$\hat D(\zeta) = e^{iuv / 2} \hat D(u\alpha/\sqrt\pi) \hat D(v\beta/\sqrt\pi) = e^{iuv / 2} e^{-iu\hat P} e^{iv \hat Q}$.
\begin{equation}\label{eq:uvdisplacement}
\hat D(\zeta) = e^{iuv / 2} e^{-iu\hat P} e^{iv \hat Q}.
\end{equation}
From~\cref{eq:gkpcodewords2} it is clear that
%$\braket{\mu_L|\hat D\dg(\zeta') \hat D(\zeta)|\nu_L} = 0$ for $\mu,\nu = 0,1$, for any $\zeta'=u'\alpha + v'\beta$, $\zeta=u\alpha+v\beta$ that satisfy $|u-u'| \bmod{1} < 1$ and $|v-v'| \bmod{1} < 1$. Thus,
the quantum error correction criteria~\cite{Nielsen10} are formally satisfied for the ideal GKP code for the set of displacement errors $\hat D(\zeta)$ such that $|u|, |v| < \sqrt \pi / 2$.

This also gives us some insight into why the approximate GKP codewords introduced in the previous section are ``good'' approximations. As long as $\eta_\Delta(u,v)$ in~\cref{eq:gkpcodewords_approx2} is sufficiently localized around zero, the ``error'' introduced in the approximate codewords is small, and as long as we ensure that all the logical operations used in our quantum computation do not amplify these errors too badly, i.e., they are ``fault-tolerant,'' then we can expect to perform quantum computation with these approximate GKP codewords with high accuracy (see~\cref{sec:errorspread} for a more precise discussion around what we mean by not ``too badly'').

%For physical GKP codes we expect that the error correction criteria are only approximately satisfied. Using e.g. \cref{eq:gkpcodewords_approx3}, one can show that to leading order in $\Delta$ and $|u|,|v| < \sqrt\pi$ \alg{These are not quite right}
%\begin{subequations}\label{eq:qec_criteria}
%\begin{align}
%&\braket{\tilde\mu_L|e^{-iu\hat P} e^{iv\hat Q}|\tilde\mu_L} \simeq e^{-\frac{u^2 + v^2 }{4\Delta^2} + (-1)^\mu \frac{i uv}{2}},\\
%&\begin{aligned}
%\braket{\tilde 1_L|e^{-iu\hat P} e^{iv\hat Q}|\tilde 0_L} \simeq{}& e^{-\frac{(u-\sqrt{\pi})^2 + v^2 }{4\Delta^2} + \frac{i (u-\sqrt\pi)v}{2}} \\
%&+ e^{-\frac{(u+\sqrt{\pi})^2 + v^2 }{4\Delta^2} - \frac{i uv}{2}}
%\end{aligned}
%\end{align}
%\end{subequations}
%Thus, we see that the error correction criteria are satisfied for small displacements, up to exponentially small corrections as $\Delta \to 0$.

For physical GKP codes and realistic error models, we expect that the error correction criteria are at best only approximately satisfied.
Realistic error models for oscillators typically include loss, heating, dephasing, unitary errors due to imperfect implementation of control Hamiltonians, etc. Since the displacement operators form an operator basis, any single-mode noise channel can be expanded in terms of displacements
\begin{equation}
    \mathcal E(\hat \rho) = \int d^2 \zeta d^2\zeta' f(\zeta,\zeta') \hat D(\zeta) \hat \rho \hat D\dg(\zeta').
    \label{eq:channel}
\end{equation}
Again, as long as $f(\zeta,\zeta')$ is sufficiently concentrated around zero, it is in principle possible to remove the noise with high fidelity. Realistic error models, however, typically have some finite support on displacements larger than $\sqrt\pi/2$, which means that the error can not be corrected perfectly, even in the limit $\Delta\to 0$. %Errors may be reduced further through concatenation with a conventional binary quantum error correction code.

\begin{figure}
    \centering
    \includegraphics{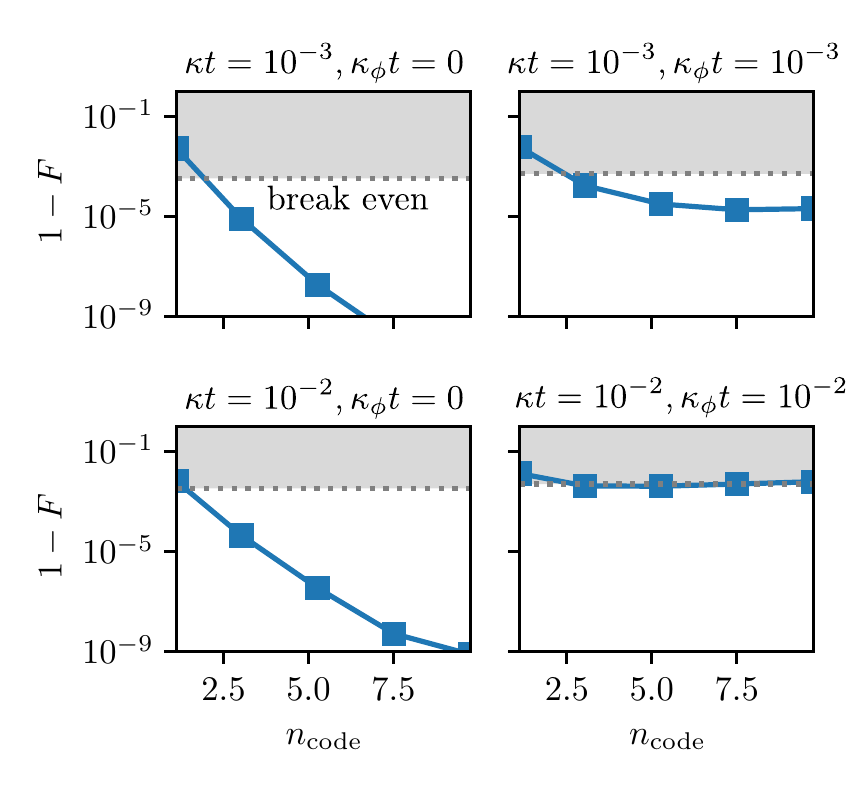}
    \caption{Average gate fidelity for an approximate square GKP code as a function of photon number $n_\text{code} = (\braket{0_L|\hat n|0_L} + \braket{1_L|\hat n|1_L})/2$, for simultaneous loss and dephasing quantified by $\kappa t$ and $\kappa_\phi t$, followed by optimal error correction. The GKP code outperforms the ``trivial encoding'' in Fock states $\ket 0$ and $\ket 1$ whenever the fidelity dips below the gray shaded region.}
    \label{fig:qecfidelity}
\end{figure}

In~\cref{fig:qecfidelity} we illustrate how the quantum error correction properties of the GKP code manifest for a practically relevant noise channel consisting of simultaneous loss and dephasing. More precisely, the noise model is given by the solution to a Lindblad master equation
\begin{equation}
    \dot{\hat \rho} = \kappa \mathcal D[\hat a]\hat \rho + \kappa_\phi \mathcal D[\hat n]\hat\rho,
\end{equation}
with $\mathcal D[\hat A]\hat\rho = \hat A \hat\rho \hat A\dg - \frac12 \hat A\dg \hat A \hat \rho - \frac12 \hat\rho \hat A\dg \hat A$, integrated up to a fixed time $t$. The noise strength in this model is thus characterized by two dimensionless numbers, $\kappa t$ and $\kappa_\phi t$, describing pure loss and dephasing, respectively. The noise is then followed by the optimal recovery channel that maximises the average gate fidelity~\cite{Nielsen02} with the identity channel. This optimal error correction map can be found numerically~\cite{Albert17}, but does not represent a practical error correction procedure---it merely puts an upper bound on the fidelity that can be achieved and illustrates the \emph{intrinsic} error correction properties of the code.

The results in~\cref{fig:qecfidelity} shows that the GKP code has an excellent potential to correct loss errors~\cite{Albert17}, but is rather poor against dephasing (there are other bosonic codes that perform far better against this latter type of noise~\cite{Grimsmo2020}). The sensitivity to dephasing is not surprising, as a rotation of phase space by a small angle gives a large displacement for large amplitudes.
In practice, dephasing might arise not only due to the intrinsic frequency fluctuations of the oscillator used to encode the GKP code (which can be made very small), but also due to off-resonant coupling to ancilliary quantum systems used to control the oscillator~\cite{Ofek16,Hu:2019aa,Campagne2020}. It is therefore crucial to minimize such residual couplings in practical implementations of GKP codes. A similar issue is likely to arise if there are over-rotations and/or unwanted residual Hamiltonian terms due to miscalibrated unitary gates, and very precise quantum control is therefore important for GKP codes.

\subsection{\label{sec:logicalops}Logical operations on GKP codes}

One of the attractive properties of GKP codes is that, apart from state preparation, all logical Clifford operations can be performed using only Gaussian operations, that is, interactions that are at most quadratic in creation and annihilation operators and homodyne measurements on the oscillator. In this section we describe how to perform logical Pauli measurements and unitary Clifford gates, leaving the more difficult topic of state preparation and error correction to~\cref{sec:stateprep}.

\subsubsection{\label{sec:measurements}Pauli quadrature measurements}

\begin{figure}
    \centering
    \includegraphics{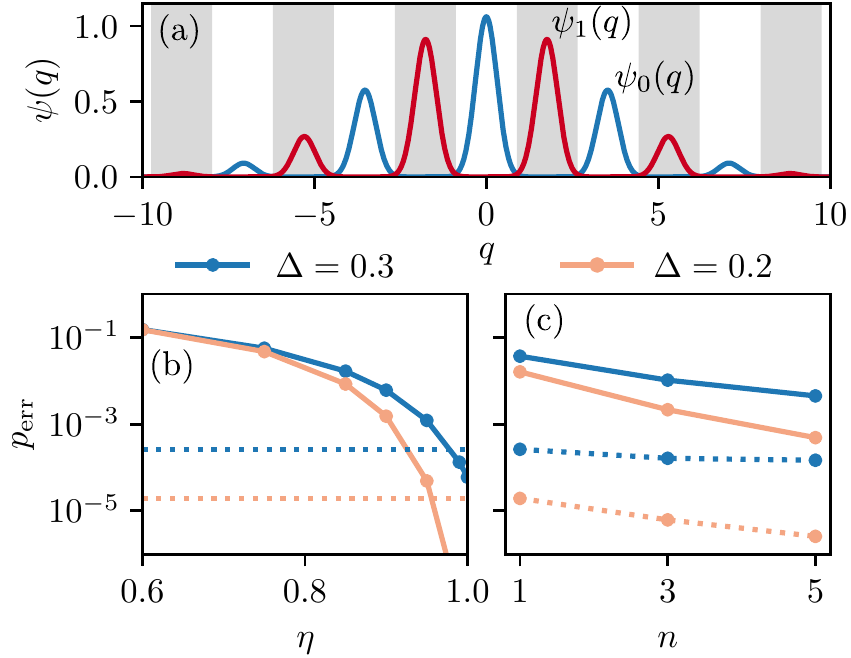}
    \caption{(a) Position wavefunctions for logical zero (blue) and one (red) for a square GKP code with $\Delta=0.3$. A logical $\mathcal M_Z$ measurement can be executed by first measuring position (homodyne measurement) and binning the result. The white (gray) regions correspond to a logical zero (one) outcome. (b) Probability of mistaking logical zero for one under a homodyne measurement with measurement efficiency $\eta$ for $\Delta=0.3$ ($\mathcal S = 10.1$ dB, blue solid) and $\Delta=0.2$ ($\mathcal S = 13.8$ dB, orange solid). Note that the bin boundaries are scaled by $\sqrt\eta$ to account for measurement inefficiency. For comparison, the dottes lines show the corresponding measurement error with one round of noiseless phase estimation using the scheme in~\cref{fig:phaseest_measure}~(b).
    (c) Same as (b) but for a majority vote over $n$ rounds of noiseless phase estimation using the schemes in~\cref{fig:phaseest_measure}~(a)~(solid) and (b)~(dotted).
    }
    \label{fig:Zmeasurement}
\end{figure}

Destructive logical measurements in any Pauli basis ($\mathcal M_{X,Y,Z}$) can be performed by measuring one of three respective quadratures
\begin{subequations}
\begin{align}
&\mathcal M_X: \text{ measure } {-}\hat P,\\
&\mathcal M_Y: \text{ measure } \hat Q - \hat P,\\
&\mathcal M_Z: \text{ measure } \hat Q,
\end{align}
\end{subequations}
and rounding the outcome to the nearest multiple of $\sqrt \pi$. If the result is an even multiple, report a $+1$ outcome, and if the result is an odd multiple, report $-1$.
That this gives a logical Pauli measurement follows from~\cref{eq:paulis}.
An attractive feature of this measurement scheme is that it is robust to small displacement errors---precisely the type of errors the GKP code is meant to be robust against---and can in this sense be said to be fault-tolerant.
The procedure is illustrated for an $\mathcal M_Z$ measurement on an approximate square GKP code in~\cref{fig:Zmeasurement} (a).

Quadrature measurements are routinely performed in both the microwave and optical domain. Performing such measurements on an encoded GKP qubit is, however, not as straight forward. {In the context of cQED and other approaches where the GKP state is encoded in a localized high-quality mode, such as the standing modes of a cavity,} the ability to rapidly perform quadrature measurements contradicts the requirement of the oscillator mode to be long-lived. It is therefore necessary to either tune the oscillator decay rate $\kappa$ from a small to a large value prior to measurement, or to map the encoded information from a high-$Q$ to a low-$Q$ mode (with $Q\sim 1/\kappa$ the quality factor)~\cite{pfaff2017controlled}.

The situation is further complicated by the fact that the measurement efficiency for homodyne detection is limited in practice. The ability to distinguish the codewords deteriorates rapidly with decreasing measurement efficiency, as shown in~\cref{fig:Zmeasurement} (b). A measurement efficiency $\eta$ below unity means that the GKP state shrinks towards vacuum, and it is important to compensate for this (assuming that $\eta$ itself is known) by rescaling the measurement bins [white and pink in~\cref{fig:Zmeasurement}~(a)] appropriately. More precisely, we {now round to the nearest integer multiple of $\sqrt{\eta\pi}$ }~\cite{Shawinprep}.
To produce the numerical results in~\cref{fig:Zmeasurement}~(b) we took an approximate $\ket{\tilde 0_L}$ state, apply a pure loss channel with $\eta = e^{-\kappa t}$, followed by an ideal measurement of the position quadrature and bin the result.

In the microwave domain, state-of-the-art measurement efficiencies are well below~$90\%$, even with the use of near quantum-limited amplifiers~\cite{Macklin2015,Touzard2019}. The results in~\cref{fig:Zmeasurement}~(b) show that measurement efficiencies will have to be improved for this approach to be promising for distinguishing GKP codewords with high fidelity. For example, at $\eta=75\%$---a high but not unreasonable value for a microwave measurement chain---the error probability is about $p_\text{err} \simeq 5.6\%$ ($4.1\%$) for $\Delta=0.3$ ($0.2$).
For comparison, at $\eta=90\%$ we find $0.61\%$ ($0.15\%$).
Note that rather large measurement efficiencies are required to have a substantial benefit from lowering $\Delta$.

The measurement efficiency may be improved if we can amplify the quadrature information \emph{prior} to releasing the GKP state to a standard microwave measurement chain~\cite{Eddins2019}.
Although most theoretical work on GKP codes assume high efficiency quadrature measurements~\cite{Vuillot2018,Noh:2019aa,Terhal2020,noh2021low}, it is an open question whether the stringent demands required for scalable, fault-tolerant quantum computing can be met with this approach. We return to this question in~\cref{sec:scaling} where we discuss concatenation with topological codes.

\subsubsection{\label{sec:phaseest_measurement}Pauli phase estimation}

An alternative, and non-destructive, way to do logical Pauli measurements is by performing phase estimation using an ancillary system.
In the simplest case this task can be performed using a discrete two-level system as an ancilla, removing the need to perform direct quadrature measurements on an encoded GKP state. 

\begin{figure}
    \centering
    \includegraphics{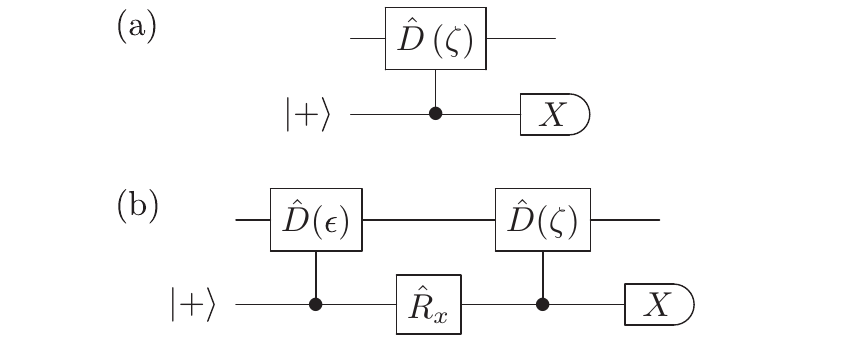}
    \caption{
    (a) One round of phase estimation. The controlled displacement is defined as $C \hat D(\zeta) = \hat D(\zeta/2) \otimes \ket{0_a}\bra{0_a} + \hat D(-\zeta/2) \otimes \ket{1_a} \bra{1_a}$, with $\ket{0_a/1_a}$ the state of the ancilla.
    For $\XL$ measurement $\zeta=\alpha$ and for $\ZL$ measurement $\zeta=\beta$.
    %For $\XL$ measurement $\hat D(\zeta) = \XL = e^{-i\sqrt\pi \hat P}$ and for $\ZL$ measurement $\hat D(\zeta) = \ZL = e^{i\sqrt\pi \hat Q}$.
    (b) Improved phase estimation scheme. Here, $\epsilon$ is a small displacement orthogonal to $\zeta$, i.e., $\arg\epsilon = \arg\zeta + \pi/2$. For $\XL$ measurement we set $\hat D(\epsilon/2) = e^{-i \lambda \hat Q}$ and for $\ZL$ measurement $\hat D(\epsilon/2) = e^{i \lambda \hat P}$, with $\lambda \in \mathbb R$.
    %For $\XL$ measurement, $\epsilon = i|\epsilon|$ and for $\ZL$ measurement, $\epsilon = - |\epsilon|$.
    The $\hat R_x$ gate is defined as {$\hat R_x = e^{-i\pi \hat \sigma_x/4}$}.
    Note that with this definition of the controlled displacement gate, the state is displaced by $\zeta/2$ from the code space after the measurement. This can be corrected for with another displacement if necessary.
    }
    \label{fig:phaseest_measure}
\end{figure}

Since the logical GKP-Paulis are unitary displacement operators, $\XL = \hat D(\alpha), \ZL = \hat D(\beta)$, their eigenvalues are of the form $e^{i\theta}$ with $\theta \in [0,2\pi)$. The task of estimating an eigenvalue of a unitary operator, or equivalently the ``phase'' $\theta$, is generally known as phase estimation.
A variety of different phase estimation protocols exist, with tradeoffs in terms of efficiency and complexity~\cite{Terhal2016,Weigand2020,Royer2020}. We here focus on a simple non-adaptive scheme using a single two-level ancilla. Specifically we consider the scheme that was used in the experiments in Refs.~\cite{Fluhmann:2019aa,Campagne2020}, illustrated in~\cref{fig:phaseest_measure}~(a), as well as a modified version that has theoretically been shown to give better performance, shown in~\cref{fig:phaseest_measure}~(b)~\cite{hastrup2020improved,Royer2020}.

The central component of these schemes is a controlled-displacement gate $C\hat{D}(\zeta)$ which applies a displacement on the GKP mode $\hat{D}(\pm \zeta/2)$ conditioned on the state of the two-level ancilla. See Fig.~\ref{fig:CD_gate} for possible implementations of this  gate in cQED. At the end of the circuit, the ancilla is measured in the $X$-basis. Consider first the simplest scheme in~\cref{fig:phaseest_measure}~(a).
The probability of getting an $X=\pm$ outcome for the ancilla measurement is given by 
\begin{equation}\label{eq:phaseest_prob}
    P(\pm) = \frac12 \left[1 \pm \frac12 \left( \braket{\hat D(\zeta)} + \braket{\hat D\dg(\zeta)}\right)\right].
\end{equation}
To be concrete, let us take $\hat D(\zeta) = \ZL$. For an ideal GKP code we have $\braket{\mu_L|\ZL|\mu_L} = \pm 1$ for $\mu=0,1$, and thus $P(+) = 1$, $P(-) = 0$ for the $\ket{0_L}$ state. For an approximate $\ket{\tilde 0_L}$ state, on the other hand, there is a non-zero probability of getting a $-1$ outcome. Using~\cref{eq:gkpcodewords_approx3} one can show that $\braket{\tilde 0_L|\ZL|\tilde 0_L} \simeq e^{-\pi \Delta^2/4}$ for small $\Delta$, such that the measurement error becomes
\begin{equation}\label{eq:perr_simple}
    p_\text{err} = P(-) \simeq \frac12\left(1 - e^{-\pi \Delta^2/4}\right) \simeq \frac{\pi\Delta^2}{8}.
\end{equation}

Interestingly, the performance can be improved significantly using the scheme shown in~\cref{fig:phaseest_measure}~(b). Here, a small controlled displacement orthogonal to the logical displacement is performed first. In the case of a $\ZL = e^{i\sqrt\pi \hat Q}$ measurement we set $\hat D(\epsilon/2) = e^{i\lambda \hat P}$ with $\lambda$ real. The intuition behind the scheme is that this gives a better approximation to a measurement of the \emph{approximate} logical Pauli operator $\ZL^\Delta$ introduced in~\cref{sec:approxgkp}~\cite{Royer2020}. In this case, one can show that the measurement error becomes
\begin{equation}\label{eq:perr_improved}
    p_\text{err} \simeq \frac12\left\{1 - e^{-\frac{\pi \Delta^2}{4}}\left[ e^{-\frac{\lambda^2}{\Delta^2}} + \sin\left(\sqrt{\pi}\lambda\right) \right]\right\}.
\end{equation}
We can treat $\lambda$ as a free parameter to be optimized. In the small $\Delta$ limit $p_\text{err}$ is minimized for $\lambda \simeq \sqrt{\pi} \Delta^2/2$ in which case one finds $p_\text{err} \simeq 0.4 \Delta^6$, a significant improvement over~\cref{eq:perr_simple}~\cite{hastrup2020improved}.

%However, since one can only get one bit of information per measurement of a two-level system, the probability of a readout error is may be large for a single round of phase estimation. %The situation can be improved by repeating the measurement and performing a majority vote.

In~\cref{fig:Zmeasurement}~(c) we show $p_\text{err}$ for a majority vote over $n$ rounds of ideal phase estimation with a noiseless two-level ancilla for the two respective schemes. 
%We use a simple, non-adaptive phase estimation scheme explained in more detail in~\cref{sec:phaseest} (``Measure'' in~\cref{fig:phaseest} with $\phi=0$).
For the simple scheme in~\cref{fig:phaseest_measure}~(a) and  $\Delta=0.3$ ($0.2$) we find $p_\text{err} \simeq 3.7\%$, $1.0\%$ and $0.4\%$ ($1.6\%$, $0.2\%$ and $0.05\%$) for $n=1$, $3$, and $5$ measurements, respectively. For the improved scheme in~\cref{fig:phaseest_measure}~(b) we find 
$p_\text{err} \simeq 2.6 \times 10^{-4}$, $1.6 \times 10^{-4}$ and $1.5 \times 10^{-4}$
($1.9 \times 10^{-5}$, $6.2 \times 10^{-6}$ and $2.5 \times 10^{-6}$) for the same parameters. These results were produced by numerically computing the probability of a $-1$ measurement outcome on a $\ket{\tilde 0_L}$ state for the circuits in~\cref{fig:phaseest_measure}, and in the case of the scheme in panel $(b)$ optimizing over the parameter $\lambda$.

These results show that phase estimation, modified to better distinguish the approximate Pauli operators of a physical GKP code, can lead to very small measurement errors when the ancilla is noiseless. In~\cref{sec:stateprep} we will discuss how this approach can form the basis for a state preparation scheme, by measuring stabilizer operators in place of logical Paulis.
A fundamental obstacle to this approach, however, is that  errors on the ancilla qubit can propagate back to the GKP mode. In particular, a bit-flip of the ancilla qubit during the controlled displacement gate $C\hat D(\zeta)$ leads to a large, random displacement of the GKP code. In~\cref{sec:ftstateprep} we discuss a potential way to make these schemes robust to such ancilla errors.

%Phase estimation will be discussed in more detail in~\cref{sec:stateprep}, and forms the basis for both state preparation and error correction with GKP codes.
%In particular, we discuss a potential way to make such a scheme robust to ancilla errors in~\cref{sec:ftstateprep}.

\subsubsection{\label{sec:gates}Clifford gates and Clifford frames}

As already mentioned, Clifford gates can be performed using interactions that are at most quadratic in the creation and annihilation operators. Specifically, the Clifford group can be generated, for example, from the Hadamard ($H$), phase ($S$) and CNOT ($C_X$) gates~\cite{Gottesman01}
\begin{equation}\label{eq:Cliffords}
    \bar H = e^{\frac{i\pi}{4} (\hat Q^2 + \hat P^2)},\quad
    \bar S = e^{\frac{i}{2} \hat Q^2},\quad
    \bar C_X = e^{-i \hat Q\otimes \hat P},
\end{equation}
where we use the generalized (code-dependent) quadratures $\hat Q, \hat P$ introduced in~\cref{sec:gkpdefs}. For the $\bar C_X$ gate the first mode is the control and the second mode the target. Together with logical basis measurements $\mathcal M_Z$ and preparation of encoded states $\ket{0_L}$ and $\ket{A_L} = \frac{1}{\sqrt 2}(\ket{0_L} + e^{i\pi/4}\ket{1_L})$, this forms a universal set.

It should be emphasized, however, that the gates in~\cref{eq:Cliffords} are in general only \emph{approximate} logical gates on approximate GKP codes, and, despite being unitary operations, may reduce the quality of the encoded information by making the codewords harder to distinguish [this can be seen from the fact that the gates do not commute with the envelope operator introduced in~\cref{eq:gkpcodewords_approx1}]~\cite{Tzitrin2020,Shawinprep}.
%(As a side note, $\bar H_\text{square} = e^{i \pi \hat a^\dagger \hat a/2}$ and $\overline{SH}_\text{hex} = e^{i\pi \hat a^\dagger \hat a/3}$ are in fact exact Cliffords on approximate square and hexagonal GKP codes, respectively, as long as the approximate states obey the rotation symmetry of the ideal GKP lattice).

\begin{figure}
    \centering
    \includegraphics{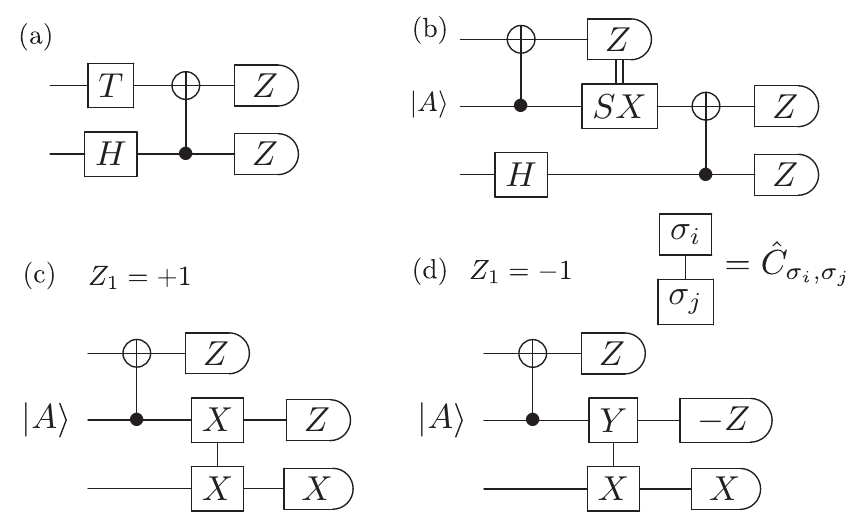}
    \caption{Example computation in the (single-qubit) Clifford frame. 
    A general quantum circuit~(a) can be rewritten using only (adaptive) Clifford gates and magic states $\ket A = T\ket +$, with $T=\diag(1,e^{i\pi/4})$~(b).
    The circuit in (b) can moreover be rewritten as either (c) or (d), depending on the measurement outcome $Z_1=\pm 1$ of the first qubit. The generalized control gates are defined in~\cref{eq:Cliffords2}. For GKP qubits, updating the Clifford frame amounts to changing the phase of local oscillators (c.f.~\cref{fig:gate1}) so that single-qubit Clifford gates are done in software.}
    \label{fig:cliffordframe}
\end{figure}

Due to the approximate nature of the logical gates on physical GKP codewords, it is desirable to minimize the number of Clifford gates in a given quantum circuit. An elegant solution to this problem is to make use of a so-called Clifford frame, where single qubit Cliffords are tracked ``in software''~\cite{Chamberland2018,aaronson2004improved}. The idea of the (single-qubit) Clifford frame is as follows: An arbitrary quantum circuit $\mathcal C$, written in terms of preparation of $\ket{0_L}$ and $\ket{A_L}$ states, gates from the set $\{H, S, C_X\}$, and Pauli measurements $\mathcal M_{Z}$, can be replaced by an equivalent circuit $\mathcal C'$, where the state preparation is identical, the measurements are in any Pauli basis, and all the gates are from the set
\begin{equation}\label{eq:Cliffords2}
    C_{\sigma_i\sigma_j} = I \otimes I - \frac12\left(I - \sigma_i\right)\otimes\left(I - \sigma_j\right),
\end{equation}
where $\sigma_{i,j} \in \{X,Y,Z\}$ runs over the usual Pauli operators. Moreover, the number of qubits, two-qubit gates, and measurements in the new circuit $\mathcal C'$ is the same as in $\mathcal C$, while all single-qubit gates have been removed. Constructing $\mathcal C'$ from $\mathcal C$ is straight forward. One simply commutes the $H$ and $S$ gates through all the $C_X$ gates, mapping them to new gates from the set~\cref{eq:Cliffords2} (and by-product single-qubit Pauli gates) in the process, and finally absorb any single-qubit Cliffords and Paulis into the measurements, mapping $\mathcal M_Z$ to general Pauli measurements~\cite{gottesman1998heisenberg}. An example is given in~\cref{fig:cliffordframe}.

For GKP-encoded qubits, the gate set~\cref{eq:Cliffords2} is particularly attractive, because switching between gates in this set is essentially a ``free'' operation in many physical platforms. More precisely, an encoded version of the gateset for GKP codes is realized by~\cite{Shawinprep}
\begin{equation}\label{eq:Cliffords3}
    \bar C_{\sigma_i\sigma_j} = e^{i \hat s_i \otimes \hat s_j},
\end{equation}
where $\hat s_1 = -\hat P$, $\hat s_2 = \hat Q - \hat P$, $\hat s_3 = \hat Q$ are the three quadratures corresponding to logical $\XL$, $\YL$ and $\ZL$, respectively.
%Note that these gates have the beneficial property that although they may propagate a displacement error from one mode to another, they do not increase the magnitude of displacement errors relative to the lattice distance of the GKP code, such that errors remain under control. The gates are in this sense fault-tolerant.

\begin{figure}
    \centering
    \includegraphics[width=\linewidth]{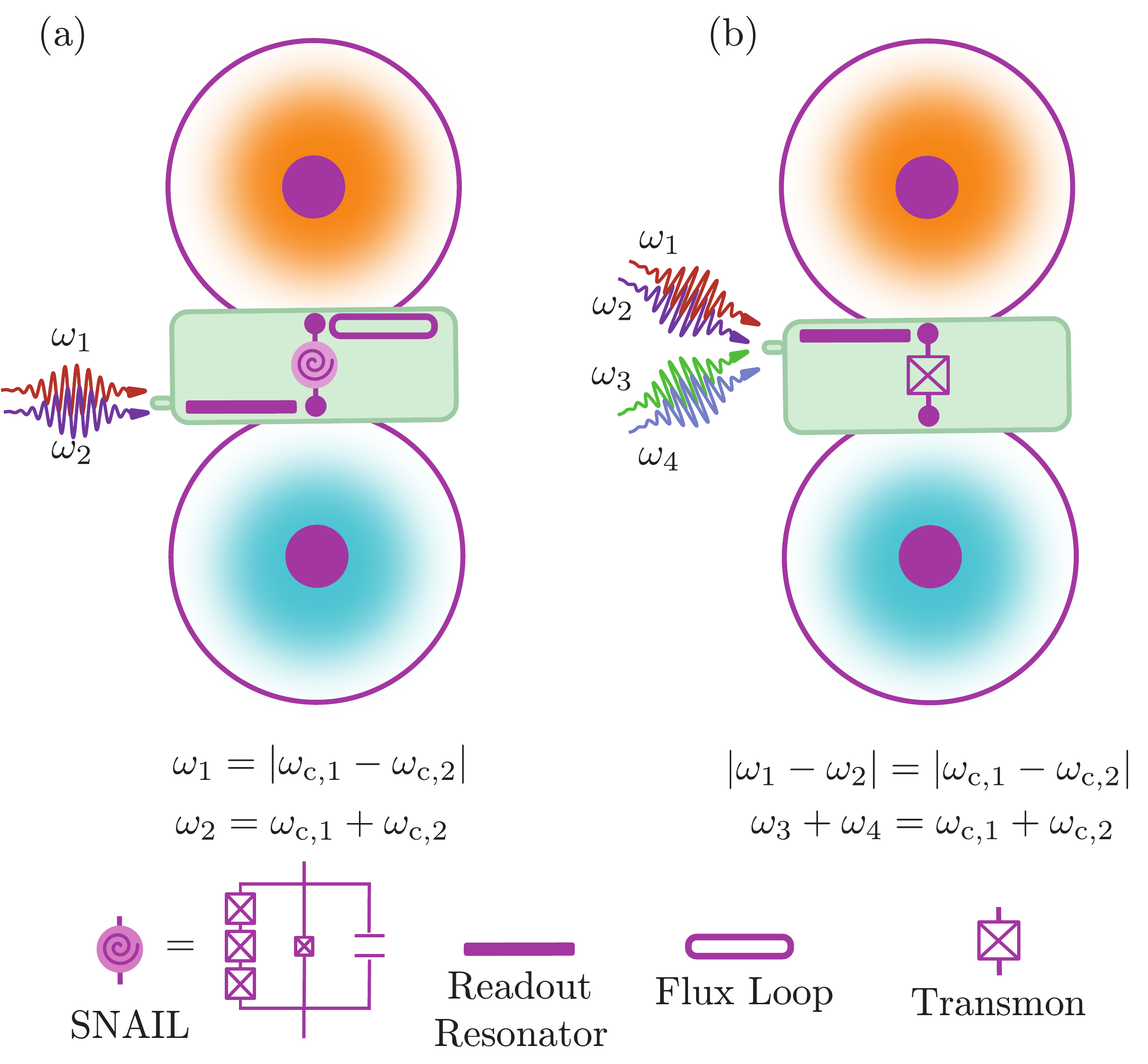}
    \caption{Illustration of how the Hamiltonian $\hat H_{\theta,\phi} \propto e^{i\theta} \hat a \hat b\dg + e^{i\phi} \hat a\hat b + \hc$, required for realization of the two-qubit Clifford gates $\bar C_{\sigma_i\sigma_j}$ in Eq.~\eqref{eq:Cliffords3}, may be realized in a cQED architecture. The two GKP resonators, shown in orange and blue, have frequencies $\omega_\mathrm{c,1}$ and $\omega_\mathrm{c,2}$, respectively. In (a), a Superconducting Nonlinear Asymmetric Inductive eLement (SNAIL) is used for the coupling~\cite{frattini20173}. At an appropriate flux-bias, the SNAIL exhibits a strong third order nonlinearity.  Two microwave drives are applied at frequencies $\omega_1$ and $\omega_2$ so that $\omega_1=|\omega_\mathrm{c,1}-\omega_\mathrm{c,2}|$ and $\omega_2=\omega_\mathrm{c,1}+\omega_\mathrm{c,2}$. Because of the third-order nonlinearity, a single photon at $\omega_1$ is consumed to convert a photon at $\omega_\mathrm{c,1}$ to that  at  $\omega_\mathrm{c,2}$. This leads to an interaction of the form $\hat{a}^\dag\hat{b} e^{i\theta}+$H.c., where the strength of the interaction and $\theta$ depend on the strength of  the microwave drive at $\omega_1$ and its phase, respectively. Similarly, a single photon at $\omega_2$ is consumed to create two photons, one at $\omega_\mathrm{c,1}$ and the other at  $\omega_\mathrm{c,2}$.  This leads to an interaction of the form $\hat{a}^\dag\hat{b}^\dag e^{i\phi}+$H.c., where, again, the strength of the interaction and $\phi$ is set by the drive strength and phase, respectively.  Alternatively, it is possible to engineer $\hat H_{\theta,\phi}$ using a transmon as shown in (b). In this case, four  pump tones can be used to generate the interaction. Due to the  fourth-order nonlinearity of the transmon, a photon each from the pumps at $\omega_3$ and $\omega_4$ such that $\omega_3+\omega_4=\omega_\mathrm{c,1}+\omega_\mathrm{c,2}$, are consumed to generate two photons at $\omega_\mathrm{c,1}$ and $\omega_\mathrm{c,2}$. On the other hand, photons from the pumps at $\omega_1$ and $\omega_2$ such that $|\omega_1-\omega_2|=|\omega_\mathrm{c,1}-\omega_\mathrm{c,2}|$, convert a photon at $\omega_\mathrm{c,1}$ to $\omega_\mathrm{c,2}$ via four-wave mixing. 
    }
    \label{fig:gate1}
\end{figure}

Any gate from the set~\cref{eq:Cliffords3} can be generated from an interaction of the form $\hat H_{\theta,\phi} \propto e^{i\theta} \hat a \hat b\dg + e^{i\phi} \hat a\hat b + \hc$, where $\hat a$ and $\hat b$ are annihilation operators for the two respective modes. In turn, $\hat H_{\theta,\phi}$ can be realized, for example, from a three-wave mixing interaction with a classical pump with two pump tone frequencies at the sum and difference of the two GKP modes, respectively, and $\theta$ and $\phi$ set by the two corresponding pump phases, see Fig.~\ref{fig:gate1}(a). Alternatively, it can be realized from  a four-wave mixing interaction, using four pump tones, as shown in Fig.~\ref{fig:gate1}(b). Updating the Clifford frame thus simply amounts to updating the programming of classical pump phases. Albeit logical gates between two GKP qubits have not yet been demonstrated at the time of writing, the ability to engineer interactions of the form $\hat H_{\theta,\phi}$ have already been used for other applications in cQED~\cite{pfaff2017controlled,Gao:2019aa,Wang2020,grimm2020stabilization,roy2016introduction}.

\subsubsection{\label{sec:errorspread}Error spread through gates}

An important consequence of the fact that Clifford gates on GKP codes can be generated by quadratic Hamiltonians is that this guarantees that errors are not amplified in a bad way by the gates. Consider, for example, the $\bar C_X \equiv \bar C_{ZX} = e^{-i\hat Q\otimes \hat P}$ gate from the set~\cref{eq:Cliffords3} (the other gates in the set behave analogously), and assume that a small displacement error $e^{-iu\hat P}e^{iv \hat Q}$ [c.f.~\cref{eq:uvdisplacement}] is present on the first (control) mode prior to performing the gate. The factor $e^{iv\hat Q}$ commutes with the $\bar C_X$ gate, but since
%\begin{equation}
%    \left(e^{-iu\hat P}e^{iv \hat Q} \otimes I\right) e^{-i\hat Q\otimes \hat P}
%    = e^{-i\hat Q\otimes \hat P} \left(e^{-iu\hat P}e^{iv \hat Q} \otimes e^{iu \hat P} \right)
%\end{equation}
\begin{equation}
    \left(e^{-iu\hat P} \otimes I\right) e^{-i\hat Q\otimes \hat P}
    = e^{-i\hat Q\otimes \hat P} \left(e^{-iu\hat P} \otimes e^{iu \hat P} \right),
\end{equation}
we see that the $\bar C_X$ gate spreads a displacement error $e^{iu\hat P}$ to the second (target) mode. Even though the error has spread, small displacements spread to small displacements, and the error can be corrected by a subsequent round of error correction. This is exactly analogous to a transversal $\bar C_X = C_X^{\otimes n}$ between two binary code blocks of $n$ qubits, where, say, $t$ $X$ errors on the control block can spread to $t$ $X$ errors on the target block.

In this sense the Clifford gates on GKP codes are fault-tolerant. This is, however, only a statement about an ideal implementation of a gate such as~\cref{eq:Cliffords3}. In a realistic implementation, where the quadratic interaction $\hat H_{\theta,\phi}$ stems from an underlying nonlinearity (c.f.~\cref{fig:gate1}), there will unavoidably be spurious higher order terms present as corrections to $\hat H_{\theta,\phi}$. Such terms might amplify and spread errors in a bad way, and it is therefore crucial that they are made as small as possible. Again, it is not expected that GKP codes can suppress errors arbitrarily. The goal is to suppress errors to sufficiently low levels that the resource overhead for the next level of protection is reduced, as discussed further in~\cref{sec:scaling}.

\section{\label{sec:stateprep}State preparation and error correction}

\subsection{\label{sec:phaseest}State preparation using two-level ancilla}

One way to prepare a GKP state is to non-destructively measure the stabilizers and a corresponding logical Pauli. For example, a measurement of $\SX$ and $\ZL$, both with $+1$ outcomes, would correspond to a preparation of the ideal $\ket{0_L}$ state.
%Since the stabilizers and logicals are displacement operators [c.f.~\cref{eq:gkp_stab_and_pauli}], and any displacement $\hat D(\zeta)$ is unitary, the eigenvalues are of the form $e^{i\theta}$ with $\theta \in [0, 2\pi)$. The task of estimating the eigenvalue of a unitary operator, or equivalently the ``phase'' $\theta$, is generally known as phase estimation.
We discussed how logical Paulis can be measured using phase estimation in~\cref{sec:phaseest_measurement}. These ideas can be extended to measuring the stabilizers $\SX, \SZ$, and introducing feedback displacements to steer the state towards the code space of an approximate GKP code.

\begin{figure}
    \centering
    \includegraphics{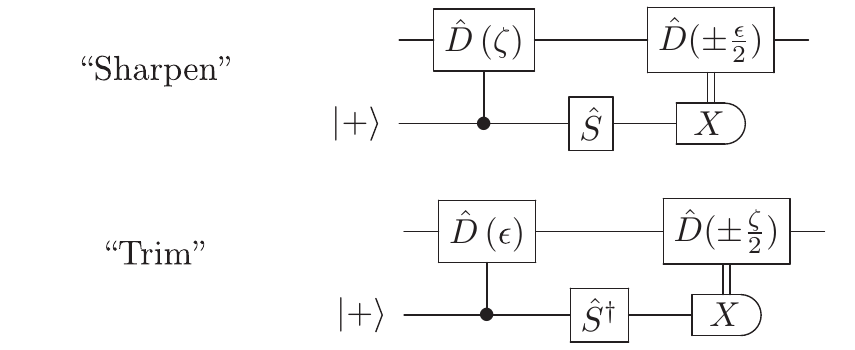}
    \caption{
    %(a) One round of phase estimation, where $\hat R_\phi = \diag(1,e^{i\phi})$ and the controlled displacement is defined as $C \hat D(\zeta) = \hat D(\zeta/2) \otimes \ket{0_a}\bra{0_a} + \hat D(-\zeta/2) \otimes \ket{1_a} \bra{1_a}$, with $\ket{0_a/1_a}$ the state of the ancilla.
    ``Sharpen'' and ``Trim'' protocols used to prepare approximate GKP codestates, the controlled displacement gate is defined as in~\cref{fig:phaseest_measure}.
    The parameter $\epsilon$ is small with $\arg\epsilon = \arg\zeta + \pi/2$, such that $\hat D(\epsilon)$ is a small displacmeent orthogonal to $\hat D(\zeta)$. The $\hat S = \diag(1,i)$ gate is the usual phase gate on the ancilla.
    }
    \label{fig:phaseest}
\end{figure}
\textbf{}

%A variety of different phase estimation protocols exist, with tradeoffs in terms of efficiency and complexity~\cite{Terhal2016,Weigand2020,Royer2020}.

%The basic primitive that underpins the scheme is the standard phase estimation circuit labeled ``Measure'' in~\cref{fig:phaseest}~(a). The central component of this circuit is a controlled-displacement gate $C\hat{D}(\zeta)$ which applies a displacement on the GKP mode $\hat{D}(\pm \zeta/2)$ conditioned on the state of the two-level ancilla. See Fig.~\ref{fig:CD_gate} for possible implementations of this  gate in cQED. At the end of the circuit, the ancilla is measured in the $X$-basis. The probability of getting an $X=\pm$ outcome for the ancilla measurement is given by 
% \begin{equation}\label{eq:phaseest_prob}
%     P_\phi(\pm) = \frac12 \left[1 \pm \frac12 \left( e^{i\phi} \braket{\hat D(\zeta)} + e^{-i\phi} \braket{\hat D\dg(\zeta)}\right)\right].
% \end{equation}
%By repeating this measurement many times for different choices of $\phi$ we can thus estimate $\braket{\hat D(\zeta)}$.

Several protocols have been developed to this end~\cite{Terhal2016,Fluhmann:2019aa,Campagne2020,Royer2020,de2020error,Weigand2020}.
To keep the discussion concrete, we here focus on the scheme illustrated in~\cref{fig:phaseest}, which is the scheme used in the experimental demonstrations of GKP codewords in Ref.~\cite{Campagne2020}.
Let us first consider in more detail the circuit labeled ``Sharpen''. This is simply a version of the standard phase estimation circuit we introduced in~\cref{fig:phaseest_measure}~(a) with a feedback displacement used to steer the state towards a $+1$ eigenstate of $\hat D(\zeta)$. 
The probability of getting $\pm$ outcomes here are respectively $P_{\pi/2}(\pm) = \frac12 \left(1 \pm \Im \braket{\hat D(\zeta)} \right)$.
%Thus by repeating this circuit many times we get an estimate of $\Im \braket{\hat D(\zeta)}$.
To prepare a state with a target phase value $\theta=0$, we introduce a measurement dependent displacement of $\hat D(\pm \epsilon/2)$ along a direction orthogonal to $\hat D(\zeta)$, so that for a $\pm$ outcome an eigenstate state with eigenvalue $e^{i\theta}$ is mapped to one with $e^{i(\theta \mp \zeta\epsilon)}$. This ensures that $\theta = 0 \pmod{2\pi}$ is a stable fixed point, while $\theta = \pi \pmod{2\pi}$ is unstable, for sufficiently small $\epsilon$.

Of course, for a fixed $\epsilon$, we can not prepare a phase arbitrarily close to zero using the above procedure, but this is also not desirable. In practice, the scheme will be limited by experimental imperfections, such as unwanted nonlinearities and dephasing, as the photon number of the state increases. It is therefore better to directly target an approximate GKP state as defined in~\cref{sec:approxgkp}, where the choice of $\Delta$ should be optimized based on experimental considerations.

It was shown in Ref.~\cite{Royer2020} that by alternating the phase estimation of $\hat D(\zeta)$ (``Sharpen'' in~\cref{fig:phaseest}) with phase estimation of a small orthogonal displacement $\hat D(\epsilon)$ (``Trim'' in~\cref{fig:phaseest}) one can prepare an approximate GKP state of the form~\cref{eq:gkpcodewords_approx1}, with
$\Delta \sim \sqrt\epsilon$.
%$\Delta^2 \simeq \epsilon/(2\sqrt\pi)$.
%$\epsilon = \sinh(\Delta^2) 2\sqrt\pi \simeq \Delta^2 2\sqrt\pi$
The intuition behind the scheme is that the first step ``sharpens the peaks'' of the target GKP state by bringing it closer to a $+1$ eigenstate of $\hat D(\zeta)$, while the second step ``trims the envelope'' of the state by weakly measuring the orthogonal quadrature~\cite{Royer2020}.

\begin{figure}
    \centering
    \includegraphics{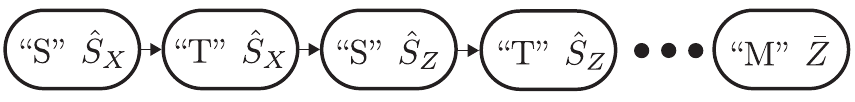}
    \caption{
    Preparation of a logical state can be done by alternating ``Sharpen'' (``S'') and ``Trim'' (``T'') cycles for the two stabilizers $\SX,\SZ$ to project onto the codespace, and finally perform a non-destructive $\ZL$ measurement (``M'') using, e.g., one of the two circuits in~\cref{fig:phaseest_measure}.
    }
    \label{fig:sharpentrimcycle}
\end{figure}

To prepare a logical state, say $\ket{\tilde 0_L}$, one can first alternate many Sharpen-and-Trim cycles of the two stabilizers $\SX,\SZ$ to project the state onto the logical subspace. Once in the codespace, a single phase estimation round of $\ZL=\hat D(\beta)$ suffices to project onto one of the two logical $Z$-basis states. This can be done using, for example, either of the two circuits in~\cref{fig:phaseest_measure}.
The full protocol is illustrated in~\cref{fig:sharpentrimcycle}. Alternatively, one can repeat the logical $\ZL$ measurement a few times and postselect on getting identical outcomes, to increase the preparation fidelity~\cite{Campagne2020,Royer2020} (see also~\cref{fig:Zmeasurement}). Finally, a Pauli correction can be applied if necessary to prepare $\ket{\tilde 0_L}$.

Various optimizations of the ``Sharpen-Trim'' scheme are possible, as well as measurement free versions. We refer the reader to Refs.~\cite{hastrup2021measurement,Royer2020,de2020error} for further details. We also note that optimal control methods have successfully been used to prepare other bosonic codes~\cite{Ofek16,Hu:2019aa}, and that similar techniques may prove useful for GKP state preparation as well.

\begin{figure}
    \centering
    \includegraphics[width=\linewidth]{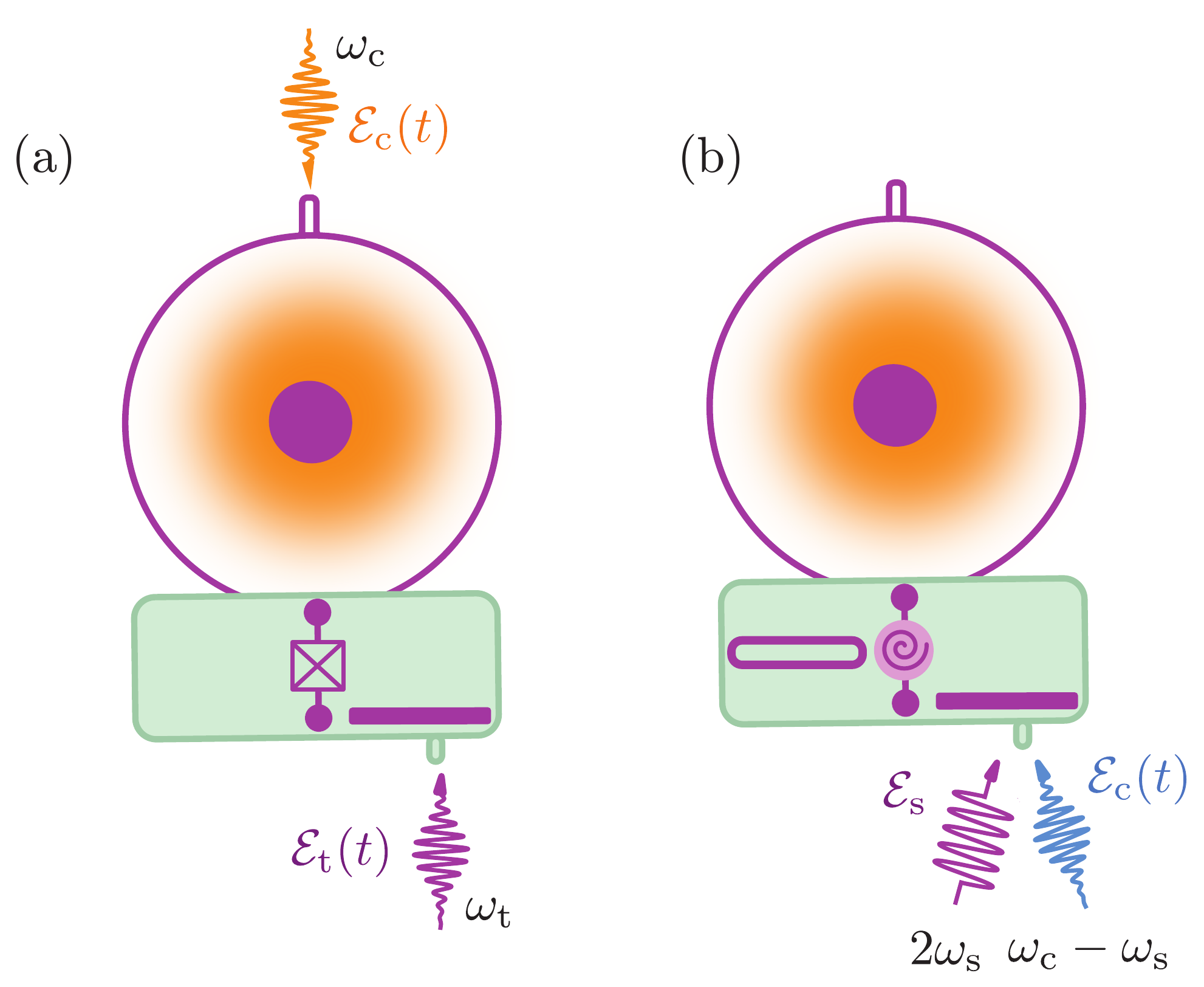}
    \caption{Implementation of controlled-displacement gate, $C\hat D(\zeta)$, between a GKP resonator and a transmon~(a) and Kerr-cat realized in a SNAIL~(b). The frequencies of the resonator, transmon, and SNAIL are $\omega_\mathrm{c}$, $\omega_\mathrm{t}$ and $\omega_\mathrm{s}$ respectively. The desired effective interaction for $C\hat D$ gate is $\propto (\hat{a}^\dag+\hat{a})\hat{Z}$, where $\hat{Z}$ is the Pauli operator of the transmon or Kerr-cat qubit. The scheme shown in (a) is from Ref.~\cite{Campagne2020} and uses the cross-Kerr coupling between the transmon and the resonator $=\chi\hat{a}^\dag\hat{a}\hat{Z}$. The underlying principle of the scheme is as follows. In the  presence of the microwave pulse  $\mathcal{E}_\mathrm{c}(t)$, the  Hamiltonian of the system (in a frame rotating at $\omega_\mathrm{c}$) is $\hat{H}=\chi\hat{a}^\dag \hat{a}\hat{Z}+\mathcal{E}_\mathrm{c}(t)(\hat{a}^\dag+\hat{a})$. In a displaced frame the effective  Hamiltonian  becomes $\hat{D}(\alpha)\hat{H}\hat{D}^\dag(\alpha)=\chi\hat{a}^\dag\hat{a}\hat{Z}+\chi(\hat{a}\alpha^*+\hat{a}^\dag\alpha)\hat{Z}+\chi|\alpha|^2\hat{Z}$, where $\alpha=\int \mathcal{E}_\mathrm{c}(t)\mathrm{d}t$. Clearly, we have obtained the important interaction for the $C\hat D(\zeta)$ gate given by the second term  in this equation. In order to cancel the effect of the other terms, the qubit can be flipped in the middle using the resonant $\pi$  pulse $\mathcal{E}_\mathrm{t}$. For more details see Ref.~\cite{Campagne2020}. In (b), a Kerr-cat qubit is realized in the SNAIL flux biased at a point where both third and fourth order nonlinearities are large. To realize the qubit Hamiltonian, a microwave drive $\mathcal{E}_\mathrm{s}$ of frequency $2\omega_\mathrm{s}$ is applied to the SNAIL. Another  microwave drive at frequency $|\omega_\mathrm{c}-\omega_\mathrm{s}|$ is applied to generate~\cref{eq:HCD_cat}. Due  to the three-wave mixing, a photon from the latter drive is consumed  to convert a photon at the SNAIL to that in the the resonator. This effectively creates the desired coupling $\propto(\hat{a}^\dag+\hat{a})\hat{Z}$ between  the resonator and the Kerr-cat qubit~\cite{Puri:2018aa}.}
    \label{fig:CD_gate}
\end{figure}

\subsection{\label{sec:ftstateprep}Fault-tolerance in state preparation}

An issue with the scheme illustrated in~\cref{fig:phaseest} (as well as the Pauli measurement schemes in~\cref{fig:phaseest_measure}) is that ancilla errors can propagate to the GKP code and lead to uncorrectable errors. In particular, an ancilla bit flip at a random time during the controlled displacement of the ``Sharpen'' step leads to a, potentially large, random displacement error. (A bit-flip during the controlled displacement of the ``Trim'' step is much less serious as it only leads to a displacement error of magnitude $\sim|\epsilon|$.) These simple circuits are, in other words, not fault-tolerant to the dominant error channels, such as relaxation, on the ancilla qubit.

It is noteworthy, however, that there is a certain amount of built-in robustness in the phase estimation circuits, in that phase flips on the ancilla qubit are relatively benign: $Z$ errors on the ancilla commute with the gates and thus only lead to measurement errors. For the ``Sharpen'' circuit a measurement error only leads to a small displacement of magnitude $|\epsilon/2|$ in the wrong direction. This will broaden the GKP peaks, but is not very harmful as long as it does not happen too often.
For the ``Trim'' circuit, a measurement error will lead to a large displacement  $\hat D(\pm\zeta/2)$ in the wrong direction, however, these displacements are equivalent up to the stabilizer $\hat D(\mp\zeta)$, and thus does not lead to a logical error. A measurement error in the final ``Measure'' step illustrated in~\cref{fig:sharpentrimcycle} is more serious, as it leads to a logical error. However, as already mentioned, we may repeat this measurement to suppress such measurement errors. In the following we outline a potential approach to increase the robustness of these circuits by exploiting the natural robustness against phase flips of the phase estimation protocols.

\subsubsection{\label{subsec:bias}Biased noise ancilla}
{
A few different protocols can be applied for more robust phase estimation. For example, the approach considered in~\cite{shi2019fault} is based on using an extra flag qubit to prevent a single ancilla error from introducing a large displacement error in the GKP state. More hardware-efficient robustness against single transmon ancilla error can also be achieved by applying the technique known as $\chi-$matching~\cite{Rosenblum2018}. In this approach, the transmon's $\ket{g}$ and $\ket{f}$ levels are used as computational states and the transmon-cavity coupling is engineered so that the GKP resonator is transparent to a single relaxation error from state $\ket{f}$ to $\ket e$ in the transmon. The approach we outline here for robust phase estimation is based on using a biased-noise ancilla qubit~\cite{Puri:2018aa}.} 

Biased-noise qubits couple asymmetrically with the environment so that one type of error, such as phase-flips or Pauli-$Z$ errors, is more common than others, such as bit-flips or Pauli-$X,Y$ errors. In such qubits it is convenient to define a quantity called the bias, which is the ratio of the dominant  error and the sum of all other errors $\eta=p_z/\left(p_x+p_y\right)$. For pure-$Z$ noise the bias  $\eta=\infty$, while for isotropic or depolarizing noise $\eta=0.5$. Many examples of such biased-noise qubits exists, including the heavy fluxonium qubit~\cite{Earnest2018}, the soft $0$--$\pi$ qubit~\cite{Gyenis2021}, and the Kerr~\cite{grimm2020stabilization} and dissipative~\cite{lescanne2020exponential} cat qubits. In the trapped ion implementation of GKP codes~\cite{Fluhmann:2019aa}, the ancillary pseudo-spin states used to control the motional mode naturally has such a strong bias.

Thanks to the robustness against phase errors, a possible path towards creating a fault-tolerant state preparation scheme is to use a biased-noise ancilla qubit where bit-flip errors are heavily suppressed~\cite{Puri:2018aa}. To be able to implement the circuits in~\cref{fig:phaseest_measure,fig:phaseest} fault-tolerantly, it is however crucial that we can perform the required controlled-displacement gates while preserving a strong suppression of bit-flip errors. {While there are several candidates for strongly biased noise qubits in the superconducting circuit platform~\cite{Earnest2018,Gyenis2021,siegele2021fault}}, here we focus on the Kerr cat qubit as an illustrative example, as this provides a particularly straight forward, hardware-efficient, realization of the operations required for the phase estimation protocols.

In the Kerr cat qubit the logical states are superpositions of coherent states $\ket\pm \propto \ket{\alpha} \pm \ket{-\alpha}$ ($\ket{0/1} \simeq \ket{\pm\alpha}$) of the electromagnetic field stored in a nonlinear oscillator. More precisely, these states are eigenstates of a Kerr-nonlinear oscillator in the presence of a two-photon pump: $\hat H_\cat = - K \hat a^{\dagger 2} \hat a^2 + K\alpha^2 (\hat a^{\dagger 2} + \hat a^2)$, where we are working in the rotating frame of the oscillator where the two-photon pump is resonant~\cite{Puri2017}. The cat states are separated from the closest eigenstates by a gap $\omega_\text{gap} \simeq 4K\alpha^2$, and we take $\alpha$ to be real for simplicity. Crucially, realistic noise channels for this system, including photon loss, heating and dephasing, are highly unlikely to cause transitions between the $\ket 0$ and $\ket 1$ states. More precisely, bit-flips are exponentially suppressed in $\alpha^2$ compared to phase-flips leading to an exponentially large bias in $\alpha^2$ ~\cite{Puri2017}.

Returning to the circuits in~\cref{fig:phaseest}, we first note that preparation and measurement in the $X$-basis, as well as the $\hat S, \hat S\dg$ phase gates, are bias-preserving operations, i.e., the effective error channel remains biased towards $Z$ errors when performing these operations. These operations have already been demonstrated experimentally for the Kerr-cat qubit~\cite{grimm2020stabilization}.
%Moreover, the $\hat S^{(\dagger)}$ gate can be implemented using a simple single-photon drive $\hat H_S = \hat H_\text{cat} + \varepsilon(\hat a\dg + \hat a)$. For the $\hat S$ ($\hat S^\dagger$) gate we use an in-phase drive $\varepsilon > 0$ ($\pi$ out-of-phase $\varepsilon < 0)$ for a time $T_S = \pi/8\alpha\varepsilon$. This gate is diagonal when projecting onto the logical codespace, and thus bias-preserving as well. \alg{Mention leakage and two-photon dissi?}
On the other hand, a controlled displacement can be implemented with a beam splitter interaction
\begin{equation}\label{eq:HCD_cat}
    \hat H_{CD} = i(g\hat a_\cat \hat a_\gkp\dg - g^*\hat a_\cat\dg \hat a_\gkp),
\end{equation}
where the subscript refers to the cat and GKP mode, respectively, and $g$ is an (in general complex) interaction strength. To show that this approximately leads to a conditional displacement, we project the Hamiltonian onto the logical cat subspace $\hat P_\cat = \ket\alpha\bra\alpha + \ket{-\alpha}\bra{-\alpha}$: $P_\cat \hat H_{CD} P_\cat = i \alpha ( g \hat a_\gkp\dg - g^* \hat a_\gkp) \ZL_\cat + \mathcal O^*(e^{-2\alpha^2})$, with $\ZL_\cat$ the logical Pauli-Z operator for the cat qubit (we use the $\mathcal O^*[f(x)]$ notation to suppress polynomial factors in $x$~\cite{woeginger2008}).
Evolution under this interaction for a time $t$ thus leads to a controlled-displacement $C\hat D(\zeta)$, with $\zeta = g\alpha t$. Interestingly, the controlled displacement gate becomes both faster and more accurate as we increase $\alpha$, and thus the bias of the ancilla.

We note that to implement the optimized measurement circuit in~\cref{fig:phaseest_measure}~(b), as well as the optimized stabilizer protocols in Ref.~\cite{Royer2020}, one also requires a rotation gate of the form $\hat R_x = \exp(-i\pi\hat \sigma_x/2)$. This gate is conveniently very simple to implement for a Kerr-cat qubit. One simply turns off the two-photon pump used to stabilize the Kerr-cat logical subspace, and let the cat evolve freely under the Kerr Hamiltonian $\hat H_K = - K\hat a^{\dagger 2} \hat a^2$ for a time $\pi/2K$~\cite{grimm2020stabilization,Puri:2019aa}. One might worry that this gate is not bias-preserving: A phase flip prior to, or during, this gate, can be rotated to a bit-flip error. However, the circuit in~\cref{fig:phaseest_measure}~(b) is constructed such that the error that propagates back to the GKP mode is precisely the logical operator we are trying to measure. Similarly, for the protocols in Ref.~\cite{Royer2020} the error is a stabilizer. This is acceptable, and we therefore expect that these improved measurement and stabilizer protocols also benefit from a biased noise qubit.

Physically, the Kerr-cat qubit can be implemented in a tunable nonlinear element such as capacitively shunted SNAIL, or ``SNAILmon''~\cite{grimm2020stabilization}. With an external magnetic flux,  the SNAIL exhibits both three-wave and four-wave mixing capabilities.
%Standard single-qubit operations on the Kerr-cat qubit have been demonstrated~\cite{grimm2020stabilization}. [ALG: I commented this out as I already sait it above.]
Moreover, the controlled-displacement~\cref{eq:HCD_cat}, which requires a three wave mixing interaction between the cat, GKP and a microwave pump, can also be activated using the same nonlinear element, thus requiring only capacitive coupling between the SNAIL device and the GKP mode [see  Fig.~\ref{fig:CD_gate}(b)].  Such an interaction between a Kerr-cat and an un-encoded oscillator has been demonstrated~\cite{grimm2020stabilization}. This proof-of-principle demonstration strongly suggests that the Kerr-cat in a SNAILmon can be used for fault-tolerant GKP state preparation. Nonetheless, the viability of this approach  depends on the  effect of
magnetic flux, required to bias the SNAIL, on the lifetime of 3D cavities and requires more experimental exploration~\cite{chapman2021mediating}. Alternatively, an approach based on high-Q oscillators in a 2D architecture must be developed.
Finally, we note that this is exactly the same type of interaction required for two-qubit gates between GKP mode, as discussed in~\cref{sec:gates} [c.f.~\cref{eq:Cliffords3}]. It is thus be possible to repurpose the same piece of hardware for both gates and state preparation [see also Fig.~\ref{fig:gate1}(a)].

\subsection{\label{sec:ec_ckts}Error correction with GKP ancillae}

In~\cref{sec:qec_criteria} we briefly discussed the ability of the GKP code to correct against realistic noise processes \emph{in principle} by looking at the quantum error correction criteria for displacement errors and performing numerical simulations using an optimal recovery map. Performing error correction \emph{in practice} requires a fault-tolerant and non-destructive way to measure the stabilizers of the GKP code. One way to do this is to perform phase estimation using a two-state ancilla, as we have already outlined above. However, this approach has the disadvantage that a single bit of information is obtained per ancilla measurement, such that several measurements is required to obtain the continuous variable GKP syndrome information with high accuracy. An alternative approach, which is the approach that has been studied most extensively in the theoretical literature on GKP codes~\cite{Gottesman01,Vuillot2018,Noh:2019aa,noh2021low}, is to use ancillae that are themselves prepared in GKP states. Here, phase estimation can be performed in a single-shot (assuming we can perform high-efficiency homodyne detection), with an accuracy set by the quality of the encoded GKP ancilla states (and the measurement efficiency).

\begin{figure}
    \centering
    \includegraphics{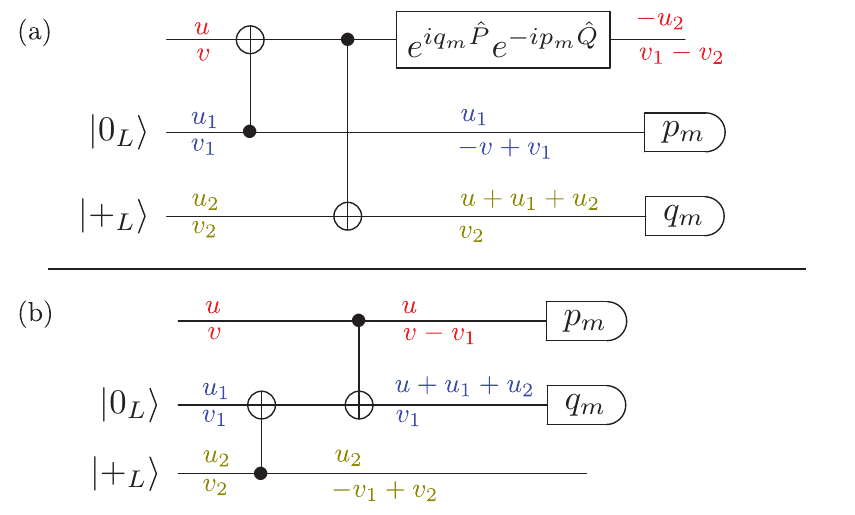}
    \caption{Illustration of ``Steane'' (a) and ``Knill'' (b) error correction circuits. The labels $\{u,v\}$ next to a rail indicates a general displacement error $e^{-iu \hat P}e^{iv\hat Q}$, and the diagram indicates how the incoming errors propagate through the circuit.
    The two measurements are of the $\hat P$ and $\hat Q$ quadratures of the GKP code, respectively, and
    for the correction shifts one uses the measurement outcomes modulo the GKP lattice spacing~$\sqrt{\pi}$~\cite{Gottesman01}.
    }
    \label{fig:qeccirc}
\end{figure}

There are two canonical ways to perform ``single-shot'' GKP error correction using GKP encoded ancillae. These are essentially bosonic versions of Steane~\cite{Steane1997} and Knill~\cite{Knill05} error correction, respectively. The two schemes are illustrated in~\cref{fig:qeccirc}. As shown in the figure, the propagation of displacement errors through the two circuits are essentially equivalent, but the Knill circuit does not require any active recovery. Here, the measurements, which contain the syndrome information, are simply used to determine whether a logical Pauli operator has been applied to the GKP state in the process of teleporting from the top to the bottom rail of the circuit~\cite{Grimsmo2020}. It is not necessary to physically apply any Pauli correction, as it can be tracked in a Pauli frame~\cite{Knill05}.
We also note that a version of the Knill circuit can be performed using beam splitter interactions in place of $\bar C_X$ gates{~\cite{Walshe2020}}, and that it was shown in Refs.~\cite{larsen2021fault,noh2021low} that this leads to a lower probability of logical error.

Although the ``single-shot'' GKP error correction schemes are highly efficient in principle, they also have two clear practical drawbacks: They require preparation of two additional GKP states per round of error correction, and they require very high efficiency quadrature measurement to be useful.
The former itself requires repeated stabilizer measurements using a two-level ancilla as discussed in~\cref{sec:phaseest}, unless some other method is developed, and 
as discussed in~\cref{sec:logicalops}, the latter is a highly nontrivial task that has not yet been demonstrated on a GKP state.
In the next section we discuss different approaches to quantum error correction with GKP codes in a large scale architecture.

\section{\label{sec:scaling}The big picture: Scalability and fault-tolerance}

%Building block: Controlled displacement

So far, we have seen that either due to finite squeezing, environmental noise, or backaction from the ancilla used in state preparation, the logical error rate in the GKP codespace cannot be decreased arbitrarily.
In order to correct for residual errors, the GKP code can be concatenated with another binary quantum error correcting code.
A particularly popular approach is concatenation of GKP code to the topological surface code~\cite{Dennis02}. Fault-tolerant error correction with GKP-surface codes are being studied in the context of both gate-based~\cite{Vuillot2018,Terhal2020,Noh:2019aa,noh2021low} and measurement-based quantum computing~\cite{Menicucci14,Fukui:2017aa,Fukui2018,Fukui:2018aa,Fukui:2019aa,bourassa2021blueprint}. {Here, we will limit our discussion to the former as it is more commonly used in the context of the superconducting-circuit and trapped ion platforms.}

%\subsection{Fault-tolerant GKP-surface code}

% in order to achieve fault-tolerance
In this approach,  each data qubit of the surface code is replaced by a single-mode GKP code. Such concatenation provides two-layers of protection. In the first layer (referred from henceforth as the \emph{inner code} and denoted $\mathcal{C}_\mathrm{GKP}$), the stabilizers $\hat{S}_X$ and $\hat{S}_Z$ are measured for each GKP mode $M$ times
%and the measurement results are used to decode errors locally 
using additional ancillae.
These additional ancillae can also be GKP-encoded, as in~\cref{sec:ec_ckts}, or discrete qubits such as transmons or Kerr-cats, as discussed in~\cref{sec:stateprep,sec:ftstateprep}. The ancilla measurement record  and details of the underlying noise model are used to estimate and correct, as accurately as possible, the noise on each GKP data mode~\cite{Vuillot2018,Noh:2019aa,noh2021low}. 

Of course, this procedure will not perfectly remove all errors.
%not all displacement errors will be corrected. For example, displacement larger than half the GKP lattice constant will be incorrectly identified and the corresponding  correction will introduce a logical $\bar{X}$ or $\bar{Z}$ error, as discussed in~\cref{sec:qec_criteria}.
The remaining
%$\bar{X}$, $\bar{Z}$
errors are instead corrected by interspersing the stabilizer measurements of $\mathcal{C}_\mathrm{GKP}$ with the parity checks of the surface code (referred to henceforth as the \emph{outer code}, and denoted $\mathcal{C}_\mathrm{surface}$). The surface code ancillae used to measure these parity checks can themselves GKP-encoded or discrete qubits.

This approach of concatenating $\mathcal{C}_\mathrm{GKP}\triangleright \mathcal{C}_\mathrm{surface}$ may seem contrary to the hardware efficiency of GKP codes argued in the Introduction. After all, at the end we are resorting to a binary surface code which incurs a substantial hardware overhead (the surface code has a vanishing code rate). Nevertheless, if it becomes possible to suppress the probability of error during logical operations far below the threshold of $\mathcal{C}_\mathrm{surface}$, 
a modest code distance might suffice to reach a target logical error rate required for useful quantum computation~\cite{noh2021low}.
In this case, the resource overhead of quantum hardware and software controls may be significantly lighter than if $\mathcal{C}_\mathrm{surface}$ is directly implemented with conventional qubits such as transmons. 

\begin{figure}
    \centering
    \includegraphics[width=0.5\textwidth]{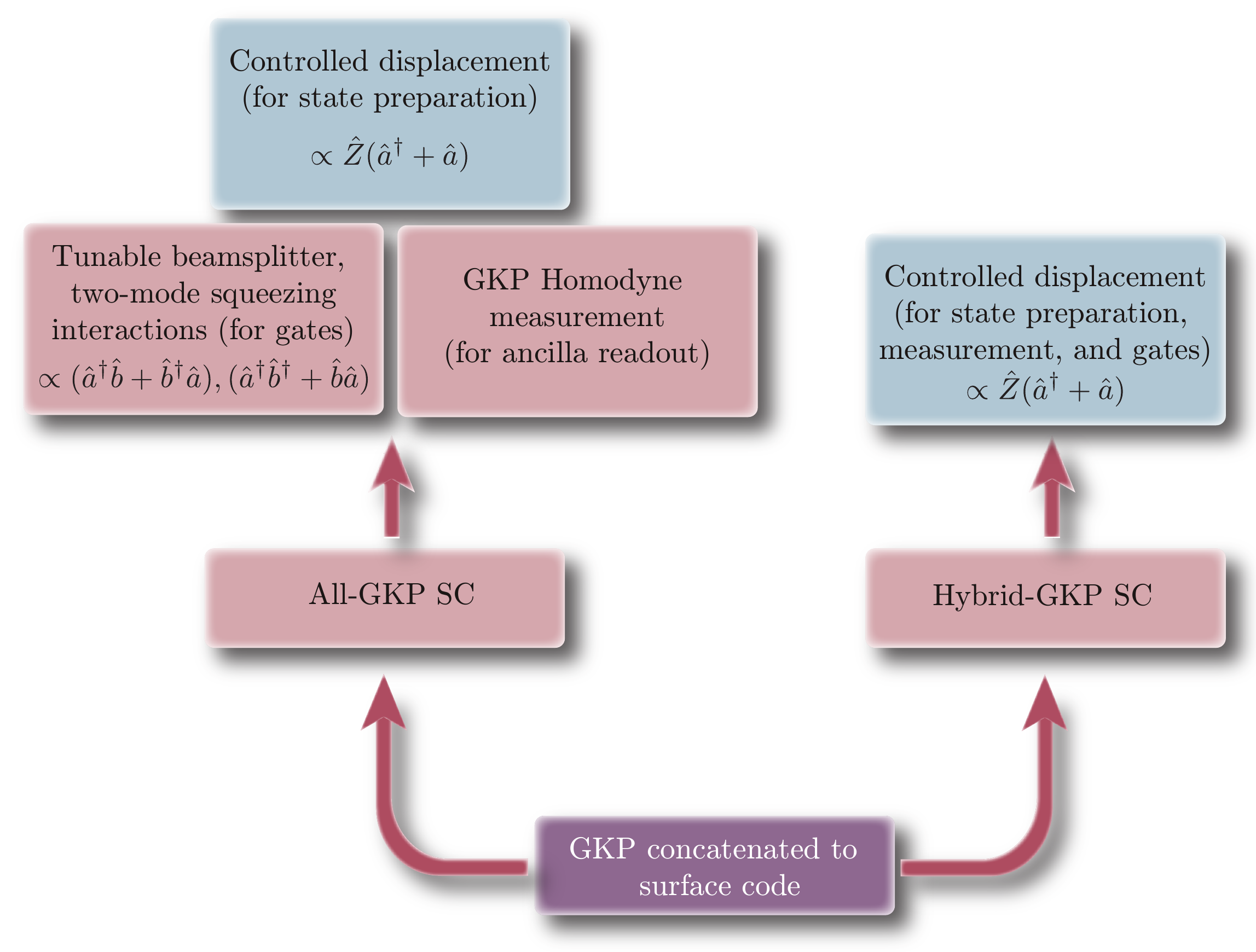}
    \caption{Building blocks of $\mathcal{C}_\mathrm{GKP}\triangleright \mathcal{C}_\mathrm{surface}$ with All-GKP surface code and Hybrid-GKP surface code. Of the required operations, only the controlled-displacement gate between a discrete qubit and a GKP-encoded qubit have been demonstrated experimentally~\cite{Fluhmann:2019aa,Campagne2020}. Not shown in the figure above are single-qubit gates and measurement of the regular ancilla qubits that may be required. We assume these are readily available.  }
    \label{fig:gkp_scov}
\end{figure}

With this overview, we now give more details on two possible constructions of the concatenated $\mathcal{C}_\mathrm{GKP}\triangleright \mathcal{C}_\mathrm{surface}$ code. In one approach, which is also the most widely studied in the theoretical literature, and is referred to as All-GKP surface code, all the data qubits and ancillae used for error correction are GKP-encoded~\cite{Vuillot2018,Noh:2019aa,noh2021low}. The other approach takes a hybrid route where the ancillae are replaced by discrete qubits (e.g., transmons or Kerr-cats), and  will be referred to as Hybrid-GKP surface code (the two schemes were referred to as Only-Surface-Code-GKP-Ancilla and All-Regular-Qubit-Ancilla, respectively, in Ref.~\cite{Terhal2020}). \Cref{fig:gkp_scov} outlines the building blocks of the two schemes.
%The terminology Hybrid-GKP-Ancilla and  All-Regular-Qubit-Ancilla was first established  in Ref.~\cite{Terhal2020}.  

%\subsection{$\mathcal{C}_\mathrm{GKP}\triangleright \mathcal{C}_\mathrm{surface}$ with All-GKP surface code }
\subsection{All-GKP surface code}

\begin{figure*}
    \includegraphics[width=\textwidth]{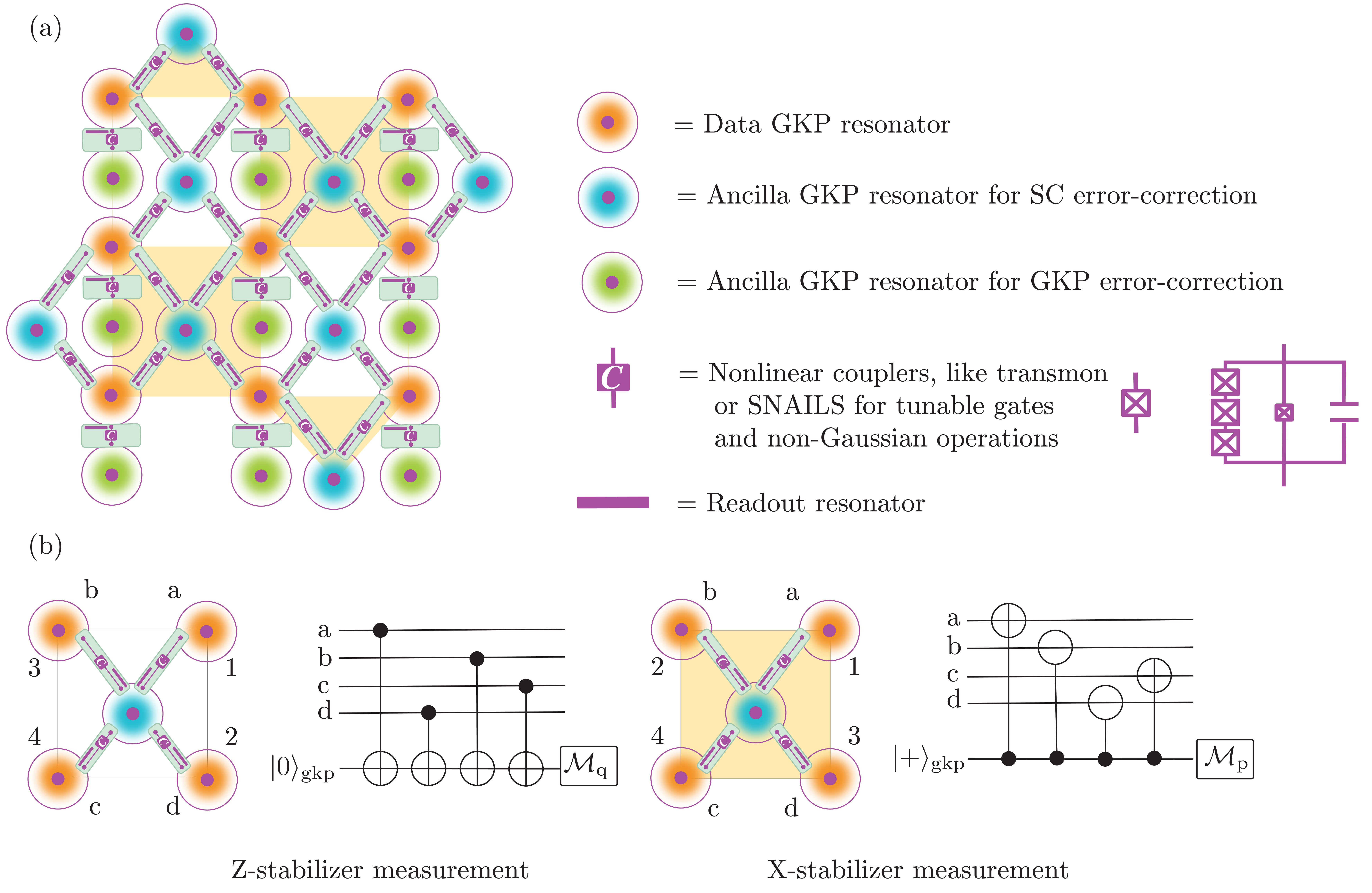}
    \caption{(a) Illustration of possible realization of $\mathcal{C}_\mathrm{GKP}\triangleright \mathcal{C}_\mathrm{surface}$ in cQED. The yellow and white regions indicate $X$ and $Z$ stabilizer checks of the surface code. Scaling up an architecture based on high-Q 3D microwave post resonators is still a very active area of research. It is also possible to have high-Q resonators on chip as an alternative to the 3D approach. The purpose of this figure is to indicate the resources that would be required to implement the widely studied All-GKP surface code. Resonators for GKP states used as data qubits in the surface code are  indicated in  orange. The ancilla GKP states required for realizing Steane-based $\mathcal{C}_\mathrm{GKP}$ and $\mathcal{C}_\mathrm{surface}$ are shown in green and blue respectively. State preparation, gates, and measurements are mediated by nonlinear couplers, such as a transmon or a SNAILmon. (b) Stabilizer measurement circuits and gate ordering for surface code error correction. The controlled-$\ominus$ gate represents the $\bar C_X^\dag$ gate between the ancilla and data GKP states.}
    \label{fig:gkp_sc1}
\end{figure*}

A possible layout of $\mathcal{C}_\mathrm{GKP}\triangleright \mathcal{C}_\mathrm{surface}$ with GKP-encoded data \emph{and} ancilla modes in the cQED architecture is illustrated in Fig.~\ref{fig:gkp_sc1}(a).
Steane or Knill based error correction circuits may be used for $\mathcal{C}_\mathrm{GKP}$ (c.f.~\cref{fig:qeccirc}), although the former is studied more widely. To be concise, we focus on the square lattice GKP code, where the probability of $\bar{X}$ errors equals that of $\bar{Z}$ errors, concatenated with the CSS surface code with the ``rotated'' layout of Ref.~\cite{bombin_optimal_2007}.

The parity check operators for $\mathcal{C}_\mathrm{surface}$ are of two types: all $\bar{X}$-type to detect $\bar{Z}$ errors and all $\bar{Z}$-type to detect $\bar{X}$ errors. 
In order to minimize amplification of displacement errors, the surface code check operators can be modified as shown in Fig.~\ref{fig:gkp_sc1})(b)~\cite{Noh:2019aa}. Each $\bar{X}$-type  parity check involves two $\bar{X}$ and two $\bar{X}^\dag$, while each $\bar{Z}$-type  parity check only involves $\bar{Z}$. 
(Note in contrast to regular discrete qubits we do not have $\bar{X}=\bar{X}^\dag$ and $\bar{Z}=\bar{Z}^\dag$ outside the GKP codespace). The parity checks of $\mathcal{C}_\mathrm{surface}$ are performed simultaneously using the circuit shown in Fig.~\ref{fig:gkp_sc1}(b). As illustrated, each GKP ancilla (shown in blue in Fig.~\ref{fig:gkp_sc1}) interacts with four neighboring GKP modes via $\bar{C}_Z=e^{i\hat{Q}\otimes\hat{Q}}$ gates for the $Z$-checks and a mix of $\bar{C}_X=e^{-i\hat{Q}\otimes\hat{P}}$ and $\bar{C}_X^\dag=e^{i\hat{Q}\otimes\hat{P}}$ for the $X$-checks.
As discussed earlier, these gates can be implemented via a nonlinear coupler such as a transmon or SNAIL.
%Here, $\bar{C}_Z=e^{i\hat{Q}\otimes\hat{Q}}$ and $\bar{C}_X=e^{-i\hat{Q}\otimes\hat{P}}$, while   $\bar{C}_X^\dag=e^{i\hat{Q}\otimes\hat{P}}$, c.f.~\cref{eq:Cliffords3}.

%\begin{widetext}
%\end{widetext}

Standard techniques of decoding, for example minimum-weight perfect matching (MWPM), can be used for error-correction in the surface code layer. {Moreover, the accuracy of the decoder can be enhanced by using analog information from the continuous variable measurement outcomes~\cite{Fukui:2017aa,Fukui2018,Fukui:2018aa,Fukui:2019aa,Vuillot2018,Noh:2019aa,Terhal2020,noh2021low}. For example, one can incorporate the conditional probabilities for Pauli errors given analog measurement outcomes into the edge weights of the MWPM problem, resulting in a dynamic matching graph.}
%in $\mathcal{C}_\mathrm{GKP}$

Various independent studies have been performed for estimating the performance of the All-GKP surface code under slightly different assumptions about noise~\cite{Vuillot2018,Noh:2019aa,Terhal2020,noh2021low}. A common feature of most of these studies is that, in order to simplify numerical analysis, noise in approximate GKP states as well as errors introduced during error-correction operations are modelled as independent Gaussian displacements with standard deviation $\sigma$, represented by the channel
\begin{align}\label{eq:displacementchannel}
    \mathcal{E}(\hat \rho)_\sigma=&\frac{1}{2\pi\sigma^2}\int \mathrm{d}^2\zeta e^{-|\zeta|^2/2\sigma^2}\hat{D}(\zeta)\hat{\rho}\hat{D}^\dag(\zeta).
\end{align}
Compared to a general noise channel as in~\cref{eq:channel}, this channel is strictly diagonal in the displacement operator basis. {An arbitrary noise channel can be brought closer to  diagonal form in the displacement basis using displacement twirling (but not necessarily following a Gaussian distribution)~\cite{Conrad2021}. \Cref{eq:displacementchannel} describes, for example, a noise process with equal loss and heating rates, given by a master equation $\dot{\hat \rho} = \kappa \left(\mathcal D[\hat a]\hat \rho + \mathcal D[\hat a\dg]\hat \rho\right)$~\cite{Albert17,Noh:2019aa}. However, in a realistic system special engineering is required to make these noise process equal. The Gaussian displacement channel does not in general represent physical noise commonly encountered in oscillator systems, such as loss, dephasing or heating at arbitrary rates.}

%one can show that a noise process with equal loss and heating rates, described by a master equation $\dot{\hat \rho} = \kappa \left(\mathcal D[\hat a]\hat \rho + \mathcal D[\hat a\dg]\hat \rho\right)$, takes this form~\cite{Albert17,Noh:2019aa}.

The standard deviation $\sigma$ in~\cref{eq:displacementchannel} determines the amount of noise in the system. Similarly to how we introduced a squeezing parameter to quantify the quality of approximate GKP states in~\cref{sec:approxgkp}, we can introduce a squeezing parameter $\mathcal S = -10\log_{10}(2\sigma^2)$ {(with the identification $\Delta^2 = 2\sigma^2$)} to quantify the noise in~\cref{eq:displacementchannel}, with large $\mathcal S$ meaning low noise.
{In the numerical studies in Refs.~\cite{Vuillot2018,Noh:2019aa,noh2021low}, approximate GKP states with squeezing $\Delta$ were modeled by applying the noise channel~\cref{eq:displacementchannel} with $\sigma = \Delta/\sqrt 2$ to an ideal GKP state. This was done in order to make the numerics tractable, but the noisy GKP states defined in this manner are still unphysical (there is no ``envelope'' in phase space). In Ref.~\cite{Terhal2020} it was shown that this model with incoherent displacements underestimates the logical error compared to the coherent superposition in~\cref{eq:gkpcodewords_approx2} with $\Delta^2 = 2\sigma^2$.}

When the Gaussian displacement channels introducing errors in the GKP codewords and every element (i.e., gates and measurements) of the error-correction circuits are assumed to be equally noisy, and Steane-based error-correction is used for $\mathcal{C}_\gkp$, then the threshold standard deviation for displacement errors has been found to be $\mathcal S = 18.6$ dB. That is, it becomes possible to realize a logical qubit with arbitrarily small probability of error with $\mathcal{C}_\mathrm{GKP}\triangleright \mathcal{C}_\mathrm{surface}$ as long as the standard deviation for displacement errors is smaller than $\sim 0.09$. On the other hand, if the Gaussian displacement channel is only applied to the data and ancilla GKP codewords, while all other operations are assumed to be noiseless, this threshold is reduced to $\mathcal S=11.2$ dB~\cite{Noh:2019aa}.

Further optimizations are possible, and a recent study showed that the latter threshold with only state preparation noise can be reduced to $\mathcal S = 9.9$ dB through better decoding and an optimized error correction protocol~\cite{noh2021low}. Beyond improving the threshold, it was shown that a decoding strategy that makes better use of the analog syndrome information can have a dramatic effect on the overhead cost to reach a certain logical error rate.

The largest GKP state prepared experimentally so far has a squeezing of $\sim 10$ dB and experimental limitations on the performance of two-qubit gates and measurement (c.f.~\cref{fig:Zmeasurement}) is largely an open question. Thus, there is opportunity for theoretical and experimental innovations in developing scalable quantum control methods for practical implementation of the GKP-surface code architecture. 

%So a lot more theoretical and experimental development is required to %%successfully implement the GKP-surface code. 

One possible path towards easing the threshold requirements is by tailoring the GKP code and the surface code to exploit structure in the noise.
In particular, the more recently tailored surface code (TSC)~\cite{tuckett2018ultrahigh,tuckett2020fault} and the XZZX surface code~\cite{bonilla2020xzzx} exhibit ultra-high thresholds when the noise channel of the underlying elementary qubits is biased. Recall the rectangular-lattice GKP states introduced in section~\ref{sec:gkpdefs} {and defined in Eq.~\eqref{eq:lattice_def}}.  Due to the phase-space structure of the rectangular lattice, the resulting Pauli errors become asymmetric even if the translation errors from the environment are isotropic resulting in a biased noise channel. If {$\lambda>1$} in Eq.~\eqref{eq:lattice_def}, then displacements along  the momentum quadrature $\hat{P}$ is  more likely to be misidentified than displacement  along the position quadrature $\hat{Q}$. Consequently, after $M$  rounds of GKP error-correction $\bar{X},\bar{Y}$-errors will be far less likely  than $\bar{Z}$ errors. Thus, the  error channel after  $\mathcal{C}_\mathrm{gkp}$ with a rectangular-GKP state  will be biased with the bias increasingly exponentially with {$\lambda$}~\cite{hanggli2020enhanced}. 
Now, the biased noise can be more accurately corrected if $\mathcal{C}_\mathrm{gkp}$ is concatenated to the TSC or the XZZX surface code. A recent work has studied the rectangular-GKP concatenated to the TSC under a simplistic noise model where only the data GKP codewords are subject to a Gaussian displacement channel, while the ancillas and error-correction circuits are perfect~\cite{hanggli2020enhanced}. It is known that with such a simplistic noise model, the threshold standard deviation when using the square lattice GKP code is $2.3$ dB. If instead a rectangular GKP code is used with $r {=\lambda^2} =3$, then as shown in~\cite{hanggli2020enhanced},  the threshold standard deviation increases and corresponds to a squeezing of $\sim$1.7 dB. Here, $r=3$, corresponds to a single-mode squeezing of $10\log_{10}(r)\sim 4.8$ dB. 
{For the regular CSS surface code, another study in contrast
found only a marginal improvement in logical error when the ancilla (but not data) qubits are rectangular~\cite{noh2021low}.}
However, more theoretical work  is required to predict if the rectangular-GKP concatenated to the  TSC, {XZZX, or another code optimized to exploit noise-bias,} provides a practical advantage when realistic circuit-level noise is considered.

%If the only source of errors is due to the finite-size of the GKP states, then this threshold corresponds to a GKP state with average photon number $\sim 4$. These numerical studies show that large-scale fault-tolerant error-correction is possible with CV GKP qubits. However, like several similar studies,  their practical use is limited for many reasons. Firstly, experimental noise sources, such as photon loss, thermal excitation, and pure-dephasing, cannot be directly modelled by the independent Gaussian displacement channel. Moreover, 

%We also show that the squeezing threshold changes to 18.6dB when both the
%GKP states and circuit elements are comparably noisy. At this threshold, each circuit component
%fails with probability 0.69 

%\subsection{ $\mathcal{C}_\mathrm{GKP}\triangleright \mathcal{C}_\mathrm{surface}$ with Hybrid-GKP surface code}
\subsection{Hybrid-GKP surface code}

The All-GKP surface code scheme may become resource intensive because auxiliary qubits are required to prepare the GKP data and ancilla modes, and additional tunable nonlinear couplers are required to implement two-qubit gates between these GKP modes.
{Each $\mathcal C_\text{GKP}$ \emph{and} $\mathcal C_\text{surface}$ syndrome measurement moreover requires preparation of fresh GKP ancilla states, which is a slow process~\cite{Campagne2020}.}
Such extra {space-time costs} can overwhelm savings in overheads that one may have otherwise expected. A more efficient approach may be to replace some or all of the ancillae with discrete qubits such as a transmon or a biased Kerr-cat qubit (see~\cref{sec:ftstateprep}). As an illustrative example, consider Fig.~\ref{fig:gkp_sc1} for counting hardware resources. In this set-up a distance $d$ surface code requires $3d^2-1$ high-Q resonators, and  $5d^2-4d$ nonlinear couplers.
%like transmons or SNAILs.
Contrast this with $d^2$ high-Q  cavities and  $2d^2-1$ nonlinear couplers that would be required to build a 
%$\mathcal{C}_\mathrm{GKP}\triangleright \mathcal{C}_\mathrm{surface}$ with 
Hybrid-GKP surface code with GKP-encoded data qubits and discrete qubit (Kerr-cat or transmon) ancillae.
%Moreover, as explained earlier, high-fidelity readout of the GKP state via homodyne measurements is challenging with near-term hardware.

Let us now look at the operations required to implement
%$\mathcal{C}_\mathrm{GKP}\triangleright \mathcal{C}_\mathrm{surface}$
such a Hybrid-GKP surface code. Recall that the controlled-displacement gate $C\hat{D}(\zeta)$ is required to measure  the  GKP stabilizers. The same gate can also be used to implement the controlled-Pauli gates between the regular qubit ancilla and GKP codewords required for surface code parity measurements.

Consider, for example, the $\bar{Z}$-type check operators. It can be measured with a discrete ancilla qubit initialized in the $\ket{+}$ state, followed by $C\hat{D}(\zeta)$ gates between the GKP data modes and the ancilla with $\zeta=\beta$, and finally an $X$-basis measurement of the ancilla. This can also be understood as one phase-estimation round of the surface code stabilizer, and in analogy with~\cref{eq:phaseest_prob}, the probability to get an $X=\pm 1$ outcome is
\begin{align}
    P(\pm)=\frac{1}{2}\left[1\pm\frac{1}{2}\left(\langle\bar{Z}\bar{Z}\bar{Z}\bar{Z}\rangle+\langle\bar{Z}^\dag\bar{Z}^\dag\bar{Z}^\dag\bar{Z}^\dag\rangle\right)\right].
\end{align}
Clearly, for ideal GKP states [\cref{eq:gkpcodewords}], $P(+)=1,0$ and $P(-)=0,1$ when $\langle\bar{Z}\bar{Z}\bar{Z}\bar{Z}\rangle=\pm 1$ respectively, and hence this procedure can be used for $\bar{Z}$-parity checks.
%In fact, this procedure can be understood as one round of phase estimation with the ancilla.  
In case of approximate GKP states however, a single round of phase estimation cannot perfectly estimate $\langle\bar{Z}\bar{Z}\bar{Z}\bar{Z}\rangle$, which leads to measurement errors even when the ancilla is noiseless.
%The qubit measurement outcome can be $X=-1$ even when there was no $\bar{X}$ error in any of the GKP data states. Similarly, the qubit measurement outcome can be $X=+1$ even when there was an $\bar{X}$ error.
%For approximate GKP states 
%such as the one in Eq.~\eqref{eq:gkpcodewords_approx1}
%this  measurement error is
For small $\Delta$ (see~\cref{sec:approxgkp}) this measurement error is $p_\mathrm{err}\sim (1-e^{-\pi\Delta^2})/2 \simeq \pi\Delta^2/2$ %which for small $\Delta$ can be approximated as $\pi\Delta^2/2$
~\cite{Terhal2020}. The $\bar{X}$-type parity checks are measured analogously using the $C\hat{D}(\zeta)$ gate with $\zeta=\alpha$.
This measurement error can in principle be reduced using the same approach as discussed in~\cref{sec:phaseest_measurement} [see~\cref{fig:phaseest_measure}~(b)] to give $p_\mathrm{err}\sim 0.8\Delta^6$ (in the small $\Delta$ limit)~\cite{hastrup2020improved,Royer2020}.

Compared to the All-GKP scheme, where GKP-encoded ancillae are used, the discrete-qubit ancillae may lead to higher fidelity parity checks when a realistic homodyne measurement efficiency is taken into account. 
To illustrate, consider a GKP state with $\sim 14$ dB of squeezing ($\Delta=0.2$). In section~\ref{sec:measurements} we saw that with a measurement efficiency of $\eta=75\%$ (which is still optimistically  high  for cQED), the error in the direct homodyne measurement of the ancilla GKP state is $\sim 4\%$. On the other hand, with a discrete qubit ancilla and modified phase estimation circuit of Ref.~\cite{hastrup2020improved}, we have $p_\mathrm{err}\sim 0.0025\%$. Of course, the ancilla readout itself is not perfect, and errors must be added to $p_\mathrm{err}$ to estimate the total error in the surface code parity check. For an ancilla such as a transmon, readout error probability $<2\%$ is standard, even with $\eta<75\%$~\cite{hatridge2013quantum,Touzard2019,walter2017rapid}. Thus, the total measurement  infidelity with the discrete qubit ancilla ($\sim 0.0025\%+2\%\simeq2\%$) can still be lower than that with a GKP-encoded ancilla.

The hardware simplicity of the Hybrid-GKP surface code architecture, which requires only $C\hat{D}(\zeta)$ gates and standard qubit operations, makes this approach very attractive. Nonetheless, a challenge with the hybrid-GKP approach is to prevent fatal propagation of errors from the standard qubit ancilla to the encoded GKP states. Indeed, due to this effect, the current performance of the controlled-displacement gate, the building block of the hybrid scheme, is limited by the relaxation time, $T_1$ of the transmon ancilla~\cite{Campagne2020,Puri:2019aa}. This limitation indicates that it will not be possible to increase the lifetime of a GKP codeword much beyond the $T_1$ of the transmon. 
Fortunately, it is possible to overcome this challenge by replacing the transmon with a biased-noise ancilla such as a Kerr-cat, as we discussed in~\cref{sec:ftstateprep}. This promises a hardware-efficient solution to the problem of ancilla-induced errors and motivates further study to quantify the performance of the hybrid setup.

%Nonetheless, more detailed study is warranted to quantify the performance of this approach.
%This will require a detailed understanding of the noise in such a hybrid setup.

Finally, it is not necessary that all the ancillae in the GKP-surface code are of the same kind, that is, all GKP-encoded or all discrete qubit ancillae. Another possibility is that both the data and syndrome qubits of $\mathcal C_\mathrm{surface}$ are GKP-encoded, while discrete qubit ancillae are used to perform the $\mathcal C_\mathrm{GKP}$ stabilizer measurements and phase estimation on the GKP-encoded syndrome qubits.
Depending  on the properties and performance  of operations, we may have an optimized code where some  ancillae are  discrete qubits while the others are GKP-encoded.

%Moreover, the $C\hat{D}(\zeta)$  gate required to measure surface code stabilizers will be the   The building block for the GKP syndrome extraction and surface code stabilizer measurements is the $C\hat{D}(\zeta)$ gate between the unencoded ancilla and the GKP mode, see~\cref{sec:stateprep}. This interaction can be used for measuring stabilizers of the GKP code, as well as implementing the controlled-Pauli gates between the regular qubit ancilla and GKP codewords required for surface code parity measurements. The controlled-displacement operation has been demonstrated in both circuit QED and trapped-ion platforms. It is implemented by exploiting the inherent nonlinearity of the ancilla without the need for additional couplers. 

\subsection{Universality}
So far, we have restricted the discussion to error correction in the $\mathcal{C}_\mathrm{GKP}\triangleright \mathcal{C}_\mathrm{surface}$ code, but have not discussed how universal computation may be performed in this concatenated architecture. An attractive feature of the surface code is that all logical Clifford operations can be implemented via lattice surgery requiring only (single or two-qubit) logical Pauli measurements~\cite{horsman2012surface,litinski2018lattice,bourassa2021blueprint}. In surface codes with regular qubits, these measurements require nearest-neighbour, physical controlled-Pauli gates between ancilla and data qubits~\cite{litinski2018lattice}. We have seen how to implement controlled-Pauli gates between two GKP codewords or between a GKP codeword and a discrete qubit [see~\cref{eq:Cliffords3}, Fig.~\ref{fig:gate1}, and Fig.\ref{fig:CD_gate}]. Hence, we can also employ lattice surgery for implementing logical Clifford gates in the $\mathcal{C}_\mathrm{GKP}\triangleright \mathcal{C}_\mathrm{surface}$ code.

\begin{figure}
    \centering
    \includegraphics{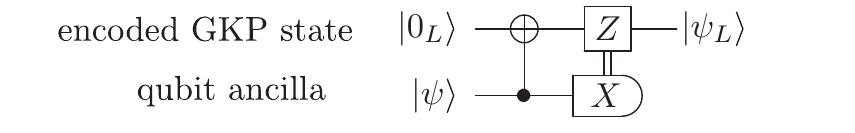}
    \caption{One-bit teleportation circuit to teleport an arbitrary state from a two-level ancilla (bottom) to an encoded GKP state (top). The $\hat C_X$ gate is implemented by a controlled displacement $C\hat D(\alpha)$. This can be used for preparation of arbitrary states, including magic states, assuming we have universal control over the ancilla.}
    \label{fig:onebitteleport}
\end{figure}

Combined with state preparation of GKP magic states $\ket{A_L} = \bar T\ket{+_L}$, logical injection and distillation~\cite{bravyi2005universal,litinski2019game} (both of which only require error correction and Clifford gates) provides the ability to perform universal quantum computation in the GKP-concatenated surface code. There are several ways to prepare GKP magic states, including using only GKP Pauli states and vacuum as a resource {by exploiting the continuous variable nature of the state space}~\cite{Baragiola:2019aa}, but the simplest approach is to use a one-bit teleportation circuit as shown in~\cref{fig:onebitteleport}, which allows us to teleport an arbitrary state from a two-level ancilla to the GKP code.

%Although we will not discuss it further here, alternative topological codes might offer some advantages over the surface code when it comes to logical gates. For example, the 2D color code has a transversal set of single-qubit Clifford gates. As single-qubit Clifford gates on GKP codes can be done in software, this implies that logical single-qubit Clifford gates on the color code level can be done in software as well. This leads to some overhead savings for lattice surgery, although it is unclear whether this will give a net benefit considering the somewhat lower threshold of the 2D color code. Nevertheless, it is a good idea to tailor the overall fault-tolerant scheme to exploit the strengths of the underlying GKP code, and further research in this direction is warranted.

%Note that the surface code is best suited to correct for Pauli error channel. It may then come as a surprise that it is able to correct for errors in a GKP code which suffers from continuous errors due to natural sources of decoherence in an oscillator such as single-photon loss, pure-dephasing, etc. This is possible because the first layer of local error correction reduces the continuous translation errors to discrete single-qubit Pauli errors, which the surface code can then handle. 

%\section{Alternative approaches: Hardware encodings of GKP-like codes}

\section{Summary and Outlook\label{sec:summary}}

Implementation of the GKP code was once considered, by many, to be beyond impossible. As pointed out by Daniel Gottesman at the Byron Bay Quantum Workshop in 2020---a workshop dedicated to the 20th anniversary of the GKP code---the authors were aware that the main challenge was going to be the first step of realizing the codewords themselves and the subsequent steps of realizing gates, measurements, etc. would be comparatively simpler. {{Technological developments since 2001 have made error correction with the GKP code a reality, and this success has inspired more exotic strategies for error correction~\cite{albert2020robust,gross2020encoding}}}.
%Technological developments since 2001 have now made these codewords a reality and no doubt multi-qubit gates will soon follow. 
Keeping current and near-future technology in mind, in this perspective article we have explored the prospect of scalable, fault-tolerant quantum error correction with GKP states in a cQED architecture. The most intriguing open question in this direction is whether error correction with GKP states can be made more resource efficient in practice compared to schemes based on conventional qubits. Below we summarize some open theoretical and experimental challenges that must be addressed to answer this question.
 
One must develop high-fidelity operations, including state-preparation, multi-qubit gates, and measurement for GKP states. The fidelity must be better than those of conventional un-encoded qubits in the same platform.
At the very least, the fidelity should not be limited by decoherence in the auxiliary discrete qubits used for initialization or couplers used for gates. 
We identify three central challenges in this respect, that can serve as milestones on the path towards a scalable and hardware efficient quantum computer with GKP-encoded qubits:

1. State preparation: One must be able to prepare approximate GKP codewords with a sufficiently small $\Delta$, and ensure that the probability of logical errors on the GKP code, e.g. propagating from the ancilla qubit used in the state preparation, is exceedingly low. Since the goal is to outperform the best physical qubit alternatives, specifically transmons and trapped ions, the probability of a logical error in state preparation must be low compared to error rates in these systems. In our view, a biased noise ancilla, such as a Kerr-cat qubit, is promising in this respect, but further analysis is needed to quantify the quality of the GKP states that can be prepared with this approach.

2. Gates: To be able to implement high fidelity gates 
as discussed in~\cref{sec:gates} in practice,
there are several targets that must be met simultaneously. One must be able to implement pristine two-mode Hamiltonians of the form $\hat H_{\theta,\phi} \propto e^{i\theta} \hat a \hat b^\dagger + e^{i\phi} \hat a \hat b + \hc$, while keeping any spurious non-linear terms minimal, and moreover, the two-mode interaction must be switched from near zero to a sufficiently large value such that the gates are fast compared to all decoherence rates. It is crucial that performing these gates does not introduce errors that the GKP code is poor against. In particular, while the GKP is expected to be excellent against loss (and heating), this is not necessarily true for other natural types of noise, such as dephasing (c.f.~\cref{fig:qecfidelity}) and spurious nonlinearities.

3: Measurements: As we have shown in~\cref{sec:measurements}, a standard homodyne measurement is unlikely to be sufficiently high-fidelity to give GKP codes an advantage. Here, new ideas are needed. Either the effective homodyne measurement efficiency has to be increased (probably past $90\%$) using an amplification step prior to release to a standard microwave measurement chain, or one can follow the route of performing phase estimation with a discrete qubit ancilla. It remains to be seen how low the GKP measurement error can be made in practice.

Along with measurement and control, efforts must be devoted to developing technology for scaling up either a 3D cQED architecture, or an architecture based on high-Q resonators on chip.
There are also open theoretical questions about how to design such a large-scale architecture. We have discussed two different approaches at a high-level, the All-GKP surface code and the Hybird-GKP surface code. Several numerical studies have been performed on the All-GKP surface code, showing promising thresholds and sub-threshold behavior~\cite{Vuillot2018,Noh:2019aa,Terhal2020,noh2021low}. However, the noise models used in these studies are rather unrealistic,
%still very simplified,
and more work is needed to model realistic noise accurately. For the Hybrid-GKP surface code, although we think it is quite promising in terms of hardware efficiency, very little quantitative analysis has been done, and its potential is largely unexplored at this stage. Arguably, the most pertinent question here is whether one will ultimately be limited by the discrete qubit ancillae used to stabilize the GKP-encoded qubits and perform syndrome extraction, and consequently whether the Hybrid-GKP approach can have a significant advantage over a more conventional scheme using only discrete qubits everywhere. For example, if biased noise qubits are used as ancillae, one should note that an approach based on using such qubits as both ancillae \emph{and} data qubits also appears very promising~\cite{guillaud_repetition_2019,chamberland2020building,Darmawan2021}.
 
% A possible approach to high-fidelity operations is the use of biased-noise ancilla. 
 
There are further avenues of research we have not touched on in this Perspective, but that nonetheless seem very promising.
Alternatively to active GKP error-correction, it may be advantageous to explore passive error-correction techniques where the GKP states are stabilized via Hamiltonian engineering~\cite{le2019doubly,rymarz2021hardware,Conrad2021}. 
%On the theory side, we must also go  beyond the Gaussian displacement channel and model the errors in the experiment more accurately. Realistically, the errors in an experiment due to, for example, single-photon loss, gain, frequency-fluctuations, cross-Kerr interactions with non-linear couplers, self-Kerr interactions, etc. can not be modeled as random displacements. So the performance  of GKP codes (and their concatenation with other codes) must be evaluated against the experimentally relevant noise for a given physical platform. Along with control techniques, efforts must also be devoted to developing technology for scaling up either a 3D cQED architecture, or an architecture based on high-Q resonators on chip.
One can also consider alternative concatenation schemes. For example, surface code variants tailored to the specific noise structure of the GKP codewords may be used. In particular, biased-noise tailored surface code such as the XZZX surface code~\cite{bonilla2020xzzx} or tailored surface code~\cite{tuckett2018ultrahigh,tuckett2020fault,hanggli2020enhanced}, are promising candidates for a rectangular-lattice or squeezed GKP code. More work is required to estimate the performance of such an architecture when realistic circuit level noise is considered. 
Other topological codes might offer some advantages over the surface code when it comes to logical gates. For example, the 2D color code has a transversal set of single-qubit Clifford gates. As we have shown that single-qubit Clifford gates on GKP codes can be done in software, this implies that logical single-qubit Clifford gates on the color code level can be done in software as well. This may lead to some overhead savings for lattice surgery~\cite{landahl2014quantum,litinski2018lattice}. In general, one should follow the design principle of tailoring the overall fault-tolerant scheme to exploit the strengths of the underlying elementary qubits, and further research in this direction is warranted.
Finally, concatenation with conventional quantum error correcting codes is not the only path towards scalability. It is possible that there is a better scheme where $k$ logical qubits are encoded in $n$ physical modes more directly, i.e., without concatenation with a binary code. {Only a small number of works have explored this avenue so far~\cite{Gottesman01,Harrington01,Noh2019capacity}.}

With challenges, come opportunities and with the accelerating pace of technological and theoretical developments the future looks bright for practical quantum computation with GKP codes.

\section*{Acknowledgements}
S.P would also like to thank numerous colleagues and students including Steve Girvin, Michel Devoret, Nicholas Frattini, and Alec Eickbusch, discussions and debates with whom have shaped her outlook on bosonic codes. In particular, discussions with Nicholas Frattini have helped in producing the conceptual figure for the GKP-surface code. A.L.G would like to thank Mackenzie Shaw for deriving the set of generalized control gates for GKP codes, Baptiste Royer for in depth discussions on preparation of GKP codewords, Ben Baragiola and Joshua Combes for help producing~\cref{fig:gkplattices} and input on~\cref{sec:intro_gkp}, and Barbara Terhal for several discussions on GKP codes that have informed this perspective article. We thank Baptiste Royer, Alec Eickbusch, and Nicholas Frattini for critical feedback on the manuscript. 
S.P. is supported by the Army Research Office (ARO) under grant number W911NF-18-1-0212. 
A.L.G. is supported by the Australian Research Council, through the Centre of Excellence for Engineered Quantum Systems (EQUS) project number CE170100009 and Discovery Early Career Research Award project number DE190100380.

%\nocite{*}

\bibliography{bosonic.bib}

\begin{thebibliography}{93}%
\makeatletter
\providecommand \@ifxundefined [1]{%
 \@ifx{#1\undefined}
}%
\providecommand \@ifnum [1]{%
 \ifnum #1\expandafter \@firstoftwo
 \else \expandafter \@secondoftwo
 \fi
}%
\providecommand \@ifx [1]{%
 \ifx #1\expandafter \@firstoftwo
 \else \expandafter \@secondoftwo
 \fi
}%
\providecommand \natexlab [1]{#1}%
\providecommand \enquote  [1]{``#1''}%
\providecommand \bibnamefont  [1]{#1}%
\providecommand \bibfnamefont [1]{#1}%
\providecommand \citenamefont [1]{#1}%
\providecommand \href@noop [0]{\@secondoftwo}%
\providecommand \href [0]{\begingroup \@sanitize@url \@href}%
\providecommand \@href[1]{\@@startlink{#1}\@@href}%
\providecommand \@@href[1]{\endgroup#1\@@endlink}%
\providecommand \@sanitize@url [0]{\catcode `\\12\catcode `\$12\catcode
  `\&12\catcode `\#12\catcode `\^12\catcode `\_12\catcode `\%12\relax}%
\providecommand \@@startlink[1]{}%
\providecommand \@@endlink[0]{}%
\providecommand \url  [0]{\begingroup\@sanitize@url \@url }%
\providecommand \@url [1]{\endgroup\@href {#1}{\urlprefix }}%
\providecommand \urlprefix  [0]{URL }%
\providecommand \Eprint [0]{\href }%
\providecommand \doibase [0]{https://doi.org/}%
\providecommand \selectlanguage [0]{\@gobble}%
\providecommand \bibinfo  [0]{\@secondoftwo}%
\providecommand \bibfield  [0]{\@secondoftwo}%
\providecommand \translation [1]{[#1]}%
\providecommand \BibitemOpen [0]{}%
\providecommand \bibitemStop [0]{}%
\providecommand \bibitemNoStop [0]{.\EOS\space}%
\providecommand \EOS [0]{\spacefactor3000\relax}%
\providecommand \BibitemShut  [1]{\csname bibitem#1\endcsname}%
\let\auto@bib@innerbib\@empty
%</preamble>
\bibitem [{\citenamefont {Gottesman}\ \emph {et~al.}(2001)\citenamefont
  {Gottesman}, \citenamefont {Kitaev},\ and\ \citenamefont
  {Preskill}}]{Gottesman01}%
  \BibitemOpen
  \bibfield  {author} {\bibinfo {author} {\bibfnamefont {D.}~\bibnamefont
  {Gottesman}}, \bibinfo {author} {\bibfnamefont {A.}~\bibnamefont {Kitaev}},\
  and\ \bibinfo {author} {\bibfnamefont {J.}~\bibnamefont {Preskill}},\
  }\bibfield  {title} {\bibinfo {title} {Encoding a qubit in an oscillator},\
  }\href {https://doi.org/10.1103/PhysRevA.64.012310} {\bibfield  {journal}
  {\bibinfo  {journal} {Phys. Rev. A}\ }\textbf {\bibinfo {volume} {64}},\
  \bibinfo {pages} {012310} (\bibinfo {year} {2001})}\BibitemShut {NoStop}%
\bibitem [{\citenamefont {Fl{\"u}hmann}\ \emph {et~al.}(2019)\citenamefont
  {Fl{\"u}hmann}, \citenamefont {Nguyen}, \citenamefont {Marinelli},
  \citenamefont {Negnevitsky}, \citenamefont {Mehta},\ and\ \citenamefont
  {Home}}]{Fluhmann:2019aa}%
  \BibitemOpen
  \bibfield  {author} {\bibinfo {author} {\bibfnamefont {C.}~\bibnamefont
  {Fl{\"u}hmann}}, \bibinfo {author} {\bibfnamefont {T.~L.}\ \bibnamefont
  {Nguyen}}, \bibinfo {author} {\bibfnamefont {M.}~\bibnamefont {Marinelli}},
  \bibinfo {author} {\bibfnamefont {V.}~\bibnamefont {Negnevitsky}}, \bibinfo
  {author} {\bibfnamefont {K.}~\bibnamefont {Mehta}},\ and\ \bibinfo {author}
  {\bibfnamefont {J.~P.}\ \bibnamefont {Home}},\ }\bibfield  {title} {\bibinfo
  {title} {Encoding a qubit in a trapped-ion mechanical oscillator},\ }\href
  {https://doi.org/10.1038/s41586-019-0960-6} {\bibfield  {journal} {\bibinfo
  {journal} {Nature}\ }\textbf {\bibinfo {volume} {566}},\ \bibinfo {pages}
  {513} (\bibinfo {year} {2019})}\BibitemShut {NoStop}%
\bibitem [{\citenamefont {Campagne-Ibarcq}\ \emph {et~al.}(2020)\citenamefont
  {Campagne-Ibarcq}, \citenamefont {Eickbusch}, \citenamefont {Touzard},
  \citenamefont {Zalys-Geller}, \citenamefont {Frattini}, \citenamefont
  {Sivak}, \citenamefont {Reinhold}, \citenamefont {Puri}, \citenamefont
  {Shankar}, \citenamefont {Schoelkopf} \emph {et~al.}}]{Campagne2020}%
  \BibitemOpen
  \bibfield  {author} {\bibinfo {author} {\bibfnamefont {P.}~\bibnamefont
  {Campagne-Ibarcq}}, \bibinfo {author} {\bibfnamefont {A.}~\bibnamefont
  {Eickbusch}}, \bibinfo {author} {\bibfnamefont {S.}~\bibnamefont {Touzard}},
  \bibinfo {author} {\bibfnamefont {E.}~\bibnamefont {Zalys-Geller}}, \bibinfo
  {author} {\bibfnamefont {N.~E.}\ \bibnamefont {Frattini}}, \bibinfo {author}
  {\bibfnamefont {V.~V.}\ \bibnamefont {Sivak}}, \bibinfo {author}
  {\bibfnamefont {P.}~\bibnamefont {Reinhold}}, \bibinfo {author}
  {\bibfnamefont {S.}~\bibnamefont {Puri}}, \bibinfo {author} {\bibfnamefont
  {S.}~\bibnamefont {Shankar}}, \bibinfo {author} {\bibfnamefont {R.~J.}\
  \bibnamefont {Schoelkopf}}, \emph {et~al.},\ }\bibfield  {title} {\bibinfo
  {title} {Quantum error correction of a qubit encoded in grid states of an
  oscillator},\ }\href@noop {} {\bibfield  {journal} {\bibinfo  {journal}
  {Nature}\ }\textbf {\bibinfo {volume} {584}},\ \bibinfo {pages} {368}
  (\bibinfo {year} {2020})}\BibitemShut {NoStop}%
\bibitem [{\citenamefont {Ofek}\ \emph {et~al.}(2016)\citenamefont {Ofek},
  \citenamefont {Petrenko}, \citenamefont {Heeres}, \citenamefont {Reinhold},
  \citenamefont {Leghtas}, \citenamefont {Vlastakis}, \citenamefont {Liu},
  \citenamefont {Frunzio}, \citenamefont {Girvin}, \citenamefont {Jiang} \emph
  {et~al.}}]{Ofek16}%
  \BibitemOpen
  \bibfield  {author} {\bibinfo {author} {\bibfnamefont {N.}~\bibnamefont
  {Ofek}}, \bibinfo {author} {\bibfnamefont {A.}~\bibnamefont {Petrenko}},
  \bibinfo {author} {\bibfnamefont {R.}~\bibnamefont {Heeres}}, \bibinfo
  {author} {\bibfnamefont {P.}~\bibnamefont {Reinhold}}, \bibinfo {author}
  {\bibfnamefont {Z.}~\bibnamefont {Leghtas}}, \bibinfo {author} {\bibfnamefont
  {B.}~\bibnamefont {Vlastakis}}, \bibinfo {author} {\bibfnamefont
  {Y.}~\bibnamefont {Liu}}, \bibinfo {author} {\bibfnamefont {L.}~\bibnamefont
  {Frunzio}}, \bibinfo {author} {\bibfnamefont {S.}~\bibnamefont {Girvin}},
  \bibinfo {author} {\bibfnamefont {L.}~\bibnamefont {Jiang}}, \emph {et~al.},\
  }\bibfield  {title} {\bibinfo {title} {Extending the lifetime of a quantum
  bit with error correction in superconducting circuits},\ }\href@noop {}
  {\bibfield  {journal} {\bibinfo  {journal} {Nature}\ }\textbf {\bibinfo
  {volume} {536}},\ \bibinfo {pages} {441} (\bibinfo {year}
  {2016})}\BibitemShut {NoStop}%
\bibitem [{\citenamefont {Hu}\ \emph {et~al.}(2019)\citenamefont {Hu},
  \citenamefont {Ma}, \citenamefont {Cai}, \citenamefont {Mu}, \citenamefont
  {Xu}, \citenamefont {Wang}, \citenamefont {Wu}, \citenamefont {Wang},
  \citenamefont {Song}, \citenamefont {Zou}, \citenamefont {Girvin},
  \citenamefont {Duan},\ and\ \citenamefont {Sun}}]{Hu:2019aa}%
  \BibitemOpen
  \bibfield  {author} {\bibinfo {author} {\bibfnamefont {L.}~\bibnamefont
  {Hu}}, \bibinfo {author} {\bibfnamefont {Y.}~\bibnamefont {Ma}}, \bibinfo
  {author} {\bibfnamefont {W.}~\bibnamefont {Cai}}, \bibinfo {author}
  {\bibfnamefont {X.}~\bibnamefont {Mu}}, \bibinfo {author} {\bibfnamefont
  {Y.}~\bibnamefont {Xu}}, \bibinfo {author} {\bibfnamefont {W.}~\bibnamefont
  {Wang}}, \bibinfo {author} {\bibfnamefont {Y.}~\bibnamefont {Wu}}, \bibinfo
  {author} {\bibfnamefont {H.}~\bibnamefont {Wang}}, \bibinfo {author}
  {\bibfnamefont {Y.~P.}\ \bibnamefont {Song}}, \bibinfo {author}
  {\bibfnamefont {C.~L.}\ \bibnamefont {Zou}}, \bibinfo {author} {\bibfnamefont
  {S.~M.}\ \bibnamefont {Girvin}}, \bibinfo {author} {\bibfnamefont {L.-M.}\
  \bibnamefont {Duan}},\ and\ \bibinfo {author} {\bibfnamefont
  {L.}~\bibnamefont {Sun}},\ }\bibfield  {title} {\bibinfo {title} {Quantum
  error correction and universal gate set operation on a binomial bosonic
  logical qubit},\ }\bibfield  {journal} {\bibinfo  {journal} {Nature Physics}\
  }\href {https://doi.org/10.1038/s41567-018-0414-3}
  {10.1038/s41567-018-0414-3} (\bibinfo {year} {2019})\BibitemShut {NoStop}%
\bibitem [{\citenamefont {Heeres}\ \emph {et~al.}(2017)\citenamefont {Heeres},
  \citenamefont {Reinhold}, \citenamefont {Ofek}, \citenamefont {Frunzio},
  \citenamefont {Jiang}, \citenamefont {Devoret},\ and\ \citenamefont
  {Schoelkopf}}]{heeres2017implementing}%
  \BibitemOpen
  \bibfield  {author} {\bibinfo {author} {\bibfnamefont {R.~W.}\ \bibnamefont
  {Heeres}}, \bibinfo {author} {\bibfnamefont {P.}~\bibnamefont {Reinhold}},
  \bibinfo {author} {\bibfnamefont {N.}~\bibnamefont {Ofek}}, \bibinfo {author}
  {\bibfnamefont {L.}~\bibnamefont {Frunzio}}, \bibinfo {author} {\bibfnamefont
  {L.}~\bibnamefont {Jiang}}, \bibinfo {author} {\bibfnamefont {M.~H.}\
  \bibnamefont {Devoret}},\ and\ \bibinfo {author} {\bibfnamefont {R.~J.}\
  \bibnamefont {Schoelkopf}},\ }\bibfield  {title} {\bibinfo {title}
  {Implementing a universal gate set on a logical qubit encoded in an
  oscillator},\ }\href@noop {} {\bibfield  {journal} {\bibinfo  {journal}
  {Nature communications}\ }\textbf {\bibinfo {volume} {8}},\ \bibinfo {pages}
  {1} (\bibinfo {year} {2017})}\BibitemShut {NoStop}%
\bibitem [{\citenamefont {Xu}\ \emph {et~al.}(2020)\citenamefont {Xu},
  \citenamefont {Ma}, \citenamefont {Cai}, \citenamefont {Mu}, \citenamefont
  {Dai}, \citenamefont {Wang}, \citenamefont {Hu}, \citenamefont {Li},
  \citenamefont {Han}, \citenamefont {Wang} \emph
  {et~al.}}]{xu2020demonstration}%
  \BibitemOpen
  \bibfield  {author} {\bibinfo {author} {\bibfnamefont {Y.}~\bibnamefont
  {Xu}}, \bibinfo {author} {\bibfnamefont {Y.}~\bibnamefont {Ma}}, \bibinfo
  {author} {\bibfnamefont {W.}~\bibnamefont {Cai}}, \bibinfo {author}
  {\bibfnamefont {X.}~\bibnamefont {Mu}}, \bibinfo {author} {\bibfnamefont
  {W.}~\bibnamefont {Dai}}, \bibinfo {author} {\bibfnamefont {W.}~\bibnamefont
  {Wang}}, \bibinfo {author} {\bibfnamefont {L.}~\bibnamefont {Hu}}, \bibinfo
  {author} {\bibfnamefont {X.}~\bibnamefont {Li}}, \bibinfo {author}
  {\bibfnamefont {J.}~\bibnamefont {Han}}, \bibinfo {author} {\bibfnamefont
  {H.}~\bibnamefont {Wang}}, \emph {et~al.},\ }\bibfield  {title} {\bibinfo
  {title} {Demonstration of controlled-phase gates between two
  error-correctable photonic qubits},\ }\href@noop {} {\bibfield  {journal}
  {\bibinfo  {journal} {Physical review letters}\ }\textbf {\bibinfo {volume}
  {124}},\ \bibinfo {pages} {120501} (\bibinfo {year} {2020})}\BibitemShut
  {NoStop}%
\bibitem [{\citenamefont {Ma}\ \emph {et~al.}(2020)\citenamefont {Ma},
  \citenamefont {Xu}, \citenamefont {Mu}, \citenamefont {Cai}, \citenamefont
  {Hu}, \citenamefont {Wang}, \citenamefont {Pan}, \citenamefont {Wang},
  \citenamefont {Song}, \citenamefont {Zou},\ and\ \citenamefont
  {Sun}}]{Ma2020}%
  \BibitemOpen
  \bibfield  {author} {\bibinfo {author} {\bibfnamefont {Y.}~\bibnamefont
  {Ma}}, \bibinfo {author} {\bibfnamefont {Y.}~\bibnamefont {Xu}}, \bibinfo
  {author} {\bibfnamefont {X.}~\bibnamefont {Mu}}, \bibinfo {author}
  {\bibfnamefont {W.}~\bibnamefont {Cai}}, \bibinfo {author} {\bibfnamefont
  {L.}~\bibnamefont {Hu}}, \bibinfo {author} {\bibfnamefont {W.}~\bibnamefont
  {Wang}}, \bibinfo {author} {\bibfnamefont {X.}~\bibnamefont {Pan}}, \bibinfo
  {author} {\bibfnamefont {H.}~\bibnamefont {Wang}}, \bibinfo {author}
  {\bibfnamefont {Y.~P.}\ \bibnamefont {Song}}, \bibinfo {author}
  {\bibfnamefont {C.~L.}\ \bibnamefont {Zou}},\ and\ \bibinfo {author}
  {\bibfnamefont {L.}~\bibnamefont {Sun}},\ }\bibfield  {title} {\bibinfo
  {title} {Error-transparent operations on a logical qubit protected by quantum
  error correction},\ }\href {https://doi.org/10.1038/s41567-020-0893-x}
  {\bibfield  {journal} {\bibinfo  {journal} {Nat. Phys.}\ }\textbf {\bibinfo
  {volume} {16}},\ \bibinfo {pages} {827} (\bibinfo {year} {2020})},\ \Eprint
  {https://arxiv.org/abs/1909.06803} {arXiv:1909.06803 [quant-ph]} \BibitemShut
  {NoStop}%
\bibitem [{\citenamefont {Gertler}\ \emph {et~al.}(2021)\citenamefont
  {Gertler}, \citenamefont {Baker}, \citenamefont {Li}, \citenamefont {Shirol},
  \citenamefont {Koch},\ and\ \citenamefont {Wang}}]{gertler2021protecting}%
  \BibitemOpen
  \bibfield  {author} {\bibinfo {author} {\bibfnamefont {J.~M.}\ \bibnamefont
  {Gertler}}, \bibinfo {author} {\bibfnamefont {B.}~\bibnamefont {Baker}},
  \bibinfo {author} {\bibfnamefont {J.}~\bibnamefont {Li}}, \bibinfo {author}
  {\bibfnamefont {S.}~\bibnamefont {Shirol}}, \bibinfo {author} {\bibfnamefont
  {J.}~\bibnamefont {Koch}},\ and\ \bibinfo {author} {\bibfnamefont
  {C.}~\bibnamefont {Wang}},\ }\bibfield  {title} {\bibinfo {title} {Protecting
  a bosonic qubit with autonomous quantum error correction},\ }\href@noop {}
  {\bibfield  {journal} {\bibinfo  {journal} {Nature}\ }\textbf {\bibinfo
  {volume} {590}},\ \bibinfo {pages} {243} (\bibinfo {year}
  {2021})}\BibitemShut {NoStop}%
\bibitem [{\citenamefont {de~Neeve}\ \emph {et~al.}(2020)\citenamefont
  {de~Neeve}, \citenamefont {Nguyen}, \citenamefont {Behrle},\ and\
  \citenamefont {Home}}]{de2020error}%
  \BibitemOpen
  \bibfield  {author} {\bibinfo {author} {\bibfnamefont {B.}~\bibnamefont
  {de~Neeve}}, \bibinfo {author} {\bibfnamefont {T.~L.}\ \bibnamefont
  {Nguyen}}, \bibinfo {author} {\bibfnamefont {T.}~\bibnamefont {Behrle}},\
  and\ \bibinfo {author} {\bibfnamefont {J.}~\bibnamefont {Home}},\ }\bibfield
  {title} {\bibinfo {title} {Error correction of a logical grid state qubit by
  dissipative pumping},\ }\href@noop {} {\bibfield  {journal} {\bibinfo
  {journal} {arXiv:2010.09681}\ } (\bibinfo {year} {2020})}\BibitemShut
  {NoStop}%
\bibitem [{\citenamefont {Chuang}\ \emph {et~al.}(1997)\citenamefont {Chuang},
  \citenamefont {Leung},\ and\ \citenamefont {Yamamoto}}]{ChuaLeunYama97}%
  \BibitemOpen
  \bibfield  {author} {\bibinfo {author} {\bibfnamefont {I.~L.}\ \bibnamefont
  {Chuang}}, \bibinfo {author} {\bibfnamefont {D.~W.}\ \bibnamefont {Leung}},\
  and\ \bibinfo {author} {\bibfnamefont {Y.}~\bibnamefont {Yamamoto}},\
  }\bibfield  {title} {\bibinfo {title} {Bosonic quantum codes for amplitude
  damping},\ }\href {https://dx.doi.org/10.1103/PhysRevA.56.1114} {\bibfield
  {journal} {\bibinfo  {journal} {Phys. Rev. A}\ }\textbf {\bibinfo {volume}
  {56}},\ \bibinfo {pages} {1114} (\bibinfo {year} {1997})}\BibitemShut
  {NoStop}%
\bibitem [{\citenamefont {Cochrane}\ \emph {et~al.}(1999)\citenamefont
  {Cochrane}, \citenamefont {Milburn},\ and\ \citenamefont
  {Munro}}]{CochMilbMunr99}%
  \BibitemOpen
  \bibfield  {author} {\bibinfo {author} {\bibfnamefont {P.~T.}\ \bibnamefont
  {Cochrane}}, \bibinfo {author} {\bibfnamefont {G.~J.}\ \bibnamefont
  {Milburn}},\ and\ \bibinfo {author} {\bibfnamefont {W.~J.}\ \bibnamefont
  {Munro}},\ }\bibfield  {title} {\bibinfo {title} {Macroscopically distinct
  quantum-superposition states as a bosonic code for amplitude damping},\
  }\href {https://doi.org/10.1103/PhysRevA.59.2631} {\bibfield  {journal}
  {\bibinfo  {journal} {Phys. Rev. A}\ }\textbf {\bibinfo {volume} {59}},\
  \bibinfo {pages} {2631} (\bibinfo {year} {1999})}\BibitemShut {NoStop}%
\bibitem [{\citenamefont {Fl\"uhmann}\ \emph {et~al.}(2018)\citenamefont
  {Fl\"uhmann}, \citenamefont {Negnevitsky}, \citenamefont {Marinelli},\ and\
  \citenamefont {Home}}]{Fluhmann:2018aa}%
  \BibitemOpen
  \bibfield  {author} {\bibinfo {author} {\bibfnamefont {C.}~\bibnamefont
  {Fl\"uhmann}}, \bibinfo {author} {\bibfnamefont {V.}~\bibnamefont
  {Negnevitsky}}, \bibinfo {author} {\bibfnamefont {M.}~\bibnamefont
  {Marinelli}},\ and\ \bibinfo {author} {\bibfnamefont {J.~P.}\ \bibnamefont
  {Home}},\ }\bibfield  {title} {\bibinfo {title} {Sequential modular position
  and momentum measurements of a trapped ion mechanical oscillator},\ }\href
  {https://doi.org/10.1103/PhysRevX.8.021001} {\bibfield  {journal} {\bibinfo
  {journal} {Phys. Rev. X}\ }\textbf {\bibinfo {volume} {8}},\ \bibinfo {pages}
  {021001} (\bibinfo {year} {2018})}\BibitemShut {NoStop}%
\bibitem [{\citenamefont {Takeda}\ and\ \citenamefont
  {Furusawa}(2019)}]{takeda2019toward}%
  \BibitemOpen
  \bibfield  {author} {\bibinfo {author} {\bibfnamefont {S.}~\bibnamefont
  {Takeda}}\ and\ \bibinfo {author} {\bibfnamefont {A.}~\bibnamefont
  {Furusawa}},\ }\bibfield  {title} {\bibinfo {title} {Toward large-scale
  fault-tolerant universal photonic quantum computing},\ }\href@noop {}
  {\bibfield  {journal} {\bibinfo  {journal} {APL Photonics}\ }\textbf
  {\bibinfo {volume} {4}},\ \bibinfo {pages} {060902} (\bibinfo {year}
  {2019})}\BibitemShut {NoStop}%
\bibitem [{\citenamefont {Walshe}\ \emph {et~al.}(2020)\citenamefont {Walshe},
  \citenamefont {Baragiola}, \citenamefont {Alexander},\ and\ \citenamefont
  {Menicucci}}]{Walshe2020}%
  \BibitemOpen
  \bibfield  {author} {\bibinfo {author} {\bibfnamefont {B.~W.}\ \bibnamefont
  {Walshe}}, \bibinfo {author} {\bibfnamefont {B.~Q.}\ \bibnamefont
  {Baragiola}}, \bibinfo {author} {\bibfnamefont {R.~N.}\ \bibnamefont
  {Alexander}},\ and\ \bibinfo {author} {\bibfnamefont {N.~C.}\ \bibnamefont
  {Menicucci}},\ }\bibfield  {title} {\bibinfo {title} {Continuous-variable
  gate teleportation and bosonic-code error correction},\ }\href
  {https://doi.org/10.1103/PhysRevA.102.062411} {\bibfield  {journal} {\bibinfo
   {journal} {Phys. Rev. A}\ }\textbf {\bibinfo {volume} {102}},\ \bibinfo
  {pages} {062411} (\bibinfo {year} {2020})}\BibitemShut {NoStop}%
\bibitem [{\citenamefont {Bourassa}\ \emph {et~al.}(2021)\citenamefont
  {Bourassa}, \citenamefont {Alexander}, \citenamefont {Vasmer}, \citenamefont
  {Patil}, \citenamefont {Tzitrin}, \citenamefont {Matsuura}, \citenamefont
  {Su}, \citenamefont {Baragiola}, \citenamefont {Guha}, \citenamefont
  {Dauphinais} \emph {et~al.}}]{bourassa2021blueprint}%
  \BibitemOpen
  \bibfield  {author} {\bibinfo {author} {\bibfnamefont {J.~E.}\ \bibnamefont
  {Bourassa}}, \bibinfo {author} {\bibfnamefont {R.~N.}\ \bibnamefont
  {Alexander}}, \bibinfo {author} {\bibfnamefont {M.}~\bibnamefont {Vasmer}},
  \bibinfo {author} {\bibfnamefont {A.}~\bibnamefont {Patil}}, \bibinfo
  {author} {\bibfnamefont {I.}~\bibnamefont {Tzitrin}}, \bibinfo {author}
  {\bibfnamefont {T.}~\bibnamefont {Matsuura}}, \bibinfo {author}
  {\bibfnamefont {D.}~\bibnamefont {Su}}, \bibinfo {author} {\bibfnamefont
  {B.~Q.}\ \bibnamefont {Baragiola}}, \bibinfo {author} {\bibfnamefont
  {S.}~\bibnamefont {Guha}}, \bibinfo {author} {\bibfnamefont {G.}~\bibnamefont
  {Dauphinais}}, \emph {et~al.},\ }\bibfield  {title} {\bibinfo {title}
  {Blueprint for a scalable photonic fault-tolerant quantum computer},\
  }\href@noop {} {\bibfield  {journal} {\bibinfo  {journal} {Quantum}\ }\textbf
  {\bibinfo {volume} {5}},\ \bibinfo {pages} {392} (\bibinfo {year}
  {2021})}\BibitemShut {NoStop}%
\bibitem [{\citenamefont {Larsen}\ \emph {et~al.}(2021)\citenamefont {Larsen},
  \citenamefont {Chamberland}, \citenamefont {Noh}, \citenamefont
  {Neergaard-Nielsen},\ and\ \citenamefont {Andersen}}]{larsen2021fault}%
  \BibitemOpen
  \bibfield  {author} {\bibinfo {author} {\bibfnamefont {M.~V.}\ \bibnamefont
  {Larsen}}, \bibinfo {author} {\bibfnamefont {C.}~\bibnamefont {Chamberland}},
  \bibinfo {author} {\bibfnamefont {K.}~\bibnamefont {Noh}}, \bibinfo {author}
  {\bibfnamefont {J.~S.}\ \bibnamefont {Neergaard-Nielsen}},\ and\ \bibinfo
  {author} {\bibfnamefont {U.~L.}\ \bibnamefont {Andersen}},\ }\bibfield
  {title} {\bibinfo {title} {A fault-tolerant continuous-variable
  measurement-based quantum computation architecture},\ }\href@noop {}
  {\bibfield  {journal} {\bibinfo  {journal} {arXiv:2101.03014}\ } (\bibinfo
  {year} {2021})}\BibitemShut {NoStop}%
\bibitem [{\citenamefont {Pirandola}\ \emph {et~al.}(2004)\citenamefont
  {Pirandola}, \citenamefont {Mancini}, \citenamefont {Vitali},\ and\
  \citenamefont {Tombesi}}]{pirandola2004constructing}%
  \BibitemOpen
  \bibfield  {author} {\bibinfo {author} {\bibfnamefont {S.}~\bibnamefont
  {Pirandola}}, \bibinfo {author} {\bibfnamefont {S.}~\bibnamefont {Mancini}},
  \bibinfo {author} {\bibfnamefont {D.}~\bibnamefont {Vitali}},\ and\ \bibinfo
  {author} {\bibfnamefont {P.}~\bibnamefont {Tombesi}},\ }\bibfield  {title}
  {\bibinfo {title} {Constructing finite-dimensional codes with optical
  continuous variables},\ }\href@noop {} {\bibfield  {journal} {\bibinfo
  {journal} {EPL (Europhysics Letters)}\ }\textbf {\bibinfo {volume} {68}},\
  \bibinfo {pages} {323} (\bibinfo {year} {2004})}\BibitemShut {NoStop}%
\bibitem [{\citenamefont {Motes}\ \emph {et~al.}(2017)\citenamefont {Motes},
  \citenamefont {Baragiola}, \citenamefont {Gilchrist},\ and\ \citenamefont
  {Menicucci}}]{MoteBaraGilc17}%
  \BibitemOpen
  \bibfield  {author} {\bibinfo {author} {\bibfnamefont {K.~R.}\ \bibnamefont
  {Motes}}, \bibinfo {author} {\bibfnamefont {B.~Q.}\ \bibnamefont
  {Baragiola}}, \bibinfo {author} {\bibfnamefont {A.}~\bibnamefont
  {Gilchrist}},\ and\ \bibinfo {author} {\bibfnamefont {N.~C.}\ \bibnamefont
  {Menicucci}},\ }\bibfield  {title} {\bibinfo {title} {Encoding qubits into
  oscillators with atomic ensembles and squeezed light},\ }\href
  {https://doi.org/10.1103/PhysRevA.95.053819} {\bibfield  {journal} {\bibinfo
  {journal} {Phys. Rev. A}\ }\textbf {\bibinfo {volume} {95}},\ \bibinfo
  {pages} {053819} (\bibinfo {year} {2017})}\BibitemShut {NoStop}%
\bibitem [{\citenamefont {Eaton}\ \emph {et~al.}(2019)\citenamefont {Eaton},
  \citenamefont {Nehra},\ and\ \citenamefont {Pfister}}]{Eaton:2019aa}%
  \BibitemOpen
  \bibfield  {author} {\bibinfo {author} {\bibfnamefont {M.}~\bibnamefont
  {Eaton}}, \bibinfo {author} {\bibfnamefont {R.}~\bibnamefont {Nehra}},\ and\
  \bibinfo {author} {\bibfnamefont {O.}~\bibnamefont {Pfister}},\ }\bibfield
  {title} {\bibinfo {title} {Gottesman-kitaev-preskill state preparation by
  photon catalysis},\ }\href {https://arxiv.org/abs/1903.01925} {\bibfield
  {journal} {\bibinfo  {journal} {arXiv:1903.01925}\ } (\bibinfo {year}
  {2019})}\BibitemShut {NoStop}%
\bibitem [{\citenamefont {Tzitrin}\ \emph {et~al.}(2020)\citenamefont
  {Tzitrin}, \citenamefont {Bourassa}, \citenamefont {Menicucci},\ and\
  \citenamefont {Sabapathy}}]{Tzitrin2020}%
  \BibitemOpen
  \bibfield  {author} {\bibinfo {author} {\bibfnamefont {I.}~\bibnamefont
  {Tzitrin}}, \bibinfo {author} {\bibfnamefont {J.~E.}\ \bibnamefont
  {Bourassa}}, \bibinfo {author} {\bibfnamefont {N.~C.}\ \bibnamefont
  {Menicucci}},\ and\ \bibinfo {author} {\bibfnamefont {K.~K.}\ \bibnamefont
  {Sabapathy}},\ }\bibfield  {title} {\bibinfo {title} {Progress towards
  practical qubit computation using approximate gottesman-kitaev-preskill
  codes},\ }\href {https://doi.org/10.1103/PhysRevA.101.032315} {\bibfield
  {journal} {\bibinfo  {journal} {Phys. Rev. A}\ }\textbf {\bibinfo {volume}
  {101}},\ \bibinfo {pages} {032315} (\bibinfo {year} {2020})}\BibitemShut
  {NoStop}%
\bibitem [{\citenamefont {Vuillot}\ \emph {et~al.}(2019)\citenamefont
  {Vuillot}, \citenamefont {Asasi}, \citenamefont {Wang}, \citenamefont
  {Pryadko},\ and\ \citenamefont {Terhal}}]{Vuillot2018}%
  \BibitemOpen
  \bibfield  {author} {\bibinfo {author} {\bibfnamefont {C.}~\bibnamefont
  {Vuillot}}, \bibinfo {author} {\bibfnamefont {H.}~\bibnamefont {Asasi}},
  \bibinfo {author} {\bibfnamefont {Y.}~\bibnamefont {Wang}}, \bibinfo {author}
  {\bibfnamefont {L.~P.}\ \bibnamefont {Pryadko}},\ and\ \bibinfo {author}
  {\bibfnamefont {B.~M.}\ \bibnamefont {Terhal}},\ }\bibfield  {title}
  {\bibinfo {title} {Quantum error correction with the toric
  gottesman-kitaev-preskill code},\ }\href
  {https://doi.org/10.1103/PhysRevA.99.032344} {\bibfield  {journal} {\bibinfo
  {journal} {Phys. Rev. A}\ }\textbf {\bibinfo {volume} {99}},\ \bibinfo
  {pages} {032344} (\bibinfo {year} {2019})}\BibitemShut {NoStop}%
\bibitem [{\citenamefont {Noh}\ and\ \citenamefont
  {Chamberland}(2019)}]{Noh:2019aa}%
  \BibitemOpen
  \bibfield  {author} {\bibinfo {author} {\bibfnamefont {K.}~\bibnamefont
  {Noh}}\ and\ \bibinfo {author} {\bibfnamefont {C.}~\bibnamefont
  {Chamberland}},\ }\bibfield  {title} {\bibinfo {title} {Fault-tolerant
  bosonic quantum error correction with the surface-gkp code},\ }\href
  {https://arxiv.org/abs/1908.03579} {\bibfield  {journal} {\bibinfo  {journal}
  {arXiv:1908.03579}\ } (\bibinfo {year} {2019})}\BibitemShut {NoStop}%
\bibitem [{\citenamefont {Terhal}\ \emph {et~al.}(2020)\citenamefont {Terhal},
  \citenamefont {Conrad},\ and\ \citenamefont {Vuillot}}]{Terhal2020}%
  \BibitemOpen
  \bibfield  {author} {\bibinfo {author} {\bibfnamefont {B.~M.}\ \bibnamefont
  {Terhal}}, \bibinfo {author} {\bibfnamefont {J.}~\bibnamefont {Conrad}},\
  and\ \bibinfo {author} {\bibfnamefont {C.}~\bibnamefont {Vuillot}},\
  }\bibfield  {title} {\bibinfo {title} {Towards scalable bosonic quantum error
  correction},\ }\href {https://doi.org/10.1088/2058-9565/ab98a5} {\bibfield
  {journal} {\bibinfo  {journal} {Quantum Science and Technology}\ }\textbf
  {\bibinfo {volume} {5}},\ \bibinfo {pages} {043001} (\bibinfo {year}
  {2020})}\BibitemShut {NoStop}%
\bibitem [{\citenamefont {Noh}\ \emph {et~al.}(2021)\citenamefont {Noh},
  \citenamefont {Chamberland},\ and\ \citenamefont {Brand{\~a}o}}]{noh2021low}%
  \BibitemOpen
  \bibfield  {author} {\bibinfo {author} {\bibfnamefont {K.}~\bibnamefont
  {Noh}}, \bibinfo {author} {\bibfnamefont {C.}~\bibnamefont {Chamberland}},\
  and\ \bibinfo {author} {\bibfnamefont {F.~G.}\ \bibnamefont {Brand{\~a}o}},\
  }\bibfield  {title} {\bibinfo {title} {Low overhead fault-tolerant quantum
  error correction with the surface-gkp code},\ }\href@noop {} {\bibfield
  {journal} {\bibinfo  {journal} {arXiv preprint arXiv:2103.06994}\ } (\bibinfo
  {year} {2021})}\BibitemShut {NoStop}%
\bibitem [{\citenamefont {Cai}\ \emph {et~al.}(2021)\citenamefont {Cai},
  \citenamefont {Ma}, \citenamefont {Wang}, \citenamefont {Zou},\ and\
  \citenamefont {Sun}}]{cai2021bosonic}%
  \BibitemOpen
  \bibfield  {author} {\bibinfo {author} {\bibfnamefont {W.}~\bibnamefont
  {Cai}}, \bibinfo {author} {\bibfnamefont {Y.}~\bibnamefont {Ma}}, \bibinfo
  {author} {\bibfnamefont {W.}~\bibnamefont {Wang}}, \bibinfo {author}
  {\bibfnamefont {C.-L.}\ \bibnamefont {Zou}},\ and\ \bibinfo {author}
  {\bibfnamefont {L.}~\bibnamefont {Sun}},\ }\bibfield  {title} {\bibinfo
  {title} {Bosonic quantum error correction codes in superconducting quantum
  circuits},\ }\href@noop {} {\bibfield  {journal} {\bibinfo  {journal}
  {Fundamental Research}\ } (\bibinfo {year} {2021})}\BibitemShut {NoStop}%
\bibitem [{\citenamefont {Ma}\ \emph {et~al.}(2021)\citenamefont {Ma},
  \citenamefont {Puri}, \citenamefont {Schoelkopf}, \citenamefont {Devoret},
  \citenamefont {Girvin},\ and\ \citenamefont {Jiang}}]{ma2021quantum}%
  \BibitemOpen
  \bibfield  {author} {\bibinfo {author} {\bibfnamefont {W.-L.}\ \bibnamefont
  {Ma}}, \bibinfo {author} {\bibfnamefont {S.}~\bibnamefont {Puri}}, \bibinfo
  {author} {\bibfnamefont {R.~J.}\ \bibnamefont {Schoelkopf}}, \bibinfo
  {author} {\bibfnamefont {M.~H.}\ \bibnamefont {Devoret}}, \bibinfo {author}
  {\bibfnamefont {S.}~\bibnamefont {Girvin}},\ and\ \bibinfo {author}
  {\bibfnamefont {L.}~\bibnamefont {Jiang}},\ }\bibfield  {title} {\bibinfo
  {title} {Quantum control of bosonic modes with superconducting circuits},\
  }\href@noop {} {\bibfield  {journal} {\bibinfo  {journal} {arXiv preprint
  arXiv:2102.09668}\ } (\bibinfo {year} {2021})}\BibitemShut {NoStop}%
\bibitem [{\citenamefont {Joshi}\ \emph {et~al.}(2021)\citenamefont {Joshi},
  \citenamefont {Noh},\ and\ \citenamefont {Gao}}]{joshi2021quantum}%
  \BibitemOpen
  \bibfield  {author} {\bibinfo {author} {\bibfnamefont {A.}~\bibnamefont
  {Joshi}}, \bibinfo {author} {\bibfnamefont {K.}~\bibnamefont {Noh}},\ and\
  \bibinfo {author} {\bibfnamefont {Y.~Y.}\ \bibnamefont {Gao}},\ }\bibfield
  {title} {\bibinfo {title} {Quantum information processing with bosonic qubits
  in circuit qed},\ }\href@noop {} {\bibfield  {journal} {\bibinfo  {journal}
  {Quantum Science and Technology}\ } (\bibinfo {year} {2021})}\BibitemShut
  {NoStop}%
\bibitem [{\citenamefont {Grimsmo}\ \emph {et~al.}(2020)\citenamefont
  {Grimsmo}, \citenamefont {Combes},\ and\ \citenamefont
  {Baragiola}}]{Grimsmo2020}%
  \BibitemOpen
  \bibfield  {author} {\bibinfo {author} {\bibfnamefont {A.~L.}\ \bibnamefont
  {Grimsmo}}, \bibinfo {author} {\bibfnamefont {J.}~\bibnamefont {Combes}},\
  and\ \bibinfo {author} {\bibfnamefont {B.~Q.}\ \bibnamefont {Baragiola}},\
  }\bibfield  {title} {\bibinfo {title} {Quantum computing with
  rotation-symmetric bosonic codes},\ }\href@noop {} {\bibfield  {journal}
  {\bibinfo  {journal} {Physical Review X}\ }\textbf {\bibinfo {volume} {10}},\
  \bibinfo {pages} {011058} (\bibinfo {year} {2020})}\BibitemShut {NoStop}%
\bibitem [{\citenamefont {Menicucci}(2014)}]{Menicucci14}%
  \BibitemOpen
  \bibfield  {author} {\bibinfo {author} {\bibfnamefont {N.~C.}\ \bibnamefont
  {Menicucci}},\ }\bibfield  {title} {\bibinfo {title} {Fault-tolerant
  measurement-based quantum computing with continuous-variable cluster
  states},\ }\href {https://doi.org/10.1103/PhysRevLett.112.120504} {\bibfield
  {journal} {\bibinfo  {journal} {Phys. Rev. Lett.}\ }\textbf {\bibinfo
  {volume} {112}},\ \bibinfo {pages} {120504} (\bibinfo {year}
  {2014})}\BibitemShut {NoStop}%
\bibitem [{\citenamefont {Royer}\ \emph {et~al.}(2020)\citenamefont {Royer},
  \citenamefont {Singh},\ and\ \citenamefont {Girvin}}]{Royer2020}%
  \BibitemOpen
  \bibfield  {author} {\bibinfo {author} {\bibfnamefont {B.}~\bibnamefont
  {Royer}}, \bibinfo {author} {\bibfnamefont {S.}~\bibnamefont {Singh}},\ and\
  \bibinfo {author} {\bibfnamefont {S.~M.}\ \bibnamefont {Girvin}},\ }\bibfield
   {title} {\bibinfo {title} {Stabilization of finite-energy
  gottesman-kitaev-preskill states},\ }\href
  {https://doi.org/10.1103/PhysRevLett.125.260509} {\bibfield  {journal}
  {\bibinfo  {journal} {Phys. Rev. Lett.}\ }\textbf {\bibinfo {volume} {125}},\
  \bibinfo {pages} {260509} (\bibinfo {year} {2020})}\BibitemShut {NoStop}%
\bibitem [{\citenamefont {Duivenvoorden}\ \emph {et~al.}(2017)\citenamefont
  {Duivenvoorden}, \citenamefont {Terhal},\ and\ \citenamefont
  {Weigand}}]{Duivenvoorden2017}%
  \BibitemOpen
  \bibfield  {author} {\bibinfo {author} {\bibfnamefont {K.}~\bibnamefont
  {Duivenvoorden}}, \bibinfo {author} {\bibfnamefont {B.~M.}\ \bibnamefont
  {Terhal}},\ and\ \bibinfo {author} {\bibfnamefont {D.}~\bibnamefont
  {Weigand}},\ }\bibfield  {title} {\bibinfo {title} {Single-mode displacement
  sensor},\ }\href {https://doi.org/10.1103/PhysRevA.95.012305} {\bibfield
  {journal} {\bibinfo  {journal} {Phys. Rev. A}\ }\textbf {\bibinfo {volume}
  {95}},\ \bibinfo {pages} {012305} (\bibinfo {year} {2017})}\BibitemShut
  {NoStop}%
\bibitem [{\citenamefont {Nielsen}\ and\ \citenamefont
  {Chuang}(2010)}]{Nielsen10}%
  \BibitemOpen
  \bibfield  {author} {\bibinfo {author} {\bibfnamefont {M.~A.}\ \bibnamefont
  {Nielsen}}\ and\ \bibinfo {author} {\bibfnamefont {I.~L.}\ \bibnamefont
  {Chuang}},\ }\href@noop {} {\emph {\bibinfo {title} {Quantum computation and
  quantum information}}}\ (\bibinfo  {publisher} {Cambridge university press},\
  \bibinfo {year} {2010})\BibitemShut {NoStop}%
\bibitem [{\citenamefont {Nielsen}(2002)}]{Nielsen02}%
  \BibitemOpen
  \bibfield  {author} {\bibinfo {author} {\bibfnamefont {M.~A.}\ \bibnamefont
  {Nielsen}},\ }\bibfield  {title} {\bibinfo {title} {A simple formula for the
  average gate fidelity of a quantum dynamical operation},\ }\href@noop {}
  {\bibfield  {journal} {\bibinfo  {journal} {Phys. Lett. A}\ }\textbf
  {\bibinfo {volume} {303}},\ \bibinfo {pages} {249} (\bibinfo {year}
  {2002})}\BibitemShut {NoStop}%
\bibitem [{\citenamefont {Albert}\ \emph {et~al.}(2018)\citenamefont {Albert},
  \citenamefont {Noh}, \citenamefont {Duivenvoorden}, \citenamefont {Young},
  \citenamefont {Brierley}, \citenamefont {Reinhold}, \citenamefont {Vuillot},
  \citenamefont {Li}, \citenamefont {Shen}, \citenamefont {Girvin},
  \citenamefont {Terhal},\ and\ \citenamefont {Jiang}}]{Albert17}%
  \BibitemOpen
  \bibfield  {author} {\bibinfo {author} {\bibfnamefont {V.~V.}\ \bibnamefont
  {Albert}}, \bibinfo {author} {\bibfnamefont {K.}~\bibnamefont {Noh}},
  \bibinfo {author} {\bibfnamefont {K.}~\bibnamefont {Duivenvoorden}}, \bibinfo
  {author} {\bibfnamefont {D.~J.}\ \bibnamefont {Young}}, \bibinfo {author}
  {\bibfnamefont {R.~T.}\ \bibnamefont {Brierley}}, \bibinfo {author}
  {\bibfnamefont {P.}~\bibnamefont {Reinhold}}, \bibinfo {author}
  {\bibfnamefont {C.}~\bibnamefont {Vuillot}}, \bibinfo {author} {\bibfnamefont
  {L.}~\bibnamefont {Li}}, \bibinfo {author} {\bibfnamefont {C.}~\bibnamefont
  {Shen}}, \bibinfo {author} {\bibfnamefont {S.~M.}\ \bibnamefont {Girvin}},
  \bibinfo {author} {\bibfnamefont {B.~M.}\ \bibnamefont {Terhal}},\ and\
  \bibinfo {author} {\bibfnamefont {L.}~\bibnamefont {Jiang}},\ }\bibfield
  {title} {\bibinfo {title} {Performance and structure of single-mode bosonic
  codes},\ }\href {https://doi.org/10.1103/PhysRevA.97.032346} {\bibfield
  {journal} {\bibinfo  {journal} {Phys. Rev. A}\ }\textbf {\bibinfo {volume}
  {97}},\ \bibinfo {pages} {032346} (\bibinfo {year} {2018})}\BibitemShut
  {NoStop}%
\bibitem [{\citenamefont {Pfaff}\ \emph {et~al.}(2017)\citenamefont {Pfaff},
  \citenamefont {Axline}, \citenamefont {Burkhart}, \citenamefont {Vool},
  \citenamefont {Reinhold}, \citenamefont {Frunzio}, \citenamefont {Jiang},
  \citenamefont {Devoret},\ and\ \citenamefont
  {Schoelkopf}}]{pfaff2017controlled}%
  \BibitemOpen
  \bibfield  {author} {\bibinfo {author} {\bibfnamefont {W.}~\bibnamefont
  {Pfaff}}, \bibinfo {author} {\bibfnamefont {C.~J.}\ \bibnamefont {Axline}},
  \bibinfo {author} {\bibfnamefont {L.~D.}\ \bibnamefont {Burkhart}}, \bibinfo
  {author} {\bibfnamefont {U.}~\bibnamefont {Vool}}, \bibinfo {author}
  {\bibfnamefont {P.}~\bibnamefont {Reinhold}}, \bibinfo {author}
  {\bibfnamefont {L.}~\bibnamefont {Frunzio}}, \bibinfo {author} {\bibfnamefont
  {L.}~\bibnamefont {Jiang}}, \bibinfo {author} {\bibfnamefont {M.~H.}\
  \bibnamefont {Devoret}},\ and\ \bibinfo {author} {\bibfnamefont {R.~J.}\
  \bibnamefont {Schoelkopf}},\ }\bibfield  {title} {\bibinfo {title}
  {Controlled release of multiphoton quantum states from a microwave cavity
  memory},\ }\href@noop {} {\bibfield  {journal} {\bibinfo  {journal} {Nature
  Physics}\ }\textbf {\bibinfo {volume} {13}},\ \bibinfo {pages} {882}
  (\bibinfo {year} {2017})}\BibitemShut {NoStop}%
\bibitem [{\citenamefont {Shaw}\ and\ \citenamefont {Grimsmo}()}]{Shawinprep}%
  \BibitemOpen
  \bibfield  {author} {\bibinfo {author} {\bibfnamefont {M.~H.}\ \bibnamefont
  {Shaw}}\ and\ \bibinfo {author} {\bibfnamefont {A.~L.}\ \bibnamefont
  {Grimsmo}},\ }\href@noop {} {\bibinfo  {journal} {in preparation}\
  }\BibitemShut {NoStop}%
\bibitem [{\citenamefont {Macklin}\ \emph {et~al.}(2015)\citenamefont
  {Macklin}, \citenamefont {O{\textquoteright}Brien}, \citenamefont {Hover},
  \citenamefont {Schwartz}, \citenamefont {Bolkhovsky}, \citenamefont {Zhang},
  \citenamefont {Oliver},\ and\ \citenamefont {Siddiqi}}]{Macklin2015}%
  \BibitemOpen
\bibfield  {journal} {  }\bibfield  {author} {\bibinfo {author} {\bibfnamefont
  {C.}~\bibnamefont {Macklin}}, \bibinfo {author} {\bibfnamefont
  {K.}~\bibnamefont {O{\textquoteright}Brien}}, \bibinfo {author}
  {\bibfnamefont {D.}~\bibnamefont {Hover}}, \bibinfo {author} {\bibfnamefont
  {M.~E.}\ \bibnamefont {Schwartz}}, \bibinfo {author} {\bibfnamefont
  {V.}~\bibnamefont {Bolkhovsky}}, \bibinfo {author} {\bibfnamefont
  {X.}~\bibnamefont {Zhang}}, \bibinfo {author} {\bibfnamefont {W.~D.}\
  \bibnamefont {Oliver}},\ and\ \bibinfo {author} {\bibfnamefont
  {I.}~\bibnamefont {Siddiqi}},\ }\bibfield  {title} {\bibinfo {title} {A
  near{\textendash}quantum-limited josephson traveling-wave parametric
  amplifier},\ }\href {https://doi.org/10.1126/science.aaa8525} {\bibfield
  {journal} {\bibinfo  {journal} {Science}\ }\textbf {\bibinfo {volume}
  {350}},\ \bibinfo {pages} {307} (\bibinfo {year} {2015})},\ \Eprint
  {https://arxiv.org/abs/https://science.sciencemag.org/content/350/6258/307.full.pdf}
  {https://science.sciencemag.org/content/350/6258/307.full.pdf} \BibitemShut
  {NoStop}%
\bibitem [{\citenamefont {Touzard}\ \emph {et~al.}(2019)\citenamefont
  {Touzard}, \citenamefont {Kou}, \citenamefont {Frattini}, \citenamefont
  {Sivak}, \citenamefont {Puri}, \citenamefont {Grimm}, \citenamefont
  {Frunzio}, \citenamefont {Shankar},\ and\ \citenamefont
  {Devoret}}]{Touzard2019}%
  \BibitemOpen
  \bibfield  {author} {\bibinfo {author} {\bibfnamefont {S.}~\bibnamefont
  {Touzard}}, \bibinfo {author} {\bibfnamefont {A.}~\bibnamefont {Kou}},
  \bibinfo {author} {\bibfnamefont {N.~E.}\ \bibnamefont {Frattini}}, \bibinfo
  {author} {\bibfnamefont {V.~V.}\ \bibnamefont {Sivak}}, \bibinfo {author}
  {\bibfnamefont {S.}~\bibnamefont {Puri}}, \bibinfo {author} {\bibfnamefont
  {A.}~\bibnamefont {Grimm}}, \bibinfo {author} {\bibfnamefont
  {L.}~\bibnamefont {Frunzio}}, \bibinfo {author} {\bibfnamefont
  {S.}~\bibnamefont {Shankar}},\ and\ \bibinfo {author} {\bibfnamefont {M.~H.}\
  \bibnamefont {Devoret}},\ }\bibfield  {title} {\bibinfo {title} {Gated
  conditional displacement readout of superconducting qubits},\ }\href
  {https://doi.org/10.1103/PhysRevLett.122.080502} {\bibfield  {journal}
  {\bibinfo  {journal} {Phys. Rev. Lett.}\ }\textbf {\bibinfo {volume} {122}},\
  \bibinfo {pages} {080502} (\bibinfo {year} {2019})}\BibitemShut {NoStop}%
\bibitem [{\citenamefont {Eddins}\ \emph {et~al.}(2019)\citenamefont {Eddins},
  \citenamefont {Kreikebaum}, \citenamefont {Toyli}, \citenamefont
  {Levenson-Falk}, \citenamefont {Dove}, \citenamefont {Livingston},
  \citenamefont {Levitan}, \citenamefont {Govia}, \citenamefont {Clerk},\ and\
  \citenamefont {Siddiqi}}]{Eddins2019}%
  \BibitemOpen
  \bibfield  {author} {\bibinfo {author} {\bibfnamefont {A.}~\bibnamefont
  {Eddins}}, \bibinfo {author} {\bibfnamefont {J.~M.}\ \bibnamefont
  {Kreikebaum}}, \bibinfo {author} {\bibfnamefont {D.~M.}\ \bibnamefont
  {Toyli}}, \bibinfo {author} {\bibfnamefont {E.~M.}\ \bibnamefont
  {Levenson-Falk}}, \bibinfo {author} {\bibfnamefont {A.}~\bibnamefont {Dove}},
  \bibinfo {author} {\bibfnamefont {W.~P.}\ \bibnamefont {Livingston}},
  \bibinfo {author} {\bibfnamefont {B.~A.}\ \bibnamefont {Levitan}}, \bibinfo
  {author} {\bibfnamefont {L.~C.~G.}\ \bibnamefont {Govia}}, \bibinfo {author}
  {\bibfnamefont {A.~A.}\ \bibnamefont {Clerk}},\ and\ \bibinfo {author}
  {\bibfnamefont {I.}~\bibnamefont {Siddiqi}},\ }\bibfield  {title} {\bibinfo
  {title} {High-efficiency measurement of an artificial atom embedded in a
  parametric amplifier},\ }\href {https://doi.org/10.1103/PhysRevX.9.011004}
  {\bibfield  {journal} {\bibinfo  {journal} {Phys. Rev. X}\ }\textbf {\bibinfo
  {volume} {9}},\ \bibinfo {pages} {011004} (\bibinfo {year}
  {2019})}\BibitemShut {NoStop}%
\bibitem [{\citenamefont {Terhal}\ and\ \citenamefont
  {Weigand}(2016)}]{Terhal2016}%
  \BibitemOpen
  \bibfield  {author} {\bibinfo {author} {\bibfnamefont {B.~M.}\ \bibnamefont
  {Terhal}}\ and\ \bibinfo {author} {\bibfnamefont {D.}~\bibnamefont
  {Weigand}},\ }\bibfield  {title} {\bibinfo {title} {Encoding a qubit into a
  cavity mode in circuit qed using phase estimation},\ }\href
  {https://doi.org/10.1103/PhysRevA.93.012315} {\bibfield  {journal} {\bibinfo
  {journal} {Phys. Rev. A}\ }\textbf {\bibinfo {volume} {93}},\ \bibinfo
  {pages} {012315} (\bibinfo {year} {2016})}\BibitemShut {NoStop}%
\bibitem [{\citenamefont {Weigand}(2020)}]{Weigand2020}%
  \BibitemOpen
  \bibfield  {author} {\bibinfo {author} {\bibfnamefont {D.~J.}\ \bibnamefont
  {Weigand}},\ }\emph {\bibinfo {title} {Encoding a qubit into an oscillator
  with near-term experimental devices}},\ \href@noop {} {Ph.D. thesis},\
  \bibinfo  {school} {Delft University of Technology}, \bibinfo {address}
  {Delft} (\bibinfo {year} {2020})\BibitemShut {NoStop}%
\bibitem [{\citenamefont {Hastrup}\ and\ \citenamefont
  {Andersen}(2020)}]{hastrup2020improved}%
  \BibitemOpen
  \bibfield  {author} {\bibinfo {author} {\bibfnamefont {J.}~\bibnamefont
  {Hastrup}}\ and\ \bibinfo {author} {\bibfnamefont {U.~L.}\ \bibnamefont
  {Andersen}},\ }\bibfield  {title} {\bibinfo {title} {Improved readout of
  qubit-coupled gottesman-kitaev-preskill states},\ }\href@noop {} {\bibfield
  {journal} {\bibinfo  {journal} {arXiv:2008.10531}\ } (\bibinfo {year}
  {2020})}\BibitemShut {NoStop}%
\bibitem [{\citenamefont {Chamberland}\ \emph {et~al.}(2018)\citenamefont
  {Chamberland}, \citenamefont {Iyer},\ and\ \citenamefont
  {Poulin}}]{Chamberland2018}%
  \BibitemOpen
  \bibfield  {author} {\bibinfo {author} {\bibfnamefont {C.}~\bibnamefont
  {Chamberland}}, \bibinfo {author} {\bibfnamefont {P.}~\bibnamefont {Iyer}},\
  and\ \bibinfo {author} {\bibfnamefont {D.}~\bibnamefont {Poulin}},\
  }\bibfield  {title} {\bibinfo {title} {Fault-tolerant quantum computing in
  the {P}auli or {C}lifford frame with slow error diagnostics},\ }\href
  {https://doi.org/10.22331/q-2018-01-04-43} {\bibfield  {journal} {\bibinfo
  {journal} {{Quantum}}\ }\textbf {\bibinfo {volume} {2}},\ \bibinfo {pages}
  {43} (\bibinfo {year} {2018})}\BibitemShut {NoStop}%
\bibitem [{\citenamefont {Aaronson}\ and\ \citenamefont
  {Gottesman}(2004)}]{aaronson2004improved}%
  \BibitemOpen
  \bibfield  {author} {\bibinfo {author} {\bibfnamefont {S.}~\bibnamefont
  {Aaronson}}\ and\ \bibinfo {author} {\bibfnamefont {D.}~\bibnamefont
  {Gottesman}},\ }\bibfield  {title} {\bibinfo {title} {Improved simulation of
  stabilizer circuits},\ }\href@noop {} {\bibfield  {journal} {\bibinfo
  {journal} {Physical Review A}\ }\textbf {\bibinfo {volume} {70}},\ \bibinfo
  {pages} {052328} (\bibinfo {year} {2004})}\BibitemShut {NoStop}%
\bibitem [{\citenamefont {Gottesman}(1998)}]{gottesman1998heisenberg}%
  \BibitemOpen
  \bibfield  {author} {\bibinfo {author} {\bibfnamefont {D.}~\bibnamefont
  {Gottesman}},\ }\bibfield  {title} {\bibinfo {title} {The heisenberg
  representation of quantum computers},\ }\href@noop {} {\bibfield  {journal}
  {\bibinfo  {journal} {quant-ph/9807006}\ } (\bibinfo {year}
  {1998})}\BibitemShut {NoStop}%
\bibitem [{\citenamefont {Gao}\ \emph {et~al.}(2019)\citenamefont {Gao},
  \citenamefont {Lester}, \citenamefont {Chou}, \citenamefont {Frunzio},
  \citenamefont {Devoret}, \citenamefont {Jiang}, \citenamefont {Girvin},\ and\
  \citenamefont {Schoelkopf}}]{Gao:2019aa}%
  \BibitemOpen
  \bibfield  {author} {\bibinfo {author} {\bibfnamefont {Y.~Y.}\ \bibnamefont
  {Gao}}, \bibinfo {author} {\bibfnamefont {B.~J.}\ \bibnamefont {Lester}},
  \bibinfo {author} {\bibfnamefont {K.~S.}\ \bibnamefont {Chou}}, \bibinfo
  {author} {\bibfnamefont {L.}~\bibnamefont {Frunzio}}, \bibinfo {author}
  {\bibfnamefont {M.~H.}\ \bibnamefont {Devoret}}, \bibinfo {author}
  {\bibfnamefont {L.}~\bibnamefont {Jiang}}, \bibinfo {author} {\bibfnamefont
  {S.~M.}\ \bibnamefont {Girvin}},\ and\ \bibinfo {author} {\bibfnamefont
  {R.~J.}\ \bibnamefont {Schoelkopf}},\ }\bibfield  {title} {\bibinfo {title}
  {Entanglement of bosonic modes through an engineered exchange interaction},\
  }\href {https://doi.org/10.1038/s41586-019-0970-4} {\bibfield  {journal}
  {\bibinfo  {journal} {Nature}\ }\textbf {\bibinfo {volume} {566}},\ \bibinfo
  {pages} {509} (\bibinfo {year} {2019})}\BibitemShut {NoStop}%
\bibitem [{\citenamefont {Wang}\ \emph {et~al.}(2020)\citenamefont {Wang},
  \citenamefont {Curtis}, \citenamefont {Lester}, \citenamefont {Zhang},
  \citenamefont {Gao}, \citenamefont {Freeze}, \citenamefont {Batista},
  \citenamefont {Vaccaro}, \citenamefont {Chuang}, \citenamefont {Frunzio},
  \citenamefont {Jiang}, \citenamefont {Girvin},\ and\ \citenamefont
  {Schoelkopf}}]{Wang2020}%
  \BibitemOpen
  \bibfield  {author} {\bibinfo {author} {\bibfnamefont {C.~S.}\ \bibnamefont
  {Wang}}, \bibinfo {author} {\bibfnamefont {J.~C.}\ \bibnamefont {Curtis}},
  \bibinfo {author} {\bibfnamefont {B.~J.}\ \bibnamefont {Lester}}, \bibinfo
  {author} {\bibfnamefont {Y.}~\bibnamefont {Zhang}}, \bibinfo {author}
  {\bibfnamefont {Y.~Y.}\ \bibnamefont {Gao}}, \bibinfo {author} {\bibfnamefont
  {J.}~\bibnamefont {Freeze}}, \bibinfo {author} {\bibfnamefont {V.~S.}\
  \bibnamefont {Batista}}, \bibinfo {author} {\bibfnamefont {P.~H.}\
  \bibnamefont {Vaccaro}}, \bibinfo {author} {\bibfnamefont {I.~L.}\
  \bibnamefont {Chuang}}, \bibinfo {author} {\bibfnamefont {L.}~\bibnamefont
  {Frunzio}}, \bibinfo {author} {\bibfnamefont {L.}~\bibnamefont {Jiang}},
  \bibinfo {author} {\bibfnamefont {S.~M.}\ \bibnamefont {Girvin}},\ and\
  \bibinfo {author} {\bibfnamefont {R.~J.}\ \bibnamefont {Schoelkopf}},\
  }\bibfield  {title} {\bibinfo {title} {Efficient multiphoton sampling of
  molecular vibronic spectra on a superconducting bosonic processor},\ }\href
  {https://doi.org/10.1103/PhysRevX.10.021060} {\bibfield  {journal} {\bibinfo
  {journal} {Phys. Rev. X}\ }\textbf {\bibinfo {volume} {10}},\ \bibinfo
  {pages} {021060} (\bibinfo {year} {2020})}\BibitemShut {NoStop}%
\bibitem [{\citenamefont {Grimm}\ \emph {et~al.}(2020)\citenamefont {Grimm},
  \citenamefont {Frattini}, \citenamefont {Puri}, \citenamefont {Mundhada},
  \citenamefont {Touzard}, \citenamefont {Mirrahimi}, \citenamefont {Girvin},
  \citenamefont {Shankar},\ and\ \citenamefont
  {Devoret}}]{grimm2020stabilization}%
  \BibitemOpen
  \bibfield  {author} {\bibinfo {author} {\bibfnamefont {A.}~\bibnamefont
  {Grimm}}, \bibinfo {author} {\bibfnamefont {N.~E.}\ \bibnamefont {Frattini}},
  \bibinfo {author} {\bibfnamefont {S.}~\bibnamefont {Puri}}, \bibinfo {author}
  {\bibfnamefont {S.~O.}\ \bibnamefont {Mundhada}}, \bibinfo {author}
  {\bibfnamefont {S.}~\bibnamefont {Touzard}}, \bibinfo {author} {\bibfnamefont
  {M.}~\bibnamefont {Mirrahimi}}, \bibinfo {author} {\bibfnamefont {S.~M.}\
  \bibnamefont {Girvin}}, \bibinfo {author} {\bibfnamefont {S.}~\bibnamefont
  {Shankar}},\ and\ \bibinfo {author} {\bibfnamefont {M.~H.}\ \bibnamefont
  {Devoret}},\ }\bibfield  {title} {\bibinfo {title} {Stabilization and
  operation of a kerr-cat qubit},\ }\href@noop {} {\bibfield  {journal}
  {\bibinfo  {journal} {Nature}\ }\textbf {\bibinfo {volume} {584}},\ \bibinfo
  {pages} {205} (\bibinfo {year} {2020})}\BibitemShut {NoStop}%
\bibitem [{\citenamefont {Roy}\ and\ \citenamefont
  {Devoret}(2016)}]{roy2016introduction}%
  \BibitemOpen
  \bibfield  {author} {\bibinfo {author} {\bibfnamefont {A.}~\bibnamefont
  {Roy}}\ and\ \bibinfo {author} {\bibfnamefont {M.}~\bibnamefont {Devoret}},\
  }\bibfield  {title} {\bibinfo {title} {Introduction to parametric
  amplification of quantum signals with josephson circuits},\ }\href@noop {}
  {\bibfield  {journal} {\bibinfo  {journal} {Comptes Rendus Physique}\
  }\textbf {\bibinfo {volume} {17}},\ \bibinfo {pages} {740} (\bibinfo {year}
  {2016})}\BibitemShut {NoStop}%
\bibitem [{\citenamefont {Frattini}\ \emph {et~al.}(2017)\citenamefont
  {Frattini}, \citenamefont {Vool}, \citenamefont {Shankar}, \citenamefont
  {Narla}, \citenamefont {Sliwa},\ and\ \citenamefont
  {Devoret}}]{frattini20173}%
  \BibitemOpen
  \bibfield  {author} {\bibinfo {author} {\bibfnamefont {N.}~\bibnamefont
  {Frattini}}, \bibinfo {author} {\bibfnamefont {U.}~\bibnamefont {Vool}},
  \bibinfo {author} {\bibfnamefont {S.}~\bibnamefont {Shankar}}, \bibinfo
  {author} {\bibfnamefont {A.}~\bibnamefont {Narla}}, \bibinfo {author}
  {\bibfnamefont {K.}~\bibnamefont {Sliwa}},\ and\ \bibinfo {author}
  {\bibfnamefont {M.}~\bibnamefont {Devoret}},\ }\bibfield  {title} {\bibinfo
  {title} {3-wave mixing josephson dipole element},\ }\href@noop {} {\bibfield
  {journal} {\bibinfo  {journal} {Applied Physics Letters}\ }\textbf {\bibinfo
  {volume} {110}},\ \bibinfo {pages} {222603} (\bibinfo {year}
  {2017})}\BibitemShut {NoStop}%
\bibitem [{\citenamefont {Hastrup}\ \emph {et~al.}(2021)\citenamefont
  {Hastrup}, \citenamefont {Park}, \citenamefont {Brask}, \citenamefont
  {Filip},\ and\ \citenamefont {Andersen}}]{hastrup2021measurement}%
  \BibitemOpen
  \bibfield  {author} {\bibinfo {author} {\bibfnamefont {J.}~\bibnamefont
  {Hastrup}}, \bibinfo {author} {\bibfnamefont {K.}~\bibnamefont {Park}},
  \bibinfo {author} {\bibfnamefont {J.~B.}\ \bibnamefont {Brask}}, \bibinfo
  {author} {\bibfnamefont {R.}~\bibnamefont {Filip}},\ and\ \bibinfo {author}
  {\bibfnamefont {U.~L.}\ \bibnamefont {Andersen}},\ }\bibfield  {title}
  {\bibinfo {title} {Measurement-free preparation of grid states},\ }\href@noop
  {} {\bibfield  {journal} {\bibinfo  {journal} {npj Quantum Information}\
  }\textbf {\bibinfo {volume} {7}},\ \bibinfo {pages} {1} (\bibinfo {year}
  {2021})}\BibitemShut {NoStop}%
\bibitem [{\citenamefont {Puri}\ \emph {et~al.}(2019)\citenamefont {Puri},
  \citenamefont {Grimm}, \citenamefont {Campagne-Ibarcq}, \citenamefont
  {Eickbusch}, \citenamefont {Noh}, \citenamefont {Roberts}, \citenamefont
  {Jiang}, \citenamefont {Mirrahimi}, \citenamefont {Devoret},\ and\
  \citenamefont {Girvin}}]{Puri:2018aa}%
  \BibitemOpen
  \bibfield  {author} {\bibinfo {author} {\bibfnamefont {S.}~\bibnamefont
  {Puri}}, \bibinfo {author} {\bibfnamefont {A.}~\bibnamefont {Grimm}},
  \bibinfo {author} {\bibfnamefont {P.}~\bibnamefont {Campagne-Ibarcq}},
  \bibinfo {author} {\bibfnamefont {A.}~\bibnamefont {Eickbusch}}, \bibinfo
  {author} {\bibfnamefont {K.}~\bibnamefont {Noh}}, \bibinfo {author}
  {\bibfnamefont {G.}~\bibnamefont {Roberts}}, \bibinfo {author} {\bibfnamefont
  {L.}~\bibnamefont {Jiang}}, \bibinfo {author} {\bibfnamefont
  {M.}~\bibnamefont {Mirrahimi}}, \bibinfo {author} {\bibfnamefont {M.~H.}\
  \bibnamefont {Devoret}},\ and\ \bibinfo {author} {\bibfnamefont {S.~M.}\
  \bibnamefont {Girvin}},\ }\bibfield  {title} {\bibinfo {title} {Stabilized
  cat in a driven nonlinear cavity: a fault-tolerant error syndrome detector},\
  }\href@noop {} {\bibfield  {journal} {\bibinfo  {journal} {Physical Review
  X}\ }\textbf {\bibinfo {volume} {9}},\ \bibinfo {pages} {041009} (\bibinfo
  {year} {2019})}\BibitemShut {NoStop}%
\bibitem [{\citenamefont {Shi}\ \emph {et~al.}(2019)\citenamefont {Shi},
  \citenamefont {Chamberland},\ and\ \citenamefont {Cross}}]{shi2019fault}%
  \BibitemOpen
  \bibfield  {author} {\bibinfo {author} {\bibfnamefont {Y.}~\bibnamefont
  {Shi}}, \bibinfo {author} {\bibfnamefont {C.}~\bibnamefont {Chamberland}},\
  and\ \bibinfo {author} {\bibfnamefont {A.~W.}\ \bibnamefont {Cross}},\
  }\bibfield  {title} {\bibinfo {title} {Fault-tolerant preparation of
  approximate gkp states},\ }\href@noop {} {\bibfield  {journal} {\bibinfo
  {journal} {arXiv:1905.00903}\ } (\bibinfo {year} {2019})}\BibitemShut
  {NoStop}%
\bibitem [{\citenamefont {Rosenblum}\ \emph {et~al.}(2018)\citenamefont
  {Rosenblum}, \citenamefont {Reinhold}, \citenamefont {Mirrahimi},
  \citenamefont {Jiang}, \citenamefont {Frunzio},\ and\ \citenamefont
  {Schoelkopf}}]{Rosenblum2018}%
  \BibitemOpen
  \bibfield  {author} {\bibinfo {author} {\bibfnamefont {S.}~\bibnamefont
  {Rosenblum}}, \bibinfo {author} {\bibfnamefont {P.}~\bibnamefont {Reinhold}},
  \bibinfo {author} {\bibfnamefont {M.}~\bibnamefont {Mirrahimi}}, \bibinfo
  {author} {\bibfnamefont {L.}~\bibnamefont {Jiang}}, \bibinfo {author}
  {\bibfnamefont {L.}~\bibnamefont {Frunzio}},\ and\ \bibinfo {author}
  {\bibfnamefont {R.~J.}\ \bibnamefont {Schoelkopf}},\ }\bibfield  {title}
  {\bibinfo {title} {Fault-tolerant detection of a quantum error},\ }\href
  {https://doi.org/10.1126/science.aat3996} {\bibfield  {journal} {\bibinfo
  {journal} {Science}\ }\textbf {\bibinfo {volume} {361}},\ \bibinfo {pages}
  {266} (\bibinfo {year} {2018})}\BibitemShut {NoStop}%
\bibitem [{\citenamefont {Earnest}\ \emph {et~al.}(2018)\citenamefont
  {Earnest}, \citenamefont {Chakram}, \citenamefont {Lu}, \citenamefont
  {Irons}, \citenamefont {Naik}, \citenamefont {Leung}, \citenamefont {Ocola},
  \citenamefont {Czaplewski}, \citenamefont {Baker}, \citenamefont {Lawrence},
  \citenamefont {Koch},\ and\ \citenamefont {Schuster}}]{Earnest2018}%
  \BibitemOpen
  \bibfield  {author} {\bibinfo {author} {\bibfnamefont {N.}~\bibnamefont
  {Earnest}}, \bibinfo {author} {\bibfnamefont {S.}~\bibnamefont {Chakram}},
  \bibinfo {author} {\bibfnamefont {Y.}~\bibnamefont {Lu}}, \bibinfo {author}
  {\bibfnamefont {N.}~\bibnamefont {Irons}}, \bibinfo {author} {\bibfnamefont
  {R.~K.}\ \bibnamefont {Naik}}, \bibinfo {author} {\bibfnamefont
  {N.}~\bibnamefont {Leung}}, \bibinfo {author} {\bibfnamefont
  {L.}~\bibnamefont {Ocola}}, \bibinfo {author} {\bibfnamefont {D.~A.}\
  \bibnamefont {Czaplewski}}, \bibinfo {author} {\bibfnamefont
  {B.}~\bibnamefont {Baker}}, \bibinfo {author} {\bibfnamefont
  {J.}~\bibnamefont {Lawrence}}, \bibinfo {author} {\bibfnamefont
  {J.}~\bibnamefont {Koch}},\ and\ \bibinfo {author} {\bibfnamefont {D.~I.}\
  \bibnamefont {Schuster}},\ }\bibfield  {title} {\bibinfo {title} {Realization
  of a $\mathrm{\ensuremath{\Lambda}}$ system with metastable states of a
  capacitively shunted fluxonium},\ }\href
  {https://doi.org/10.1103/PhysRevLett.120.150504} {\bibfield  {journal}
  {\bibinfo  {journal} {Phys. Rev. Lett.}\ }\textbf {\bibinfo {volume} {120}},\
  \bibinfo {pages} {150504} (\bibinfo {year} {2018})}\BibitemShut {NoStop}%
\bibitem [{\citenamefont {Gyenis}\ \emph {et~al.}(2021)\citenamefont {Gyenis},
  \citenamefont {Mundada}, \citenamefont {Di~Paolo}, \citenamefont {Hazard},
  \citenamefont {You}, \citenamefont {Schuster}, \citenamefont {Koch},
  \citenamefont {Blais},\ and\ \citenamefont {Houck}}]{Gyenis2021}%
  \BibitemOpen
  \bibfield  {author} {\bibinfo {author} {\bibfnamefont {A.}~\bibnamefont
  {Gyenis}}, \bibinfo {author} {\bibfnamefont {P.~S.}\ \bibnamefont {Mundada}},
  \bibinfo {author} {\bibfnamefont {A.}~\bibnamefont {Di~Paolo}}, \bibinfo
  {author} {\bibfnamefont {T.~M.}\ \bibnamefont {Hazard}}, \bibinfo {author}
  {\bibfnamefont {X.}~\bibnamefont {You}}, \bibinfo {author} {\bibfnamefont
  {D.~I.}\ \bibnamefont {Schuster}}, \bibinfo {author} {\bibfnamefont
  {J.}~\bibnamefont {Koch}}, \bibinfo {author} {\bibfnamefont {A.}~\bibnamefont
  {Blais}},\ and\ \bibinfo {author} {\bibfnamefont {A.~A.}\ \bibnamefont
  {Houck}},\ }\bibfield  {title} {\bibinfo {title} {Experimental realization of
  a protected superconducting circuit derived from the $0$--$\ensuremath{\pi}$
  qubit},\ }\href {https://doi.org/10.1103/PRXQuantum.2.010339} {\bibfield
  {journal} {\bibinfo  {journal} {PRX Quantum}\ }\textbf {\bibinfo {volume}
  {2}},\ \bibinfo {pages} {010339} (\bibinfo {year} {2021})}\BibitemShut
  {NoStop}%
\bibitem [{\citenamefont {Lescanne}\ \emph {et~al.}(2020)\citenamefont
  {Lescanne}, \citenamefont {Villiers}, \citenamefont {Peronnin}, \citenamefont
  {Sarlette}, \citenamefont {Delbecq}, \citenamefont {Huard}, \citenamefont
  {Kontos}, \citenamefont {Mirrahimi},\ and\ \citenamefont
  {Leghtas}}]{lescanne2020exponential}%
  \BibitemOpen
  \bibfield  {author} {\bibinfo {author} {\bibfnamefont {R.}~\bibnamefont
  {Lescanne}}, \bibinfo {author} {\bibfnamefont {M.}~\bibnamefont {Villiers}},
  \bibinfo {author} {\bibfnamefont {T.}~\bibnamefont {Peronnin}}, \bibinfo
  {author} {\bibfnamefont {A.}~\bibnamefont {Sarlette}}, \bibinfo {author}
  {\bibfnamefont {M.}~\bibnamefont {Delbecq}}, \bibinfo {author} {\bibfnamefont
  {B.}~\bibnamefont {Huard}}, \bibinfo {author} {\bibfnamefont
  {T.}~\bibnamefont {Kontos}}, \bibinfo {author} {\bibfnamefont
  {M.}~\bibnamefont {Mirrahimi}},\ and\ \bibinfo {author} {\bibfnamefont
  {Z.}~\bibnamefont {Leghtas}},\ }\bibfield  {title} {\bibinfo {title}
  {Exponential suppression of bit-flips in a qubit encoded in an oscillator},\
  }\href@noop {} {\bibfield  {journal} {\bibinfo  {journal} {Nature Physics}\
  }\textbf {\bibinfo {volume} {16}},\ \bibinfo {pages} {509} (\bibinfo {year}
  {2020})}\BibitemShut {NoStop}%
\bibitem [{\citenamefont {Siegele}\ \emph {et~al.}(2021)\citenamefont
  {Siegele}, \citenamefont {Mirrahimi},\ and\ \citenamefont
  {Campagne-Ibarcq}}]{siegele2021fault}%
  \BibitemOpen
  \bibfield  {author} {\bibinfo {author} {\bibfnamefont {C.}~\bibnamefont
  {Siegele}}, \bibinfo {author} {\bibfnamefont {M.}~\bibnamefont {Mirrahimi}},\
  and\ \bibinfo {author} {\bibfnamefont {P.}~\bibnamefont {Campagne-Ibarcq}},\
  }\bibfield  {title} {\bibinfo {title} {Fault-tolerant error syndrome
  detection in the gkp code},\ }\href@noop {} {\bibfield  {journal} {\bibinfo
  {journal} {Bulletin of the American Physical Society}\ } (\bibinfo {year}
  {2021})}\BibitemShut {NoStop}%
\bibitem [{\citenamefont {Puri}\ \emph {et~al.}(2017)\citenamefont {Puri},
  \citenamefont {Boutin},\ and\ \citenamefont {Blais}}]{Puri2017}%
  \BibitemOpen
  \bibfield  {author} {\bibinfo {author} {\bibfnamefont {S.}~\bibnamefont
  {Puri}}, \bibinfo {author} {\bibfnamefont {S.}~\bibnamefont {Boutin}},\ and\
  \bibinfo {author} {\bibfnamefont {A.}~\bibnamefont {Blais}},\ }\bibfield
  {title} {\bibinfo {title} {Engineering the quantum states of light in a
  kerr-nonlinear resonator by two-photon driving},\ }\href@noop {} {\bibfield
  {journal} {\bibinfo  {journal} {npj Quantum Information}\ }\textbf {\bibinfo
  {volume} {3}},\ \bibinfo {pages} {18} (\bibinfo {year} {2017})}\BibitemShut
  {NoStop}%
\bibitem [{\citenamefont {Woeginger}(2008)}]{woeginger2008}%
  \BibitemOpen
  \bibfield  {author} {\bibinfo {author} {\bibfnamefont {G.~J.}\ \bibnamefont
  {Woeginger}},\ }\bibfield  {title} {\bibinfo {title} {Open problems around
  exact algorithms},\ }\href@noop {} {\bibfield  {journal} {\bibinfo  {journal}
  {Discrete Applied Mathematics}\ }\textbf {\bibinfo {volume} {156}},\ \bibinfo
  {pages} {397} (\bibinfo {year} {2008})}\BibitemShut {NoStop}%
\bibitem [{\citenamefont {Puri}\ \emph {et~al.}(2020)\citenamefont {Puri},
  \citenamefont {St-Jean}, \citenamefont {Gross}, \citenamefont {Grimm},
  \citenamefont {Frattini}, \citenamefont {Iyer}, \citenamefont {Krishna},
  \citenamefont {Touzard}, \citenamefont {Jiang}, \citenamefont {Blais},
  \citenamefont {Flammia},\ and\ \citenamefont {Girvin}}]{Puri:2019aa}%
  \BibitemOpen
  \bibfield  {author} {\bibinfo {author} {\bibfnamefont {S.}~\bibnamefont
  {Puri}}, \bibinfo {author} {\bibfnamefont {L.}~\bibnamefont {St-Jean}},
  \bibinfo {author} {\bibfnamefont {J.~A.}\ \bibnamefont {Gross}}, \bibinfo
  {author} {\bibfnamefont {A.}~\bibnamefont {Grimm}}, \bibinfo {author}
  {\bibfnamefont {N.~E.}\ \bibnamefont {Frattini}}, \bibinfo {author}
  {\bibfnamefont {P.~S.}\ \bibnamefont {Iyer}}, \bibinfo {author}
  {\bibfnamefont {A.}~\bibnamefont {Krishna}}, \bibinfo {author} {\bibfnamefont
  {S.}~\bibnamefont {Touzard}}, \bibinfo {author} {\bibfnamefont
  {L.}~\bibnamefont {Jiang}}, \bibinfo {author} {\bibfnamefont
  {A.}~\bibnamefont {Blais}}, \bibinfo {author} {\bibfnamefont {S.~T.}\
  \bibnamefont {Flammia}},\ and\ \bibinfo {author} {\bibfnamefont {S.~M.}\
  \bibnamefont {Girvin}},\ }\href@noop {} {\bibfield  {journal} {\bibinfo
  {journal} {Science advances}\ }\textbf {\bibinfo {volume} {6}},\ \bibinfo
  {pages} {eaay5901} (\bibinfo {year} {2020})}\BibitemShut {NoStop}%
\bibitem [{\citenamefont {Chapman}\ \emph {et~al.}(2021)\citenamefont
  {Chapman}, \citenamefont {de~Graaf}, \citenamefont {Zhang}, \citenamefont
  {Mundhada}, \citenamefont {Frunzio}, \citenamefont {Girvin},\ and\
  \citenamefont {Schoelkopf}}]{chapman2021mediating}%
  \BibitemOpen
  \bibfield  {author} {\bibinfo {author} {\bibfnamefont {B.}~\bibnamefont
  {Chapman}}, \bibinfo {author} {\bibfnamefont {S.}~\bibnamefont {de~Graaf}},
  \bibinfo {author} {\bibfnamefont {Y.}~\bibnamefont {Zhang}}, \bibinfo
  {author} {\bibfnamefont {S.}~\bibnamefont {Mundhada}}, \bibinfo {author}
  {\bibfnamefont {L.}~\bibnamefont {Frunzio}}, \bibinfo {author} {\bibfnamefont
  {S.}~\bibnamefont {Girvin}},\ and\ \bibinfo {author} {\bibfnamefont
  {R.}~\bibnamefont {Schoelkopf}},\ }\bibfield  {title} {\bibinfo {title}
  {Mediating high-fidelity interactions between superconducting microwave
  cavities, part ii.},\ }\href@noop {} {\bibfield  {journal} {\bibinfo
  {journal} {Bulletin of the American Physical Society}\ } (\bibinfo {year}
  {2021})}\BibitemShut {NoStop}%
\bibitem [{\citenamefont {Steane}(1997)}]{Steane1997}%
  \BibitemOpen
  \bibfield  {author} {\bibinfo {author} {\bibfnamefont {A.~M.}\ \bibnamefont
  {Steane}},\ }\bibfield  {title} {\bibinfo {title} {Active stabilization,
  quantum computation, and quantum state synthesis},\ }\href@noop {} {\bibfield
   {journal} {\bibinfo  {journal} {Phys. Rev. Lett.}\ }\textbf {\bibinfo
  {volume} {78}},\ \bibinfo {pages} {2252} (\bibinfo {year}
  {1997})}\BibitemShut {NoStop}%
\bibitem [{\citenamefont {Knill}(2005)}]{Knill05}%
  \BibitemOpen
  \bibfield  {author} {\bibinfo {author} {\bibfnamefont {E.}~\bibnamefont
  {Knill}},\ }\bibfield  {title} {\bibinfo {title} {Quantum computing with
  realistically noisy devices},\ }\href {http://dx.doi.org/10.1038/nature03350}
  {\bibfield  {journal} {\bibinfo  {journal} {Nature}\ }\textbf {\bibinfo
  {volume} {434}},\ \bibinfo {pages} {39} (\bibinfo {year} {2005})}\BibitemShut
  {NoStop}%
\bibitem [{\citenamefont {Dennis}\ \emph {et~al.}(2002)\citenamefont {Dennis},
  \citenamefont {Kitaev}, \citenamefont {Landahl},\ and\ \citenamefont
  {Preskill}}]{Dennis02}%
  \BibitemOpen
  \bibfield  {author} {\bibinfo {author} {\bibfnamefont {E.}~\bibnamefont
  {Dennis}}, \bibinfo {author} {\bibfnamefont {A.}~\bibnamefont {Kitaev}},
  \bibinfo {author} {\bibfnamefont {A.}~\bibnamefont {Landahl}},\ and\ \bibinfo
  {author} {\bibfnamefont {J.}~\bibnamefont {Preskill}},\ }\bibfield  {title}
  {\bibinfo {title} {Topological quantum memory},\ }\href
  {https://doi.org/10.1063/1.1499754} {\bibfield  {journal} {\bibinfo
  {journal} {J. Math. Phys. (N.Y.)}\ }\textbf {\bibinfo {volume} {43}},\
  \bibinfo {pages} {4452} (\bibinfo {year} {2002})},\ \Eprint
  {https://arxiv.org/abs/quant-ph/0110143} {arXiv:quant-ph/0110143}
  \BibitemShut {NoStop}%
\bibitem [{\citenamefont {Fukui}\ \emph {et~al.}(2017)\citenamefont {Fukui},
  \citenamefont {Tomita},\ and\ \citenamefont {Okamoto}}]{Fukui:2017aa}%
  \BibitemOpen
  \bibfield  {author} {\bibinfo {author} {\bibfnamefont {K.}~\bibnamefont
  {Fukui}}, \bibinfo {author} {\bibfnamefont {A.}~\bibnamefont {Tomita}},\ and\
  \bibinfo {author} {\bibfnamefont {A.}~\bibnamefont {Okamoto}},\ }\bibfield
  {title} {\bibinfo {title} {Analog quantum error correction with encoding a
  qubit into an oscillator},\ }\href
  {https://doi.org/10.1103/PhysRevLett.119.180507} {\bibfield  {journal}
  {\bibinfo  {journal} {Phys. Rev. Lett.}\ }\textbf {\bibinfo {volume} {119}},\
  \bibinfo {pages} {180507} (\bibinfo {year} {2017})}\BibitemShut {NoStop}%
\bibitem [{\citenamefont {Fukui}\ \emph
  {et~al.}(2018{\natexlab{a}})\citenamefont {Fukui}, \citenamefont {Tomita},
  \citenamefont {Okamoto},\ and\ \citenamefont {Fujii}}]{Fukui2018}%
  \BibitemOpen
  \bibfield  {author} {\bibinfo {author} {\bibfnamefont {K.}~\bibnamefont
  {Fukui}}, \bibinfo {author} {\bibfnamefont {A.}~\bibnamefont {Tomita}},
  \bibinfo {author} {\bibfnamefont {A.}~\bibnamefont {Okamoto}},\ and\ \bibinfo
  {author} {\bibfnamefont {K.}~\bibnamefont {Fujii}},\ }\bibfield  {title}
  {\bibinfo {title} {High-threshold fault-tolerant quantum computation with
  analog quantum error correction},\ }\href
  {https://doi.org/10.1103/PhysRevX.8.021054} {\bibfield  {journal} {\bibinfo
  {journal} {Phys. Rev. X}\ }\textbf {\bibinfo {volume} {8}},\ \bibinfo {pages}
  {021054} (\bibinfo {year} {2018}{\natexlab{a}})}\BibitemShut {NoStop}%
\bibitem [{\citenamefont {Fukui}\ \emph
  {et~al.}(2018{\natexlab{b}})\citenamefont {Fukui}, \citenamefont {Tomita},\
  and\ \citenamefont {Okamoto}}]{Fukui:2018aa}%
  \BibitemOpen
  \bibfield  {author} {\bibinfo {author} {\bibfnamefont {K.}~\bibnamefont
  {Fukui}}, \bibinfo {author} {\bibfnamefont {A.}~\bibnamefont {Tomita}},\ and\
  \bibinfo {author} {\bibfnamefont {A.}~\bibnamefont {Okamoto}},\ }\bibfield
  {title} {\bibinfo {title} {Tracking quantum error correction},\ }\href
  {https://doi.org/10.1103/PhysRevA.98.022326} {\bibfield  {journal} {\bibinfo
  {journal} {Phys. Rev. A}\ }\textbf {\bibinfo {volume} {98}},\ \bibinfo
  {pages} {022326} (\bibinfo {year} {2018}{\natexlab{b}})}\BibitemShut
  {NoStop}%
\bibitem [{\citenamefont {Fukui}(2019)}]{Fukui:2019aa}%
  \BibitemOpen
  \bibfield  {author} {\bibinfo {author} {\bibfnamefont {K.}~\bibnamefont
  {Fukui}},\ }\bibfield  {title} {\bibinfo {title} {High-threshold
  fault-tolerant quantum computation with the gkp qubit and realistically noisy
  devices},\ }\href {https://arxiv.org/abs/1906.09767} {\bibfield  {journal}
  {\bibinfo  {journal} {arXiv:1906.09767}\ } (\bibinfo {year}
  {2019})}\BibitemShut {NoStop}%
\bibitem [{\citenamefont {Bombin}\ and\ \citenamefont
  {Martin-Delgado}(2007)}]{bombin_optimal_2007}%
  \BibitemOpen
  \bibfield  {author} {\bibinfo {author} {\bibfnamefont {H.}~\bibnamefont
  {Bombin}}\ and\ \bibinfo {author} {\bibfnamefont {M.~A.}\ \bibnamefont
  {Martin-Delgado}},\ }\bibfield  {title} {\bibinfo {title} {Optimal resources
  for topological two-dimensional stabilizer codes: {Comparative} study},\
  }\href {https://doi.org/10.1103/PhysRevA.76.012305} {\bibfield  {journal}
  {\bibinfo  {journal} {Phys. Rev. A}\ }\textbf {\bibinfo {volume} {76}},\
  \bibinfo {pages} {012305} (\bibinfo {year} {2007})},\ \Eprint
  {https://arxiv.org/abs/quant-ph/0703272} {arXiv:quant-ph/0703272}
  \BibitemShut {NoStop}%
\bibitem [{\citenamefont {Conrad}(2021)}]{Conrad2021}%
  \BibitemOpen
  \bibfield  {author} {\bibinfo {author} {\bibfnamefont {J.}~\bibnamefont
  {Conrad}},\ }\bibfield  {title} {\bibinfo {title} {Twirling and hamiltonian
  engineering via dynamical decoupling for gottesman-kitaev-preskill quantum
  computing},\ }\href {https://doi.org/10.1103/PhysRevA.103.022404} {\bibfield
  {journal} {\bibinfo  {journal} {Phys. Rev. A}\ }\textbf {\bibinfo {volume}
  {103}},\ \bibinfo {pages} {022404} (\bibinfo {year} {2021})}\BibitemShut
  {NoStop}%
\bibitem [{\citenamefont {Tuckett}\ \emph {et~al.}(2018)\citenamefont
  {Tuckett}, \citenamefont {Bartlett},\ and\ \citenamefont
  {Flammia}}]{tuckett2018ultrahigh}%
  \BibitemOpen
  \bibfield  {author} {\bibinfo {author} {\bibfnamefont {D.~K.}\ \bibnamefont
  {Tuckett}}, \bibinfo {author} {\bibfnamefont {S.~D.}\ \bibnamefont
  {Bartlett}},\ and\ \bibinfo {author} {\bibfnamefont {S.~T.}\ \bibnamefont
  {Flammia}},\ }\bibfield  {title} {\bibinfo {title} {Ultrahigh error threshold
  for surface codes with biased noise},\ }\href@noop {} {\bibfield  {journal}
  {\bibinfo  {journal} {Physical review letters}\ }\textbf {\bibinfo {volume}
  {120}},\ \bibinfo {pages} {050505} (\bibinfo {year} {2018})}\BibitemShut
  {NoStop}%
\bibitem [{\citenamefont {Tuckett}\ \emph {et~al.}(2020)\citenamefont
  {Tuckett}, \citenamefont {Bartlett}, \citenamefont {Flammia},\ and\
  \citenamefont {Brown}}]{tuckett2020fault}%
  \BibitemOpen
  \bibfield  {author} {\bibinfo {author} {\bibfnamefont {D.~K.}\ \bibnamefont
  {Tuckett}}, \bibinfo {author} {\bibfnamefont {S.~D.}\ \bibnamefont
  {Bartlett}}, \bibinfo {author} {\bibfnamefont {S.~T.}\ \bibnamefont
  {Flammia}},\ and\ \bibinfo {author} {\bibfnamefont {B.~J.}\ \bibnamefont
  {Brown}},\ }\bibfield  {title} {\bibinfo {title} {Fault-tolerant thresholds
  for the surface code in excess of 5\% under biased noise},\ }\href@noop {}
  {\bibfield  {journal} {\bibinfo  {journal} {Physical review letters}\
  }\textbf {\bibinfo {volume} {124}},\ \bibinfo {pages} {130501} (\bibinfo
  {year} {2020})}\BibitemShut {NoStop}%
\bibitem [{\citenamefont {Ataides}\ \emph {et~al.}(2021)\citenamefont
  {Ataides}, \citenamefont {Tuckett}, \citenamefont {Bartlett}, \citenamefont
  {Flammia},\ and\ \citenamefont {Brown}}]{bonilla2020xzzx}%
  \BibitemOpen
  \bibfield  {author} {\bibinfo {author} {\bibfnamefont {J.~P.~B.}\
  \bibnamefont {Ataides}}, \bibinfo {author} {\bibfnamefont {D.~K.}\
  \bibnamefont {Tuckett}}, \bibinfo {author} {\bibfnamefont {S.~D.}\
  \bibnamefont {Bartlett}}, \bibinfo {author} {\bibfnamefont {S.~T.}\
  \bibnamefont {Flammia}},\ and\ \bibinfo {author} {\bibfnamefont {B.~J.}\
  \bibnamefont {Brown}},\ }\bibfield  {title} {\bibinfo {title} {The xzzx
  surface code},\ }\href@noop {} {\bibfield  {journal} {\bibinfo  {journal}
  {Nature Communications}\ }\textbf {\bibinfo {volume} {12}},\ \bibinfo {pages}
  {1} (\bibinfo {year} {2021})}\BibitemShut {NoStop}%
\bibitem [{\citenamefont {H{\"a}nggli}\ \emph {et~al.}(2020)\citenamefont
  {H{\"a}nggli}, \citenamefont {Heinze},\ and\ \citenamefont
  {K{\"o}nig}}]{hanggli2020enhanced}%
  \BibitemOpen
  \bibfield  {author} {\bibinfo {author} {\bibfnamefont {L.}~\bibnamefont
  {H{\"a}nggli}}, \bibinfo {author} {\bibfnamefont {M.}~\bibnamefont
  {Heinze}},\ and\ \bibinfo {author} {\bibfnamefont {R.}~\bibnamefont
  {K{\"o}nig}},\ }\bibfield  {title} {\bibinfo {title} {Enhanced noise
  resilience of the surface--gottesman-kitaev-preskill code via designed
  bias},\ }\href@noop {} {\bibfield  {journal} {\bibinfo  {journal} {Physical
  Review A}\ }\textbf {\bibinfo {volume} {102}},\ \bibinfo {pages} {052408}
  (\bibinfo {year} {2020})}\BibitemShut {NoStop}%
\bibitem [{\citenamefont {Hatridge}\ \emph {et~al.}(2013)\citenamefont
  {Hatridge}, \citenamefont {Shankar}, \citenamefont {Mirrahimi}, \citenamefont
  {Schackert}, \citenamefont {Geerlings}, \citenamefont {Brecht}, \citenamefont
  {Sliwa}, \citenamefont {Abdo}, \citenamefont {Frunzio}, \citenamefont
  {Girvin} \emph {et~al.}}]{hatridge2013quantum}%
  \BibitemOpen
  \bibfield  {author} {\bibinfo {author} {\bibfnamefont {M.}~\bibnamefont
  {Hatridge}}, \bibinfo {author} {\bibfnamefont {S.}~\bibnamefont {Shankar}},
  \bibinfo {author} {\bibfnamefont {M.}~\bibnamefont {Mirrahimi}}, \bibinfo
  {author} {\bibfnamefont {F.}~\bibnamefont {Schackert}}, \bibinfo {author}
  {\bibfnamefont {K.}~\bibnamefont {Geerlings}}, \bibinfo {author}
  {\bibfnamefont {T.}~\bibnamefont {Brecht}}, \bibinfo {author} {\bibfnamefont
  {K.}~\bibnamefont {Sliwa}}, \bibinfo {author} {\bibfnamefont
  {B.}~\bibnamefont {Abdo}}, \bibinfo {author} {\bibfnamefont {L.}~\bibnamefont
  {Frunzio}}, \bibinfo {author} {\bibfnamefont {S.~M.}\ \bibnamefont {Girvin}},
  \emph {et~al.},\ }\bibfield  {title} {\bibinfo {title} {Quantum back-action
  of an individual variable-strength measurement},\ }\href@noop {} {\bibfield
  {journal} {\bibinfo  {journal} {Science}\ }\textbf {\bibinfo {volume}
  {339}},\ \bibinfo {pages} {178} (\bibinfo {year} {2013})}\BibitemShut
  {NoStop}%
\bibitem [{\citenamefont {Walter}\ \emph {et~al.}(2017)\citenamefont {Walter},
  \citenamefont {Kurpiers}, \citenamefont {Gasparinetti}, \citenamefont
  {Magnard}, \citenamefont {Poto{\v{c}}nik}, \citenamefont {Salath{\'e}},
  \citenamefont {Pechal}, \citenamefont {Mondal}, \citenamefont {Oppliger},
  \citenamefont {Eichler} \emph {et~al.}}]{walter2017rapid}%
  \BibitemOpen
  \bibfield  {author} {\bibinfo {author} {\bibfnamefont {T.}~\bibnamefont
  {Walter}}, \bibinfo {author} {\bibfnamefont {P.}~\bibnamefont {Kurpiers}},
  \bibinfo {author} {\bibfnamefont {S.}~\bibnamefont {Gasparinetti}}, \bibinfo
  {author} {\bibfnamefont {P.}~\bibnamefont {Magnard}}, \bibinfo {author}
  {\bibfnamefont {A.}~\bibnamefont {Poto{\v{c}}nik}}, \bibinfo {author}
  {\bibfnamefont {Y.}~\bibnamefont {Salath{\'e}}}, \bibinfo {author}
  {\bibfnamefont {M.}~\bibnamefont {Pechal}}, \bibinfo {author} {\bibfnamefont
  {M.}~\bibnamefont {Mondal}}, \bibinfo {author} {\bibfnamefont
  {M.}~\bibnamefont {Oppliger}}, \bibinfo {author} {\bibfnamefont
  {C.}~\bibnamefont {Eichler}}, \emph {et~al.},\ }\bibfield  {title} {\bibinfo
  {title} {Rapid high-fidelity single-shot dispersive readout of
  superconducting qubits},\ }\href@noop {} {\bibfield  {journal} {\bibinfo
  {journal} {Physical Review Applied}\ }\textbf {\bibinfo {volume} {7}},\
  \bibinfo {pages} {054020} (\bibinfo {year} {2017})}\BibitemShut {NoStop}%
\bibitem [{\citenamefont {Horsman}\ \emph {et~al.}(2012)\citenamefont
  {Horsman}, \citenamefont {Fowler}, \citenamefont {Devitt},\ and\
  \citenamefont {Van~Meter}}]{horsman2012surface}%
  \BibitemOpen
  \bibfield  {author} {\bibinfo {author} {\bibfnamefont {C.}~\bibnamefont
  {Horsman}}, \bibinfo {author} {\bibfnamefont {A.~G.}\ \bibnamefont {Fowler}},
  \bibinfo {author} {\bibfnamefont {S.}~\bibnamefont {Devitt}},\ and\ \bibinfo
  {author} {\bibfnamefont {R.}~\bibnamefont {Van~Meter}},\ }\bibfield  {title}
  {\bibinfo {title} {Surface code quantum computing by lattice surgery},\
  }\href@noop {} {\bibfield  {journal} {\bibinfo  {journal} {New Journal of
  Physics}\ }\textbf {\bibinfo {volume} {14}},\ \bibinfo {pages} {123011}
  (\bibinfo {year} {2012})}\BibitemShut {NoStop}%
\bibitem [{\citenamefont {Litinski}\ and\ \citenamefont {von
  Oppen}(2018)}]{litinski2018lattice}%
  \BibitemOpen
  \bibfield  {author} {\bibinfo {author} {\bibfnamefont {D.}~\bibnamefont
  {Litinski}}\ and\ \bibinfo {author} {\bibfnamefont {F.}~\bibnamefont {von
  Oppen}},\ }\bibfield  {title} {\bibinfo {title} {Lattice surgery with a
  twist: Simplifying clifford gates of surface codes},\ }\href@noop {}
  {\bibfield  {journal} {\bibinfo  {journal} {Quantum}\ }\textbf {\bibinfo
  {volume} {2}},\ \bibinfo {pages} {62} (\bibinfo {year} {2018})}\BibitemShut
  {NoStop}%
\bibitem [{\citenamefont {Bravyi}\ and\ \citenamefont
  {Kitaev}(2005)}]{bravyi2005universal}%
  \BibitemOpen
  \bibfield  {author} {\bibinfo {author} {\bibfnamefont {S.}~\bibnamefont
  {Bravyi}}\ and\ \bibinfo {author} {\bibfnamefont {A.}~\bibnamefont
  {Kitaev}},\ }\bibfield  {title} {\bibinfo {title} {Universal quantum
  computation with ideal clifford gates and noisy ancillas},\ }\href@noop {}
  {\bibfield  {journal} {\bibinfo  {journal} {Physical Review A}\ }\textbf
  {\bibinfo {volume} {71}},\ \bibinfo {pages} {022316} (\bibinfo {year}
  {2005})}\BibitemShut {NoStop}%
\bibitem [{\citenamefont {Litinski}(2019)}]{litinski2019game}%
  \BibitemOpen
  \bibfield  {author} {\bibinfo {author} {\bibfnamefont {D.}~\bibnamefont
  {Litinski}},\ }\bibfield  {title} {\bibinfo {title} {A game of surface codes:
  Large-scale quantum computing with lattice surgery},\ }\href@noop {}
  {\bibfield  {journal} {\bibinfo  {journal} {Quantum}\ }\textbf {\bibinfo
  {volume} {3}},\ \bibinfo {pages} {128} (\bibinfo {year} {2019})}\BibitemShut
  {NoStop}%
\bibitem [{\citenamefont {Baragiola}\ \emph {et~al.}(2019)\citenamefont
  {Baragiola}, \citenamefont {Pantaleoni}, \citenamefont {Alexander},
  \citenamefont {Karanjai},\ and\ \citenamefont
  {Menicucci}}]{Baragiola:2019aa}%
  \BibitemOpen
  \bibfield  {author} {\bibinfo {author} {\bibfnamefont {B.~Q.}\ \bibnamefont
  {Baragiola}}, \bibinfo {author} {\bibfnamefont {G.}~\bibnamefont
  {Pantaleoni}}, \bibinfo {author} {\bibfnamefont {R.~N.}\ \bibnamefont
  {Alexander}}, \bibinfo {author} {\bibfnamefont {A.}~\bibnamefont
  {Karanjai}},\ and\ \bibinfo {author} {\bibfnamefont {N.~C.}\ \bibnamefont
  {Menicucci}},\ }\bibfield  {title} {\bibinfo {title} {All-gaussian
  universality and fault tolerance with the gottesman-kitaev-preskill code},\
  }\href {https://arxiv.org/abs/1903.00012} {\bibfield  {journal} {\bibinfo
  {journal} {arXiv:1903.00012}\ } (\bibinfo {year} {2019})}\BibitemShut
  {NoStop}%
\bibitem [{\citenamefont {Albert}\ \emph {et~al.}(2020)\citenamefont {Albert},
  \citenamefont {Covey},\ and\ \citenamefont {Preskill}}]{albert2020robust}%
  \BibitemOpen
  \bibfield  {author} {\bibinfo {author} {\bibfnamefont {V.~V.}\ \bibnamefont
  {Albert}}, \bibinfo {author} {\bibfnamefont {J.~P.}\ \bibnamefont {Covey}},\
  and\ \bibinfo {author} {\bibfnamefont {J.}~\bibnamefont {Preskill}},\
  }\bibfield  {title} {\bibinfo {title} {Robust encoding of a qubit in a
  molecule},\ }\href
  {https://doi.org/https://doi.org/10.1103/PhysRevX.10.031050} {\bibfield
  {journal} {\bibinfo  {journal} {Physical Review X}\ }\textbf {\bibinfo
  {volume} {10}},\ \bibinfo {pages} {031050} (\bibinfo {year}
  {2020})}\BibitemShut {NoStop}%
\bibitem [{\citenamefont {Gross}(2020)}]{gross2020encoding}%
  \BibitemOpen
  \bibfield  {author} {\bibinfo {author} {\bibfnamefont {J.~A.}\ \bibnamefont
  {Gross}},\ }\bibfield  {title} {\bibinfo {title} {Encoding a qubit in a
  spin},\ }\href@noop {} {\bibfield  {journal} {\bibinfo  {journal} {arXiv
  preprint arXiv:2005.10910}\ } (\bibinfo {year} {2020})}\BibitemShut {NoStop}%
\bibitem [{\citenamefont {Guillaud}\ and\ \citenamefont
  {Mirrahimi}(2019)}]{guillaud_repetition_2019}%
  \BibitemOpen
  \bibfield  {author} {\bibinfo {author} {\bibfnamefont {J.}~\bibnamefont
  {Guillaud}}\ and\ \bibinfo {author} {\bibfnamefont {M.}~\bibnamefont
  {Mirrahimi}},\ }\bibfield  {title} {\bibinfo {title} {Repetition {Cat}
  {Qubits} for {Fault}-{Tolerant} {Quantum} {Computation}},\ }\href
  {https://doi.org/10.1103/PhysRevX.9.041053} {\bibfield  {journal} {\bibinfo
  {journal} {Phys. Rev. X}\ }\textbf {\bibinfo {volume} {9}},\ \bibinfo {pages}
  {041053} (\bibinfo {year} {2019})},\ \bibinfo {note} {publisher: American
  Physical Society}\BibitemShut {NoStop}%
\bibitem [{\citenamefont {Chamberland}\ \emph {et~al.}(2020)\citenamefont
  {Chamberland}, \citenamefont {Noh}, \citenamefont {Arrangoiz-Arriola},
  \citenamefont {Campbell}, \citenamefont {Hann}, \citenamefont {Iverson},
  \citenamefont {Putterman}, \citenamefont {Bohdanowicz}, \citenamefont
  {Flammia}, \citenamefont {Keller}, \citenamefont {Refael}, \citenamefont
  {Preskill}, \citenamefont {Jiang}, \citenamefont {Safavi-Naeini},
  \citenamefont {Painter},\ and\ \citenamefont
  {Brandão}}]{chamberland2020building}%
  \BibitemOpen
  \bibfield  {author} {\bibinfo {author} {\bibfnamefont {C.}~\bibnamefont
  {Chamberland}}, \bibinfo {author} {\bibfnamefont {K.}~\bibnamefont {Noh}},
  \bibinfo {author} {\bibfnamefont {P.}~\bibnamefont {Arrangoiz-Arriola}},
  \bibinfo {author} {\bibfnamefont {E.~T.}\ \bibnamefont {Campbell}}, \bibinfo
  {author} {\bibfnamefont {C.~T.}\ \bibnamefont {Hann}}, \bibinfo {author}
  {\bibfnamefont {J.}~\bibnamefont {Iverson}}, \bibinfo {author} {\bibfnamefont
  {H.}~\bibnamefont {Putterman}}, \bibinfo {author} {\bibfnamefont {T.~C.}\
  \bibnamefont {Bohdanowicz}}, \bibinfo {author} {\bibfnamefont {S.~T.}\
  \bibnamefont {Flammia}}, \bibinfo {author} {\bibfnamefont {A.}~\bibnamefont
  {Keller}}, \bibinfo {author} {\bibfnamefont {G.}~\bibnamefont {Refael}},
  \bibinfo {author} {\bibfnamefont {J.}~\bibnamefont {Preskill}}, \bibinfo
  {author} {\bibfnamefont {L.}~\bibnamefont {Jiang}}, \bibinfo {author}
  {\bibfnamefont {A.~H.}\ \bibnamefont {Safavi-Naeini}}, \bibinfo {author}
  {\bibfnamefont {O.}~\bibnamefont {Painter}},\ and\ \bibinfo {author}
  {\bibfnamefont {F.~G. S.~L.}\ \bibnamefont {Brandão}},\ }\href@noop {}
  {\bibinfo {title} {Building a fault-tolerant quantum computer using
  concatenated cat codes}} (\bibinfo {year} {2020}),\ \Eprint
  {https://arxiv.org/abs/2012.04108} {arXiv:2012.04108 [quant-ph]} \BibitemShut
  {NoStop}%
\bibitem [{\citenamefont {Darmawan}\ \emph {et~al.}(2021)\citenamefont
  {Darmawan}, \citenamefont {Brown}, \citenamefont {Grimsmo}, \citenamefont
  {Tuckett},\ and\ \citenamefont {Puri}}]{Darmawan2021}%
  \BibitemOpen
  \bibfield  {author} {\bibinfo {author} {\bibfnamefont {A.~S.}\ \bibnamefont
  {Darmawan}}, \bibinfo {author} {\bibfnamefont {B.~J.}\ \bibnamefont {Brown}},
  \bibinfo {author} {\bibfnamefont {A.~L.}\ \bibnamefont {Grimsmo}}, \bibinfo
  {author} {\bibfnamefont {D.~K.}\ \bibnamefont {Tuckett}},\ and\ \bibinfo
  {author} {\bibfnamefont {S.}~\bibnamefont {Puri}},\ }\bibfield  {title}
  {\bibinfo {title} {Practical quantum error correction with the xzzx code and
  kerr-cat qubits},\ }\href@noop {} {\bibfield  {journal} {\bibinfo  {journal}
  {arXiv:2104.09539}\ } (\bibinfo {year} {2021})}\BibitemShut {NoStop}%
\bibitem [{\citenamefont {Le}\ \emph {et~al.}(2019)\citenamefont {Le},
  \citenamefont {Grimsmo}, \citenamefont {M{\"u}ller},\ and\ \citenamefont
  {Stace}}]{le2019doubly}%
  \BibitemOpen
  \bibfield  {author} {\bibinfo {author} {\bibfnamefont {D.~T.}\ \bibnamefont
  {Le}}, \bibinfo {author} {\bibfnamefont {A.}~\bibnamefont {Grimsmo}},
  \bibinfo {author} {\bibfnamefont {C.}~\bibnamefont {M{\"u}ller}},\ and\
  \bibinfo {author} {\bibfnamefont {T.}~\bibnamefont {Stace}},\ }\bibfield
  {title} {\bibinfo {title} {Doubly nonlinear superconducting qubit},\
  }\href@noop {} {\bibfield  {journal} {\bibinfo  {journal} {Physical Review
  A}\ }\textbf {\bibinfo {volume} {100}},\ \bibinfo {pages} {062321} (\bibinfo
  {year} {2019})}\BibitemShut {NoStop}%
\bibitem [{\citenamefont {Rymarz}\ \emph {et~al.}(2021)\citenamefont {Rymarz},
  \citenamefont {Bosco}, \citenamefont {Ciani},\ and\ \citenamefont
  {DiVincenzo}}]{rymarz2021hardware}%
  \BibitemOpen
  \bibfield  {author} {\bibinfo {author} {\bibfnamefont {M.}~\bibnamefont
  {Rymarz}}, \bibinfo {author} {\bibfnamefont {S.}~\bibnamefont {Bosco}},
  \bibinfo {author} {\bibfnamefont {A.}~\bibnamefont {Ciani}},\ and\ \bibinfo
  {author} {\bibfnamefont {D.~P.}\ \bibnamefont {DiVincenzo}},\ }\bibfield
  {title} {\bibinfo {title} {Hardware-encoding grid states in a nonreciprocal
  superconducting circuit},\ }\href@noop {} {\bibfield  {journal} {\bibinfo
  {journal} {Physical Review X}\ }\textbf {\bibinfo {volume} {11}},\ \bibinfo
  {pages} {011032} (\bibinfo {year} {2021})}\BibitemShut {NoStop}%
\bibitem [{\citenamefont {Landahl}\ and\ \citenamefont
  {Ryan-Anderson}(2014)}]{landahl2014quantum}%
  \BibitemOpen
  \bibfield  {author} {\bibinfo {author} {\bibfnamefont {A.~J.}\ \bibnamefont
  {Landahl}}\ and\ \bibinfo {author} {\bibfnamefont {C.}~\bibnamefont
  {Ryan-Anderson}},\ }\bibfield  {title} {\bibinfo {title} {Quantum computing
  by color-code lattice surgery},\ }\href@noop {} {\bibfield  {journal}
  {\bibinfo  {journal} {arXiv:1407.5103}\ } (\bibinfo {year}
  {2014})}\BibitemShut {NoStop}%
\bibitem [{\citenamefont {Harrington}\ and\ \citenamefont
  {Preskill}(2001)}]{Harrington01}%
  \BibitemOpen
  \bibfield  {author} {\bibinfo {author} {\bibfnamefont {J.}~\bibnamefont
  {Harrington}}\ and\ \bibinfo {author} {\bibfnamefont {J.}~\bibnamefont
  {Preskill}},\ }\bibfield  {title} {\bibinfo {title} {Achievable rates for the
  gaussian quantum channel},\ }\href
  {https://doi.org/10.1103/PhysRevA.64.062301} {\bibfield  {journal} {\bibinfo
  {journal} {Phys. Rev. A}\ }\textbf {\bibinfo {volume} {64}},\ \bibinfo
  {pages} {062301} (\bibinfo {year} {2001})}\BibitemShut {NoStop}%
\bibitem [{\citenamefont {Noh}\ \emph {et~al.}(2019)\citenamefont {Noh},
  \citenamefont {Albert},\ and\ \citenamefont {Jiang}}]{Noh2019capacity}%
  \BibitemOpen
  \bibfield  {author} {\bibinfo {author} {\bibfnamefont {K.}~\bibnamefont
  {Noh}}, \bibinfo {author} {\bibfnamefont {V.~V.}\ \bibnamefont {Albert}},\
  and\ \bibinfo {author} {\bibfnamefont {L.}~\bibnamefont {Jiang}},\ }\bibfield
   {title} {\bibinfo {title} {Quantum capacity bounds of gaussian thermal loss
  channels and achievable rates with gottesman-kitaev-preskill codes},\ }\href
  {https://doi.org/10.1109/TIT.2018.2873764} {\bibfield  {journal} {\bibinfo
  {journal} {IEEE Transactions on Information Theory}\ }\textbf {\bibinfo
  {volume} {65}},\ \bibinfo {pages} {2563} (\bibinfo {year}
  {2019})}\BibitemShut {NoStop}%
\end{thebibliography}%

\end{document}